\documentclass[twocolumn]{29onr_art}
\usepackage{natbib}
\usepackage{ulem}
\usepackage{graphicx}
\usepackage{fixltx2e}
\usepackage{amsmath,float}
\usepackage{graphicx}
\usepackage{amsmath,amssymb}
\usepackage[english]{babel}
\usepackage{epsfig}
\usepackage{color}
\usepackage{epstopdf}
\usepackage{dblfloatfix}
\usepackage{hyperref}
\usepackage{overpic}

\newcount\ndots
\def\drwln#1#2{\raise 2.5pt\vbox{\hrule width #1pt height #2pt}}

\def\square   {${\vcenter{\hrule height .4pt
			\hbox{\vrule width .4pt height 3pt \kern 3pt
				\vrule width .4pt}
			\hrule height .4pt}}$\nobreak\ }

\def\diam     {$\diamond$\nobreak\ }

\def\filsqr   {${\vcenter{\hrule height 2pt
			\hbox{\vrule width 2.2pt height 0.2pt \kern 0.1pt
				\vrule width 2.2pt}
			\hrule height 2.2pt}}$\nobreak\ }
                          
\definecolor{docgreen}{rgb}{0,.5,0}

\pdfoutput = 1

\begin{document}
	

	\title{Air Entrainment and Surface Fluctuations in a Turbulent Ship Hull Boundary Layer}
	
	\author{
		Naeem Masnadi$^{1a}$, Martin A. Erinin$^1$, Nathan Washuta$^{1b}$, Farshad Nasiri$^2$, \\Elias Balaras$^2$ and  James H. Duncan$^1$,}
	
	\affiliation{( 
		$^1$University of Maryland, College Park, Maryland, USA.\\
		$^2$The George Washington University, DC, USA\\
		$^a$Present address: University of Washington, Seattle, Washington\\
		$^b$Present address: The Citadel, Charleston, South Carolina)}
	\maketitle
	
	\begin{abstract}
		The air entrainment due to the turbulence in a free surface boundary layer shear flow created by a horizontally moving vertical  surface-piercing wall is studied through experiments and direct numerical simulations. 
		
		In the experiments, a laboratory-scale device was built that utilizes a surface-piercing stainless steel belt that travels in a loop around two vertical rollers, with one length of the belt between the rollers acting as a horizontally-moving flat wall. The belt is accelerated suddenly from rest until reaching constant speed in order to create a temporally-evolving boundary layer analogous to the spatially-evolving boundary layer that would exist along a surface-piercing towed flat plate. To complement the experiments, Direct Numerical Simulations (DNS) of the two-phase boundary layer problem were carried out with the domain including a streamwise belt section simulated with  periodic boundary conditions. Cinematic Laser-Induced Fluorescence (LIF) measurements of water surface profiles  in two vertical planes oriented parallel to the belt surface (wall-parallel profiles) are presented and compared to previous measurements of profiles in  a vertical plane oriented normal to the belt surface (wall-normal profiles).  Additionally, photographic observations of air entrainment and measurements of air bubble size distributions and motions are reported herein. The bubble entrainment mechanisms are studied in detail through the results obtained by the DNS simulations. Free surface features resembling breaking waves and traveling parallel to the belt are observed in the wall-parallel LIF movies. These free surface features travel up to 3 times faster than the free surface features moving away from the belt in the wall-normal LIF movies. These breaking events are thought to be one of the mechanisms by which the air is entrained into the underlying flow. The bubble size distribution is found to have a characteristic break in slope, similar to the Hinze scale previously observed in breaking waves \cite[]{Deane2002}. The number of bubbles, their velocity, and size are reported versus depth from the experimental data. These results are qualitatively similar to results obtained by the simulations. Finally, several entrainment mechanisms are found in the simulations and their prevalence in the free surface boundary layer is assessed.
		
	\end{abstract}
	
	\section{INTRODUCTION}
	
\begin{figure}[!htb]
		\begin{center}
			\includegraphics[trim=0 0.1in 0 0.1in, clip=true,width=3in]{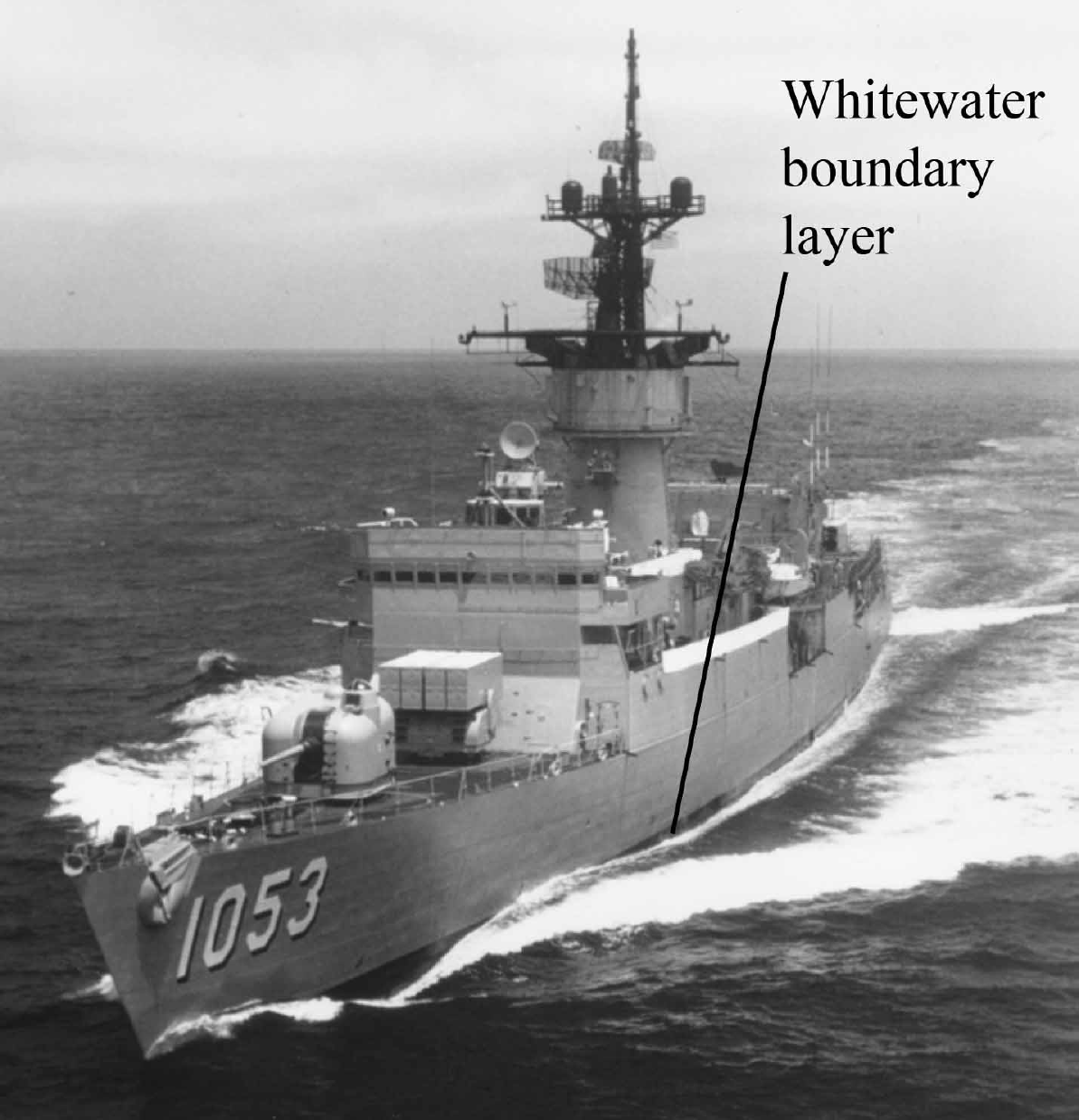}
		\end{center}
		\vspace*{-0.17in} \caption{Photograph of a naval combatant ship showing zone of white water next to the hull.} \label{fig:ship}
	\end{figure}

	Turbulent boundary layers near the free surface along ship hulls and surface-piercing flat plates have been explored by a number of authors, see for example \cite{Stern1989}, \cite{Longo1998}, \cite{Sreedhar1998},   and \cite{Stern1993}.  Air entrainment and bubble distributions in these free surface flows have been explored by 
	\cite{CarricaEtAl1999}, \cite{MoragaEtAl2008}, 
	\cite{PerretCarrica2015}, \cite{CastroEtAl2016} and \cite{LiEtAl2016}. 
	One obvious region of two-phase flow in the vicinity of  a ship is the 
	layer of white water next to the hull, see for example the photograph in Figure~\ref{fig:ship}.  The mechanisms by which air enters this region of the flow is poorly understood.  
	In particular, it is not known to what degree this white water is the result of active spray generation and air entrainment due to turbulence in the boundary layer along the ship hull and to what degree the white water is the result of spray and air bubbles that are generated upstream in the breaking bow wave and then swept downstream with the flow. In the free surface boundary layer, the air entrainment process is controlled by the ratios of the turbulent kinetic energy to the gravitational potential energy and the turbulent kinetic energy to the surface tension energy. The ratio of the turbulent kinetic energy to the gravitational potential energy is given by the square of the turbulent Froude number ($Fr^2 = q^2 / (g L)$) and the ratio of turbulent kinetic energy to surface tension energy is given by the Weber number ($We = \rho q^2 L/ \sigma$), where $g$ is the acceleration of gravity, $\rho$ is the density of water, $\sigma$  is the surface tension of water, $q$ is the characteristic magnitude of the turbulent velocity fluctuations and $L$ is the length scale of this turbulence. 
	
\begin{figure}[!htb]
		\begin{center}
			\includegraphics[width=3in]{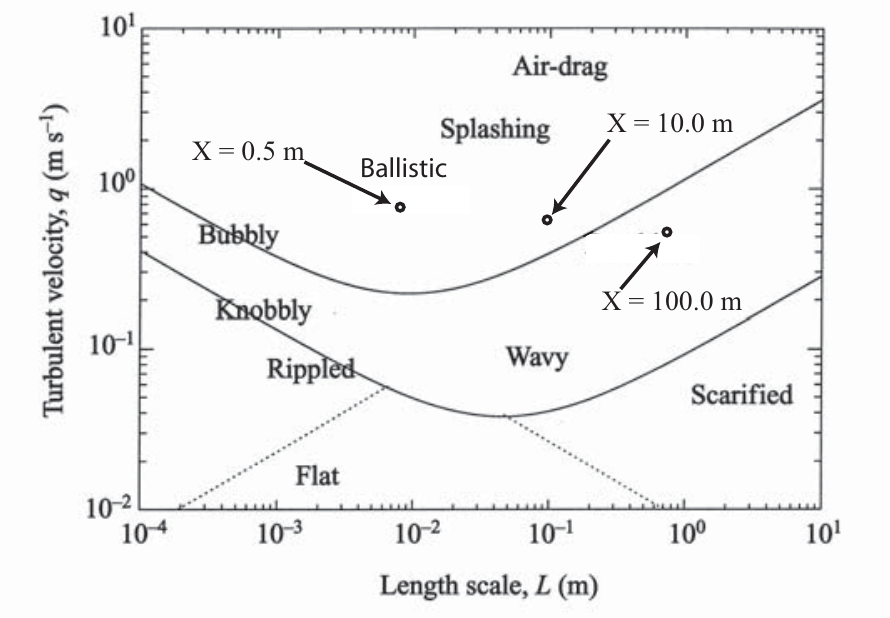}
		\end{center}
		\vspace*{-0.2in} \caption{Regions of various types of surface motions for free surface turbulence with velocity fluctuation magnitude $q$ (vertical axis) and length scale $L$ (horizontal axis), from \cite{broc:2001}. Air entrainment and spray production occur in the upper region, above the uppermost curved line. The three data points are values obtained for the turbulent boundary layer on a flat plate with $q$ taken as the rms of the spanwise (which is vertical for the boundary layer along a ship hull) velocity fluctuation ($w'$) and L taken as the boundary layer thickness ($\delta$).} \label{fig:brocchiniperegrine}
	\end{figure}

	Several authors have applied theory and numerical methods to explore the interaction of turbulence and a free surface, see for example \cite{Shen2001}, \cite{Guo2009}, \cite{kim:2013} and \cite{broc:2001}. \cite{broc:2001} have used scaling arguments to predict the critical Froude and Weber numbers above which air entrainment and spray generation will occur due to strong free-surface turbulence. Figure~\ref{fig:brocchiniperegrine}, which is from their paper, shows the boundaries of various types of surface undulations on a plot of $q$ versus $L$. The upper region of the plot is the region of air entrainment and droplet generation. We have used classical boundary layer correlations to make estimates of $q$ (taken as the root-mean-square vertical component of the turbulent velocity fluctuations) and $L$ (taken as the boundary layer thickness) at three streamwise positions in a ship boundary layer and plotted these points on the $q$-$L$ map in Figure~\ref{fig:brocchiniperegrine}.  As can be seen from the figure, the points are clearly in the air entrainment region of the plot, especially the points near the bow. Thus, air entrainment due to strong turbulent fluid motions in the hull boundary layer at the free surface is a likely cause of the layer of white water. 
	
	The difficulty with laboratory experiments on bubble entrainment and spray stems from the fact that the experiments are performed in the same gravitational field as found in ship flows and that the only practical liquid available is water, as is also found in the ocean. Thus, with $g$, $\rho$ , and $\sigma$ the same in the field and in the laboratory, one must attempt to achieve full-scale flow speeds in order to obtain Froude, Weber, and Reynolds similarity with field conditions. Also, even if full scale-values of $q$ and $L$ were obtained by towing a surface piercing flat plate with the length of the ship at high speed in a ship model basin, the free surface flow would include a bow wave which would obfuscate the source of the bubbles and spray. Another problem is that in order to obtain realistic entrainment/spray conditions and bubble/droplet size distributions, these experiments should be performed in salt water which is not typically used in ship model basins (note that though the experiments presented in this paper were performed in fresh water, we hope to repeat the experiments in salt water in the furture).
	
	In view of the above difficulties in simulating air entrainment due to the turbulent boundary layer, we have built a novel device that produces an approximation of a full scale ship boundary layer in the laboratory. This device, called the Ship Boundary Layer (SBL) simulator generates a temporally evolving boundary layer on a vertical, surface-piercing flat wall. This vertical wall consists of a stainless steel belt loop that is 1.0~m wide and about 15 m long. The belt is mounted on two vertically oriented rollers as shown in Figure~\ref{fig:TankSchem}. The rollers are driven by hydraulic motors and the entire device is placed in a large open-surface water tank as shown in the figure. Before each experimental run, the belt and the water in the tank are stationary. The water level is set below the top edge of the belt and the flow outside the belt loop on one of the long lengths between the rollers is studied. The belt is accelerated from rest using a hydraulic control system, which is able to create a highly repeatable belt motion.
	
	In the experiments discussed herein, the belt is accelerated suddenly from rest until it reaches a pre-defined speed which is held steady for a short time. The flow on the surface of the belt in this case is a simulation of the flow seen by a stationary observer in the ocean as a ship, that makes no waves, passes by at constant speed. The temporally-evolving boundary layer created along the belt can be considered equivalent to the spatially-developing boundary layer along a flat ship hull, with the distance along the ship hull corresponding to the distance traveled by the belt at any time $t$. The primary objective of the experimental study is to gain insight into the different entrainment mechanisms via quantitative and qualitative observations of the water free surface as the belt is launched from rest and to quantify the statistics of the entrained bubbles including their diameters, positions and velocities.  
	
	In addition to the experiments, a Direct Numerical Simulation (DNS) that reproduces the main features of the above-described experiments is performed.  The computations consider a small streamwise section of the flow along the belt in the experiments, and apply periodic boundary conditions along the direction of motion of the belt. The Navier-Stokes equations for incompressible flow are solved in both the air and water portions of the flow, allowing us to examine the entire three-dimensional velocity field.  The primary objective of the computational study is to identify and understand the behavior of turbulent structures immediately below the free surface and their impact on air entrainment. 
	
	The remainder of the paper is divided into four sections.  The experimental setup for the ship boundary layer, surface profile measurements, and bubble measurements, are described in the second section of the paper and the numerical setup for the DNS is reported in the third section. This is followed by the presentation and discussion of the results in the forth section of the paper. Finally, the conclusions of this study are presented in the fifth section.
	
	
	\section{EXPERIMENTAL DETAILS}
	
	\begin{figure*}[!htb]
		\begin{center}
			\includegraphics[trim=0 0.7in 0 0.00in,clip=true,width = 6in]{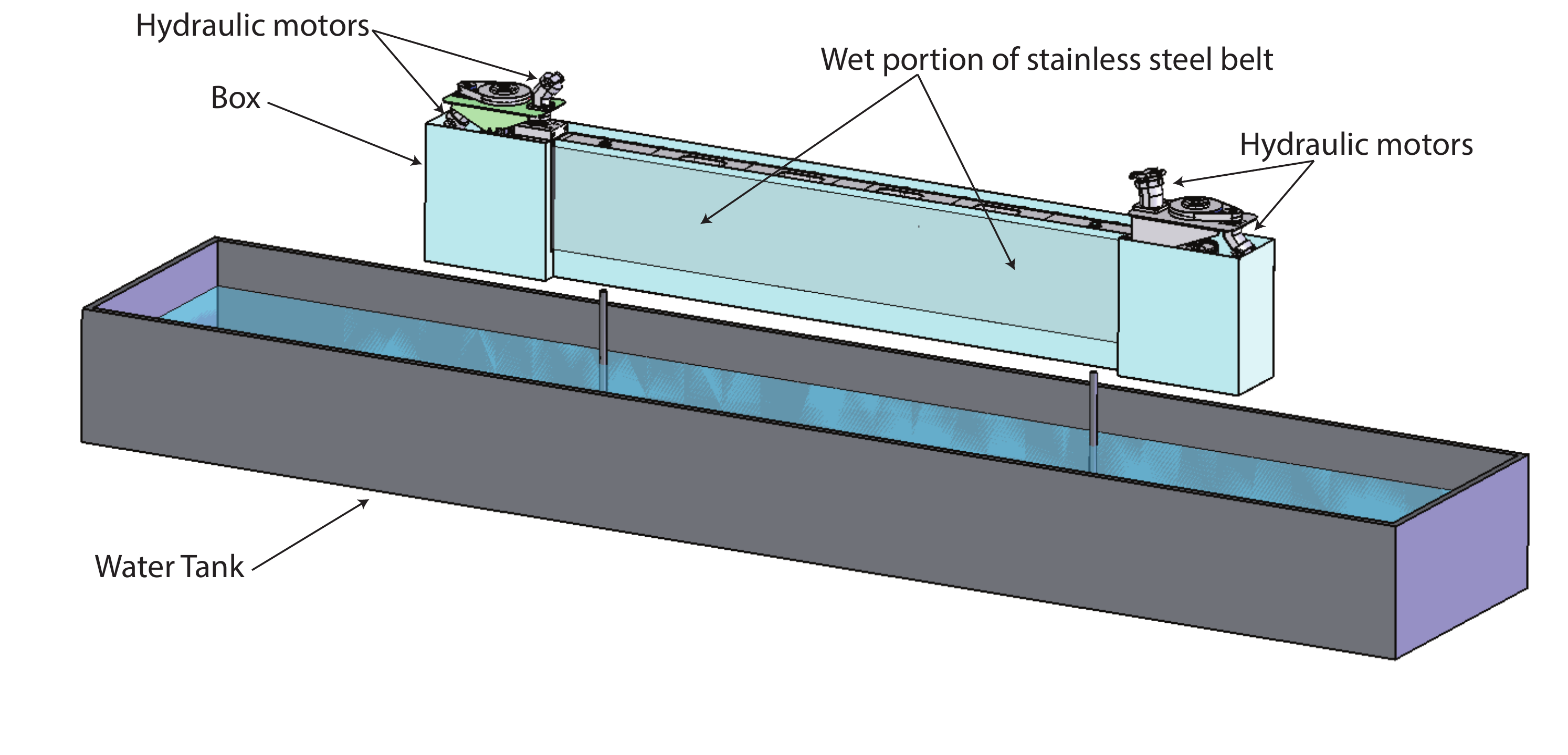}
		\end{center}
		\vspace*{-0.1in} \caption{Perspective view of the Ship Boundary Layer (SBL) simulator and the water tank.} \label{fig:TankSchem}
	\end{figure*}
	
	The experiments  were performed in the same tank and with similar measurement techniques as those described in \cite{WashutaThesis} and \cite{Washuta2016}.  A brief overview of these facilities and techniques are given below; the interested reader is referred to the original references for further details.
	
	The experiments were performed in an open-surface water tank that is 13.34 m long, 2.37 m wide and 1.32 m deep, see Figure~\ref{fig:TankSchem}.  The tank walls and bottom are made of clear plastic panels for optical access. The top of the tank is open, offering an unobstructed view of the water surface. 
	The main functional component of the Ship Boundary Layer (SBL) simulator is a one-meter-wide 0.8-mm-thick  stainless steel belt loop that is driven by two 0.46-meter-diameter, 1.1-meter-long rollers whose  rotation axes are vertically oriented and separated by a horizontal distance of approximately 7.5 meters. The rollers are each driven by two bent-axis hydraulic motors via toothed-belt-and-pulley systems.  Each roller along with the motors and drive systems form single drive units that are attached to a welded steel frame that maintains the separation between, and relative parallel orientation of the rollers.
	
	
	
	The assembled SBL device is placed in a stainless steel sheet metal box (called the dry box). The dry box keeps the assembly essentially dry, while one of the two straight sections of the belt exits the dry box through a set of seals near the roller on the left and travels through the water to the second set of seals near the opposite roller where the belt re-enters the dry box. The lone straight section exposed to water is approximately 6 meters long and pierces the free surface with approximately 0.33 meters of freeboard for the water level used in the experiments presented in this paper. At the location where the belt leaves the dry box and enters the water, a sheet metal fairing is installed to reduce the flow separation caused by the backwards-facing step associated with the shape of the dry box at this location.
	
	When performing experiments, the belt is launched from rest and accelerates until reaching constant speed. Throughout these transient experiments, the belt travel is analogous to the passage of a flat-sided ship that makes no bow waves; the length along the hull is equivalent to the total distance traveled by the belt. Belt speeds ranging from 3 to 5~m/s were used and measurements were continued until a specified belt length had passed by the measurement site. The time to accelerate varies depending on the final belt speeds and independent measurements of the belt travel show that during launch the belt travels 0.85, 1.45, and 2.29~m at belt speeds of 3, 4, and 5~m/s, respectively.
	
	\begin{figure*}[!htb]
		\begin{center}
			\begin{tabular}{cc}
				\includegraphics[trim=0in 0.0in 0 0.00in,clip=true,scale=0.6]{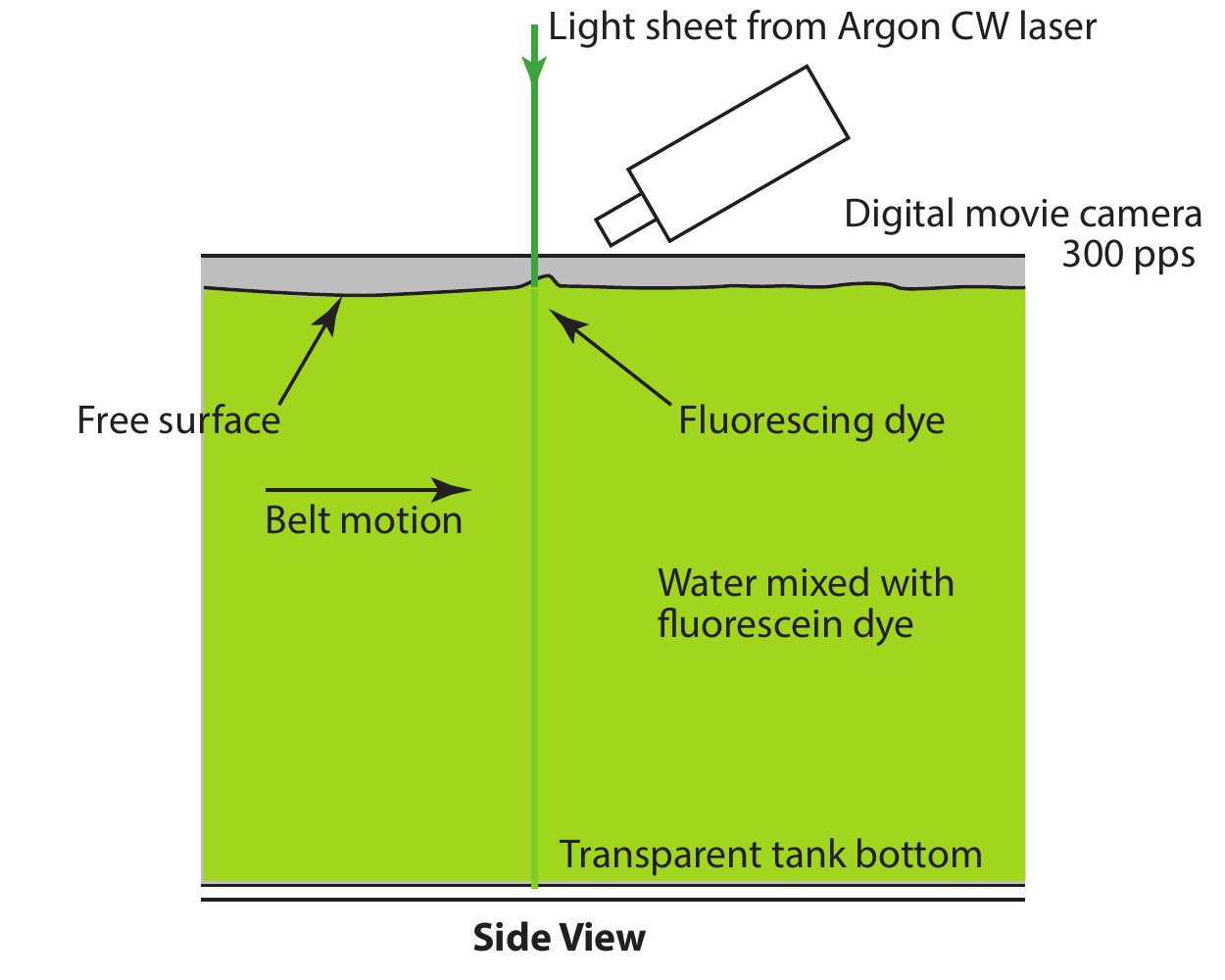}
				\includegraphics[trim=0in 0.0in 0 0.00in,clip=true,scale=0.6]{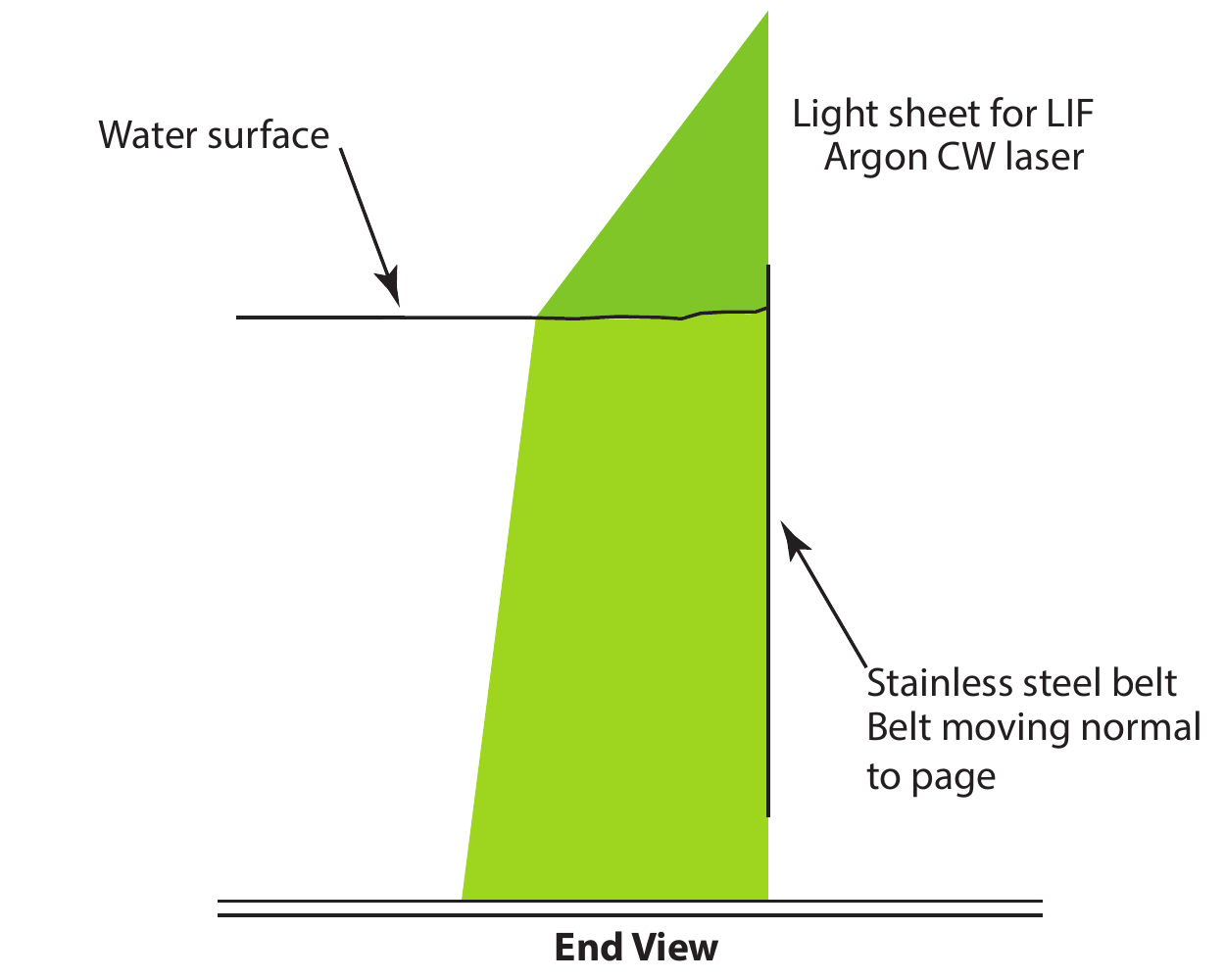}
			\end{tabular}
		\end{center}
		\vspace*{-0.1in} \caption{Schematic drawing showing the set up for the cinematic LIF measurements of the free surface shape.  The setup in the figure is for measurements of profiles in a vertical plane oriented normal to the belt surface.  The cameras and light sheet were rotated by 90 degrees about a vertical axis for measurements of profiles in vertical planes oriented parallel to the belt surface.} \label{fig:LIFSchem}
	\end{figure*} 
	
	A cinematic Laser Induced Fluorescence (LIF) technique, see Figure~\ref{fig:LIFSchem}, was used to measure the temporally evolving water surface deformation pattern.  In this technique, a continuous-wave Argon Ion laser beam is converted to a thin sheet using a system of spherical and cylindrical lenses. This sheet is projected vertically down onto the water surface in two orientations; one with the plane of the light sheet parallel to the plane of the belt and one with the light sheet perpendicular to the belt. The laser emits light primarily at wavelengths of 488~nm and 512~nm. The water in the tank is mixed with fluorescein dye at a concentration of about 5~ppm and dye within the light sheet fluoresces. High-speed cameras viewed the intersection of the light sheet and the water surface from the side with viewing angles of approximately 20 degrees from horizontal. The images seen by the  cameras show a sharp line at the intersection of the light sheet with the free surface. Using image processing, instantaneous surface profiles are extracted from these images.
	
	\begin{figure}[!htb]
		\begin{center}
			\includegraphics[trim=0 0.0in 0 0.00in,clip=true,width=3in]{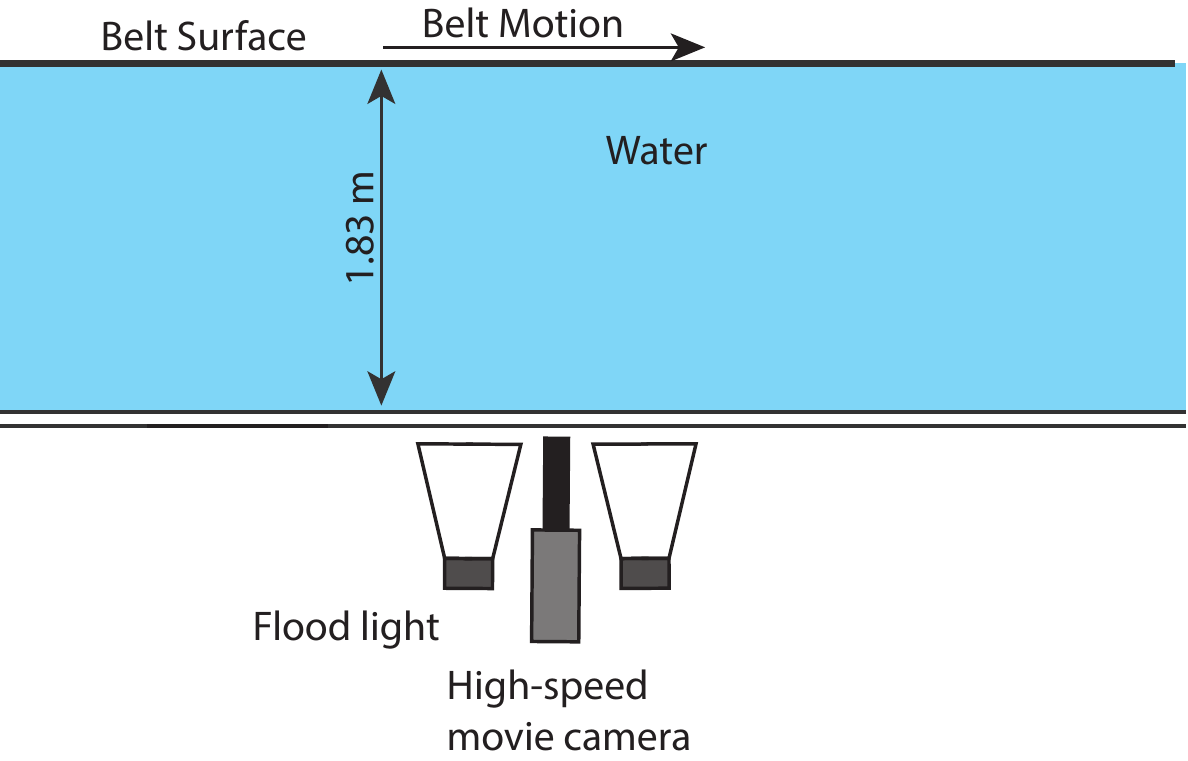}
		\end{center}
		\caption{Top view schematic of the single camera bubble shadowgraph measurement setup. The camera, with a lens attached to it (colored gray), is viewing the surface of the belt from outside the tank and from just below the free surface. The camera is focused on a plane close to the belt's surface and is flanked by two flood lights which are illuminating the portion of the belt that is in the field of view of the camera.
			\label{fig:bubble_setup_planar}}
	\end{figure}
	
	\begin{figure}[!htb]
		\begin{center}
			\includegraphics[trim=0 0.0in 0 0.00in,clip=true,width=2in]{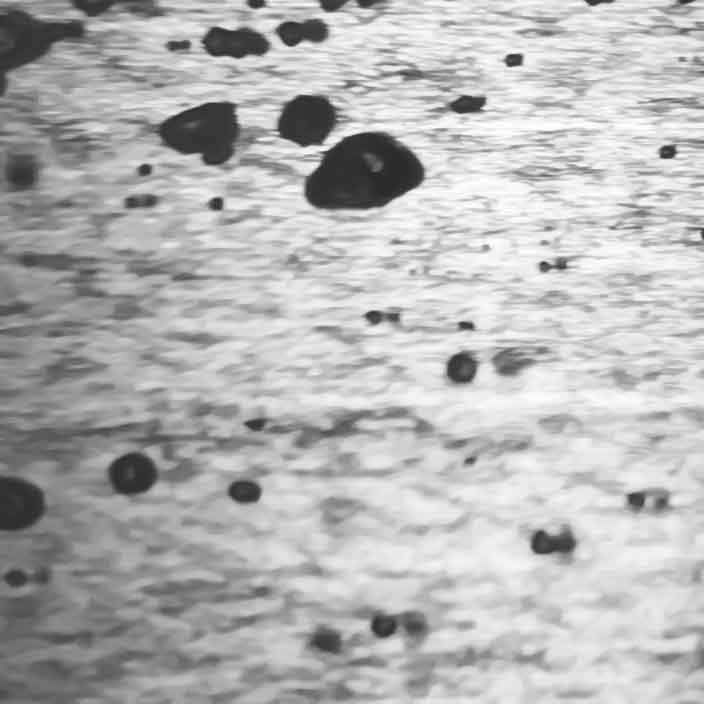}
		\end{center}
		\caption{Sample bubble image from a single camera shadowgraph measurement. The surface of the belt is seen in the background as a random intensity pattern while bubbles are seen as black blobs. \label{fig:BubblePhotos}}
	\end{figure}
	
	The present research was focused on preliminary measurements of bubbles under the above-described experimental conditions.  In these measurements, a single camera viewed the boundary layer region of the flow from underwater as shown in Figure~\ref{fig:bubble_setup_planar}.  A sample image from this setup is shown in Figure~\ref{fig:BubblePhotos}.  Analysis of the images allows for quantitative  measurement of the bubble diameters, their two-dimensional positions, and trajectories for radii ranging down to 0.5~mm.  In future experiments, more accurate measurements of bubbles are planned using cinematic stereo photography and cinematic inline holography.
	In the stereo measurements, the cameras and lights are mounted in underwater boxes close to the water free surface and the surface of the belt.   Each box contains a camera, that is mounted on a Scheimpflug mount and oriented so that the camera looks down at a mirror that turns the camera's line of sight to horizontal.  The lines of sight of  the two cameras are oriented to view the belt at $\pm45$ degrees from the normal to the belt surface.  
	Both cameras are calibrated and focused to look at the same portion of the belt. 
	The system is calibrated through images of a known 3D target and yields the 3D positions and equivalent diameters of the bubbles in any image pair.
	Illumination is provided by LED light sources that are placed in each underwater box.
	
\begin{figure}[!htb]
		\begin{center}
			\includegraphics[trim=0 0.0in 0 0.00in,clip=true,width=2.5in]{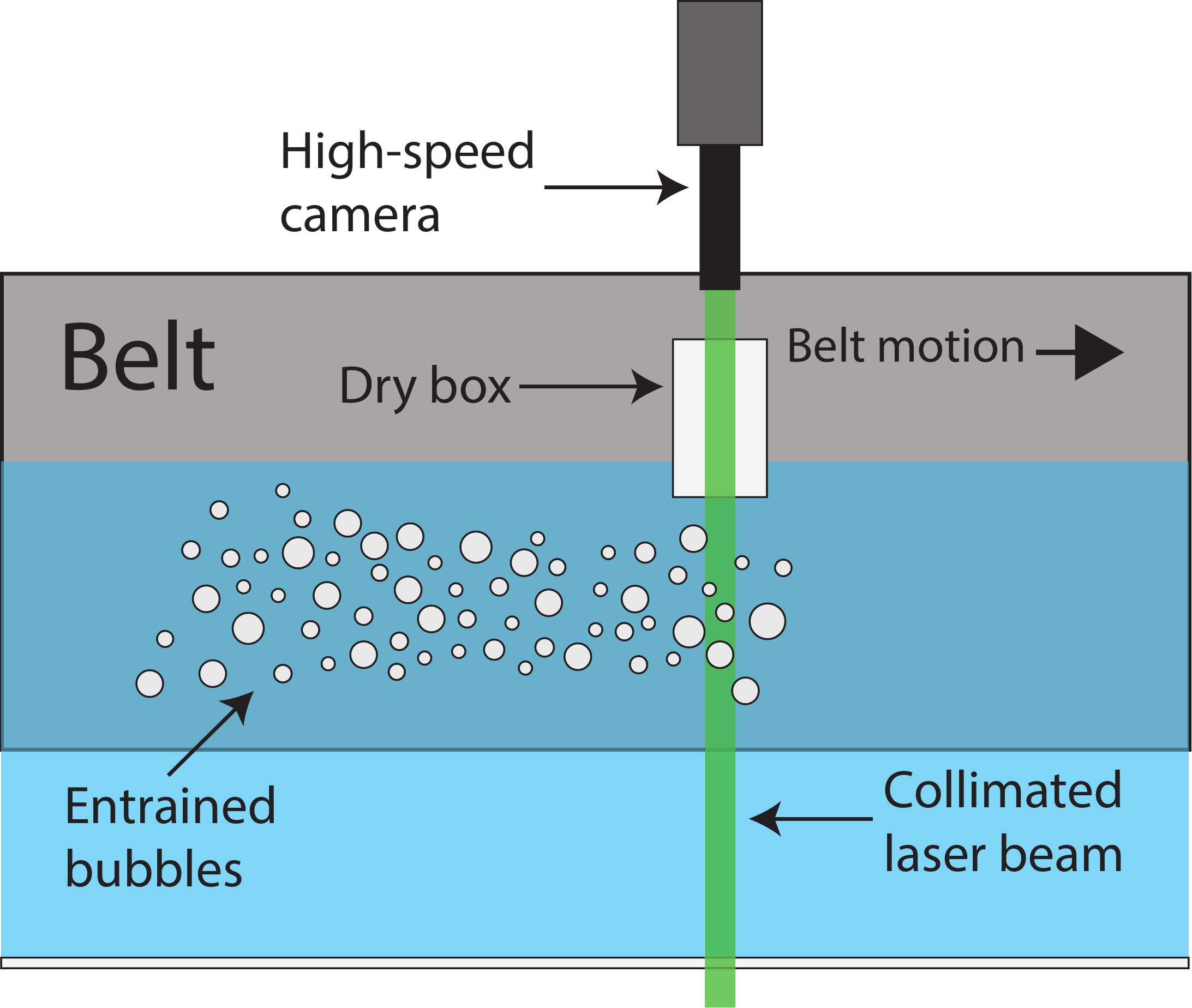}
		\end{center}
		\caption{A side view schematic of the proposed digital inline holographic setup to measure bubbles down to a size of 20 $\mu$m. The high-speed camera is oriented vertically close to the belt and is looking down into the water. A partially submerged dry box is present to prevent any light distortion from the rippled water free surface. A collimated laser beam is directed into the camera sensor from below the water tank.\label{fig:bubble_holography}}
	\end{figure}

\begin{figure}[!htb]
		\begin{center}
			\includegraphics[trim=0 0.0in 0 0.00in,clip=true,width=2.5in]{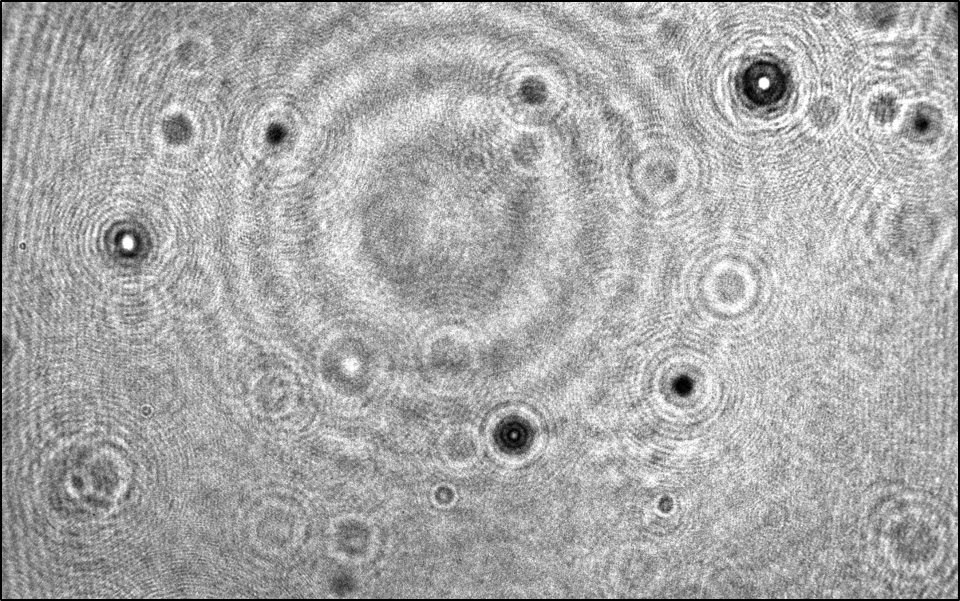}
		\end{center}
		\caption{A  hologram of droplets generated by a breaking wave. The holograms of each droplet are dark circles with characteristic round Fresnel patterns. The holograms are reconstructed digitally and the size and three-dimensional position of the drops can be measured. The processing method of bubbles is similar to that of droplets. \label{fig:droplet_hologram_example}}
	\end{figure}

	In future experiments we are also planning to use a digital inline holography system to measure the size, velocity, and position of bubbles down to a radius of 20 $\mu$m. The experimental setup consists of a camera fitted with a long distance microfocus lens, oriented vertically next to the belt, looking down into a dry box, as seen in Figure \ref{fig:bubble_holography}. The dry box is always partially submerged so as to negate the light distorting properties of the rough water free surface. A collimated laser beam from a pulsed ND:YLF laser is directed upwards from the bottom of the tank into the camera lens and sensor. When a bubble is in the path of the collimated laser beam a hologram is recorded by the camera sensor. This hologram can then be reconstructed digitally and the size and three-dimensional position of the bubble can be measured. Once the size and location of the bubbles is obtained, the bubbles are tracked in time and their velocities can be obtained from the resulting trajectories. This system has been successfully implemented in our laboratory to measure droplets generated by breaking waves \cite[]{Erinin2017}.  An example of a hologram from these droplet measurements is shown in Figure \ref{fig:droplet_hologram_example}.
	
	\section{COMPUTATIONAL SETUP}
\begin{figure}[!htb]
		\centering
		\begin{overpic}[width=\linewidth]{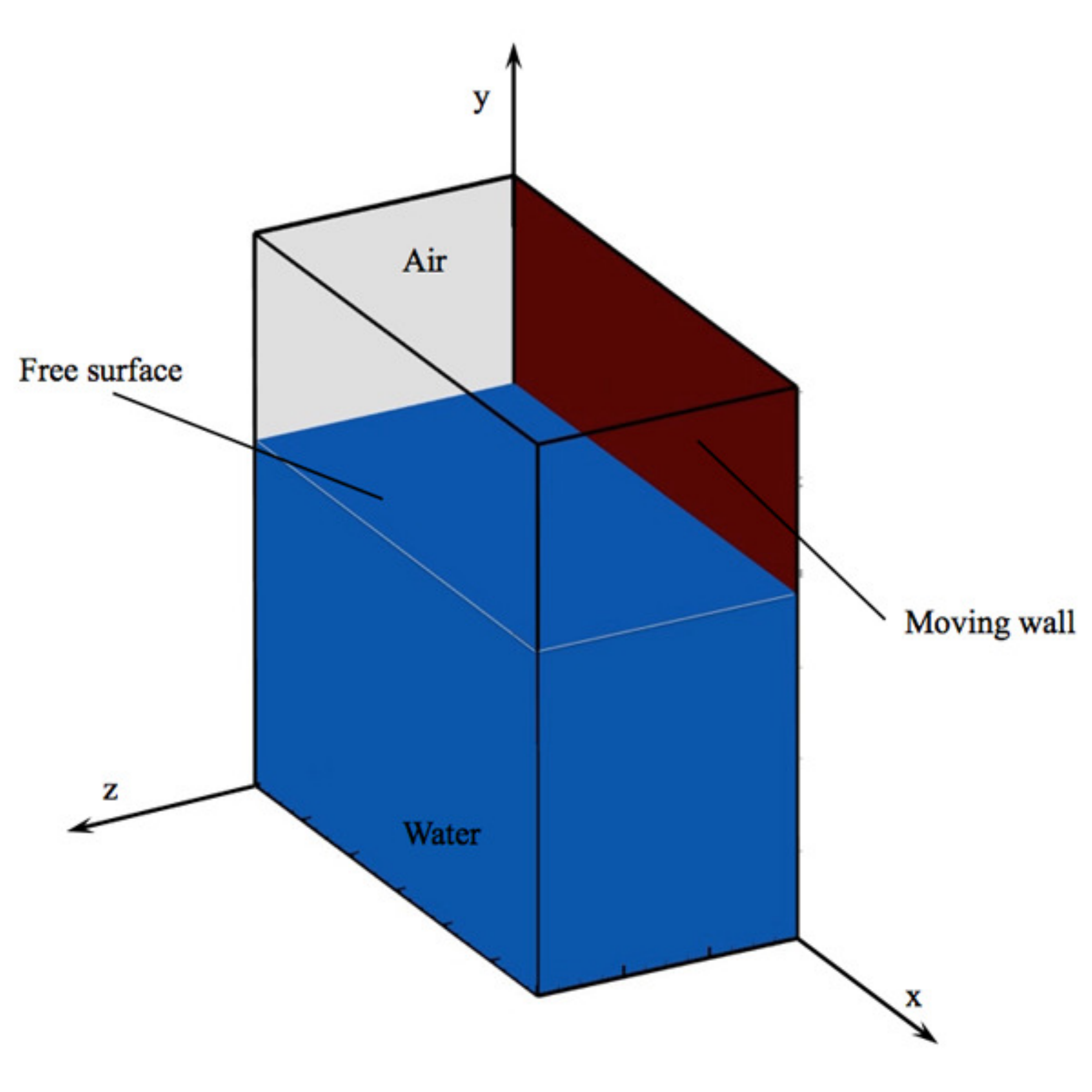}
			\put(80,6){\colorbox{white}{-x}}
			\put(41,90){\colorbox{white}{z}}
			\put(8,25){\colorbox{white}{y}}
		\end{overpic}
		\caption{Computational domain}
		\label{fig:domain}
	\end{figure}

	The computational simulations were designed to mimic the conditions in the experiments. The main challenges for the computations are to properly resolve the boundary layer along the moving belt and, at the same time, capture the complex free-surface deformations and air-entrainment phenomena. We only simulate a small part of the moving belt and apply periodic boundary conditions along the direction of motion of the belt.  A schematic of a typical computational box is shown in Figure \ref{fig:domain}, where a portion of the air above the free-surface is also considered. We define the Cartesian domain $(x,y,z)$ in such way that $x$ is the streamwise direction, making $y-z$ the cross-stream plane. The moving wall is located at $y=0$, the undisturbed free surface is at $z=0$ (parallel to $x-y$ plane) and gravity is imposed in the $-z$ direction. The Navier-Stokes equations for incompressible flow are solved in both the air and water portions of the domain and the interface is implicitly advected and tracked using a geometric reconstruction approach \citep{QIN2015219}.  The governing equations are solved on a block-structured Cartesian grid with Adaptive Mesh Refinement (AMR) \citep{Vanella2010JCP,Vanella2014}. AMR allows us to cluster grid points at the dynamically evolving interface, as well as the boundary layer in a cost-efficient manner.  The equations are advanced in time using an exact projection method. All spatial derivatives are discretized using second-order, central finite-differences. The jump conditions at the interface are imposed in a sharp manner using a variant of the ghost-fluid method \citep{FEDKIW1999457}.  Details on the overall formulation together with a detailed validation in a series of problems of increasing complexity can be found in \citet{DelaneyPhD2014}.\par

	In the experiments, the belt starts from rest and quickly reaches its terminal speed. The boundary layer, undergoes transition and gradually thickens as a function of time. The critical Froude number based on the local momentum thickness, $\theta$, and the belt velocity when air entrainment is initiated is $Fr \sim O(10)$. The corresponding Reynolds number at this time instant is $Re_\theta\sim O(10^4)$.  Simulating this process starting from the belt at rest and arriving to post-entrainment Reynolds and Froude number has the advantage of well defined initial and boundary conditions but it is prohibitively expensive even on leadership parallel computing platforms. Due to this limitation, and in order to keep the computational cost at reasonable levels, we considered significantly lower Reynolds numbers, but kept the Froude number in the same regime as in the experiments, where high deformations of the free-surface are observed, leading to air-entrainment. This is a significant advantage of the simulations where we can independently change the surface tension and gravity to replicate the conditions in the air entrainment regime in the experiment at lower Reynolds numbers. In particular we consider the boundary layer to evolve from $Re_\theta=900$ to $1400$ and will discuss two Froude numbers: $Fr=4$ and $Fr=12$, for the same Reynolds number, $Re_\theta=1400$. As an additional cost reduction, we started from the fully turbulent regime, bypassing the transition phase. Despite these approximations, which only enable qualitative comparisons to the experimental results, the computations are well positioned to quantify the effects of the Froude number on the flow physics.
	
	In addition to these considerations the grid was dynamically refined to capture the dynamics of the triple contact point, while the resolution at any location in the domain containing the interface was kept at the highest refinement level. We use periodic boundary conditions in the streamwise direction. At the moving wall the impermeability condition is enforced and the velocity in the streamwise direction is set to the reference value. We impose a Sommerfeld radiation condition on the boundary opposite the moving wall in order to convect the surface waves out of the domain. The convective velocity is calculated using the average water phase wall-normal velocity at the boundary. Details can be found in \citet{DelaneyPhD2014}. Slip-wall conditions were used at the two remaining boundaries. The domain dimensions were driven by the maximum Reynolds number we wanted to achieve, and were selected based on prior computations of turbulent boundary layers in the literature.\par
	In all computations, we define the \textit{midsection} as the depth range where the effects of the free-surface are not felt and the velocity statistics are identical to the ones in a zero pressure gradient boundary layer.  Also, unless otherwise stated, the midsection quantities used to normalize the results are taken from the flow field at $Re_\theta=1400$.  
	\begin{figure}[!htb]
		\centering
		\begin{overpic}[width=1\linewidth]{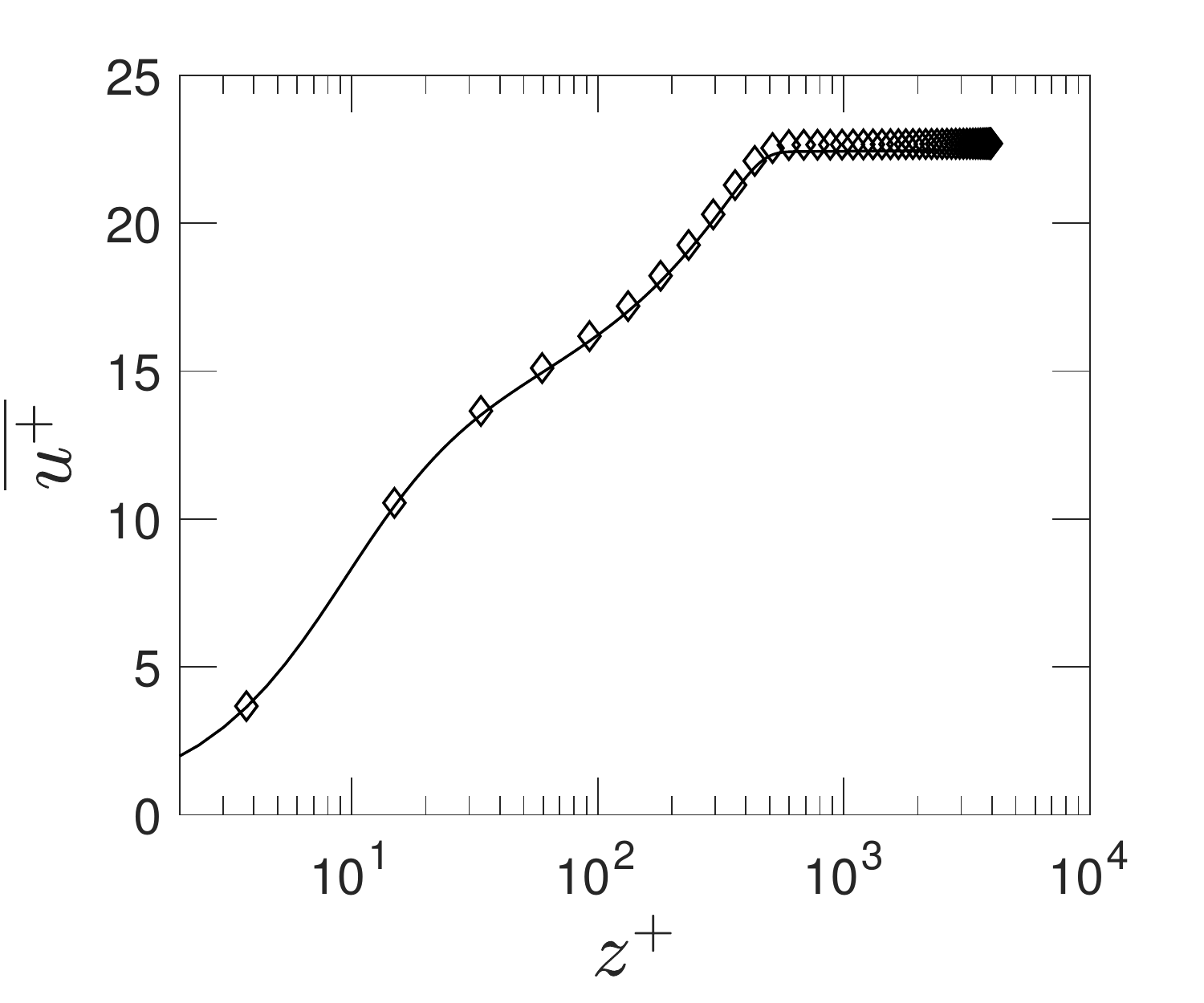}
			\put(20,70){\text{(a)}}
			\put(48,3){\colorbox{white}{$y^+$}}
		\end{overpic}
		\begin{overpic}[width=1\linewidth]{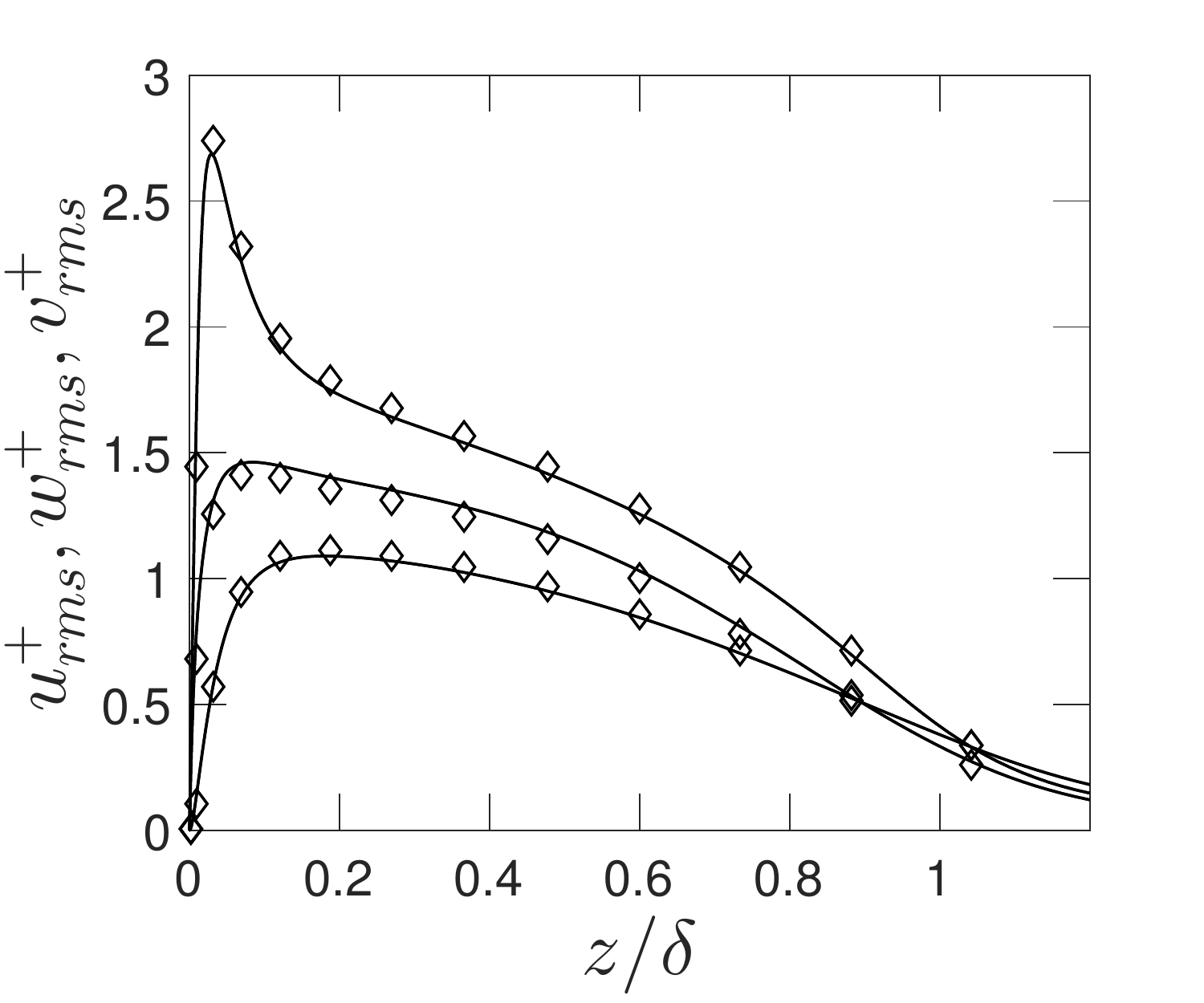}
			\put(80,70){\text{(b)}}
			\put(48,2){\colorbox{white}{$y/\delta$}}
		\end{overpic}
		\caption{Midsection statistics. (a) Mean
			streamwise velocity at $Re_\theta=1400$; (b) Turbulent intensities at $Re_\theta=1400$.  
			\hbox{\drwln{12}{.5}}, present DNS; \diam, Spalart 1988.}
		\label{fig:valid}
	\end{figure}
	Quantitative comparisons at the midsection to reference data in the literature are shown in Figure \ref{fig:valid}. The velocity statistics at $Re_\theta=1400$ are compared to the DNS by \citet{Spalart1988}. The agreement for both the mean velocity and the turbulent intensities is excellent.\par
	The numerical investigation is primarily deployed to study the mechanisms of air entrainment due to the turbulence field beneath the surface.
	In the remainder of this study, we report on the numerical findings where comparison to experiments is possible, giving us greater confidence in the turbulent entrainment analysis.\par
	%
	\section{RESULTS AND DISCUSSION}
	
	\subsection{Surface Profiles}
	
\begin{figure}[!htb]
		\begin{center}
			\begin{tabular}{c}
				\includegraphics[trim=0 0.0in 0 0.0in,clip=true,width=3in]{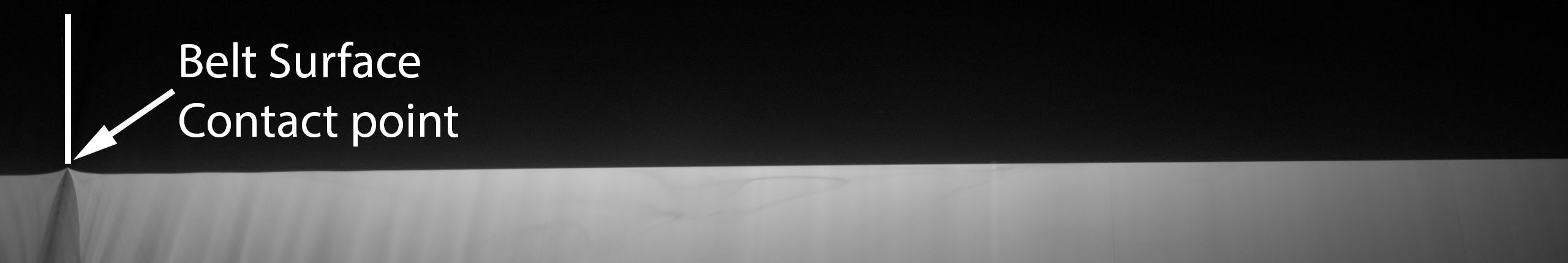}\\
				(a)\\
				\includegraphics[trim=0 0.0in 0 0.0in,clip=true,width=3in]{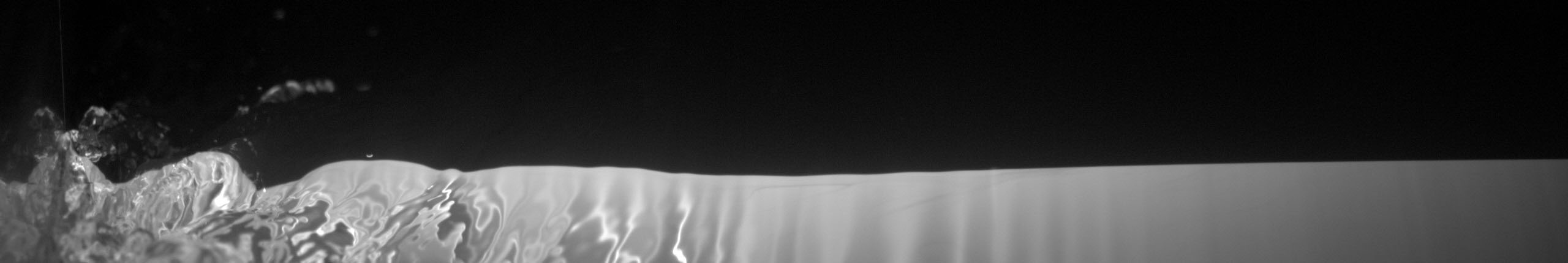}\\
				(b)\\
				\includegraphics[trim=0 0.0in 0 0.0in,clip=true,width=3in]{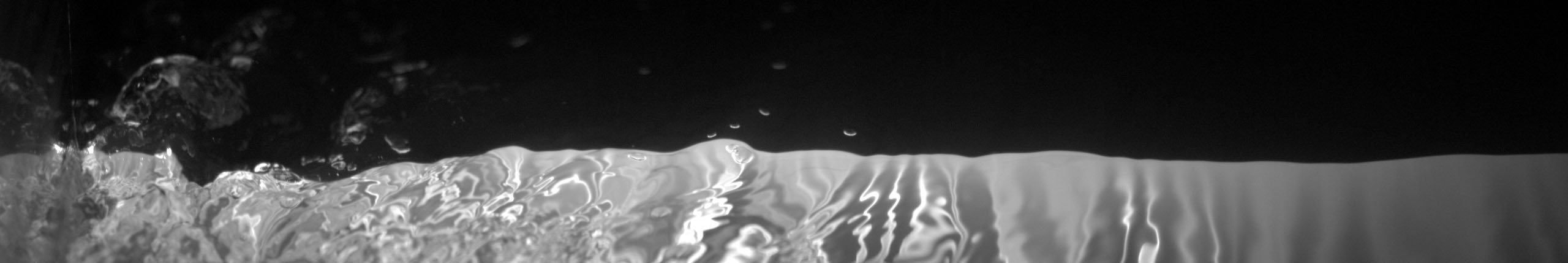}\\
				(c)\\
				\includegraphics[trim=0 0.0in 0 0.0in,clip=true,width=3in]{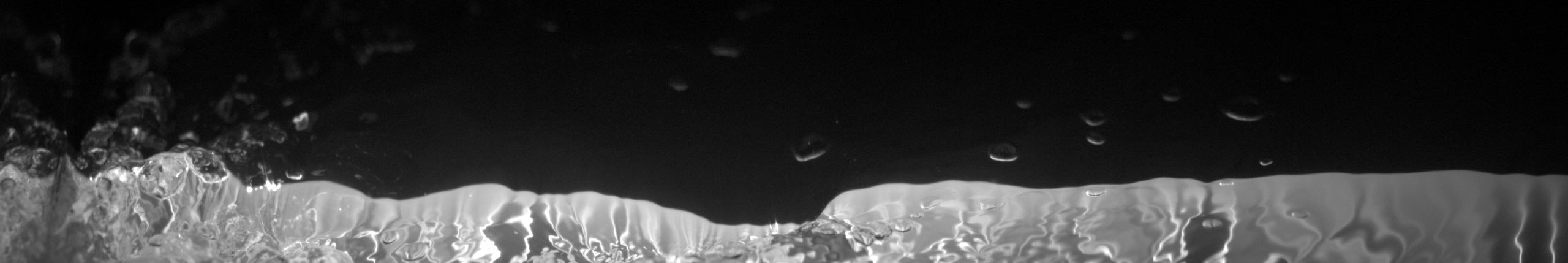}\\
				(d)\\
				\includegraphics[trim=0 0.0in 0 0.0in,clip=true,width=3in]{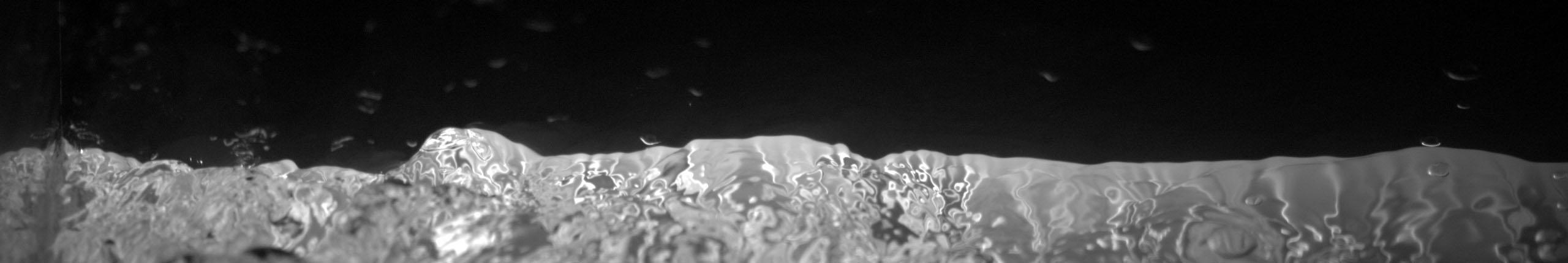}\\
				(e)\\
			\end{tabular}
		\end{center}
		\vspace{-0.2in} \caption{A sequence of five images from a high speed movie of the free surface during a belt launch to 5~m/s with the light sheet perpendicular to the belt. These images are taken at equivalent belt lengths of (a) 0~m (b) 5~m (c) 10~m (d) 15~m and (e) 20~m from the bow of the ship. The high reflectivity of the stainless steel belt makes it appear as a symmetry plane on the left side of the images. Thus, in all of the images, the part of the image to the left of the location of water surface contact point on the belt, as noted in image (\textit{a}), is a reflection of the part of the image to the right of the contact point.  The horizontal field of view for these images is approximately 31~cm.}\label{fig:overall}
	\end{figure}

	
	Surface profile measurements were performed at belt speeds of $U=3$, $4$, and $5$~m/s using the cinematic Laser Induced Fluorescence (LIF) technique. Through initial trials, it was determined that a frame rate of 1000~fps was necessary to provide a sufficient temporal resolution so that surface features perpendicular to the belt could be identified and tracked smoothly in successive frames. LIF images of the water free surface with the light sheet oriented perpendicular to the belt in an experimental run with $U$ = 5~m/s are shown in Figure~\ref{fig:overall}. The five images in the figure are spaced out equally by distance of belt travel, with the first image (a) taken at 0.0~s, the time when the belt first starts to move. The instantaneous belt speed from the beginning of belt motion through the acceleration portion until the belt reaches constant speed has been measured separately and is used to correlate the time of each frame to the belt travel distance. Here and in the following, rather than refer to images and data by the time after the belt has started moving, we refer to them by the distance, $x$, from the leading edge of an equivalent flat plate, which is also the distance that the belt has traveled
	
	\[
	x = \int_0^t U(t^\prime )dt^\prime,
	\]
	where $t=0$ is the instant that the belt starts to move.
	This integral is performed numerically with the measured function $U(t)$, which includes an initial phase of nearly constant acceleration, i.e., $dU/dt\approx$ a constant,  followed by a longer period of constant speed, i.e., $U = $ a constant.
	Thus, the images in Figure~\ref{fig:overall} depict a portion of a run, with images (a), (b), (c), (d) and (e) captured at 0~s, 1.35~s, 2.35~s, 3.35~s, and 4.35~s, respectively, corresponding to $x=$ 0.0, 5.0, 10.0, 15.0 and 20.0~m. 
	
	In this subsection, we will quantitatively examine the free surface profiles parallel and perpendicular to the belt surface for  $U= 5$~m/s and report some observed entrainment events from the parallel free surface profiles. Then we will look at processed free surface profiles perpendicular and parallel to the belt at belt speed $3$ m/s and compare them to the computational results qualitatively. Finally, the free surface height from the experimental results is analyzed qualitatively for $3$, $4$, and $5$ m/s.
	
	
	\begin{figure}[!htb]
		\begin{center}
			\begin{tabular}{c}
				\includegraphics[trim=0 0.0in 0 0.00in,clip=true,width=3in]{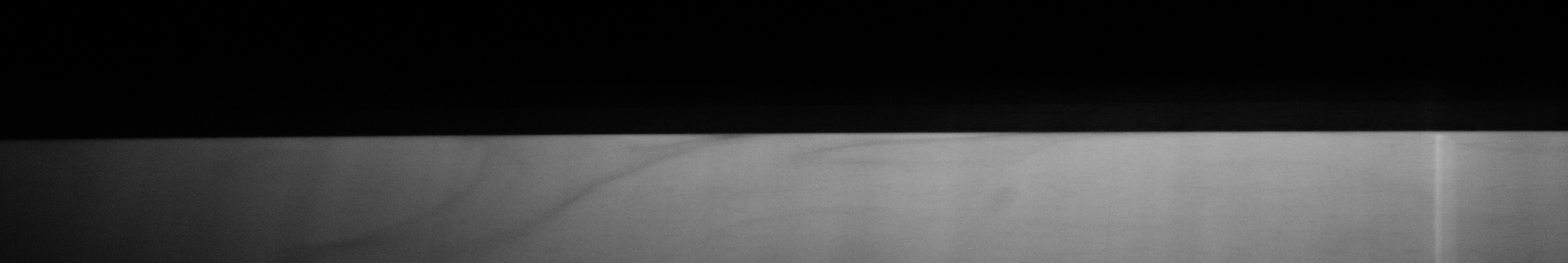}\\
				(a)\\
				\includegraphics[trim=0 0.0in 0 0.00in,clip=true,width=3in]{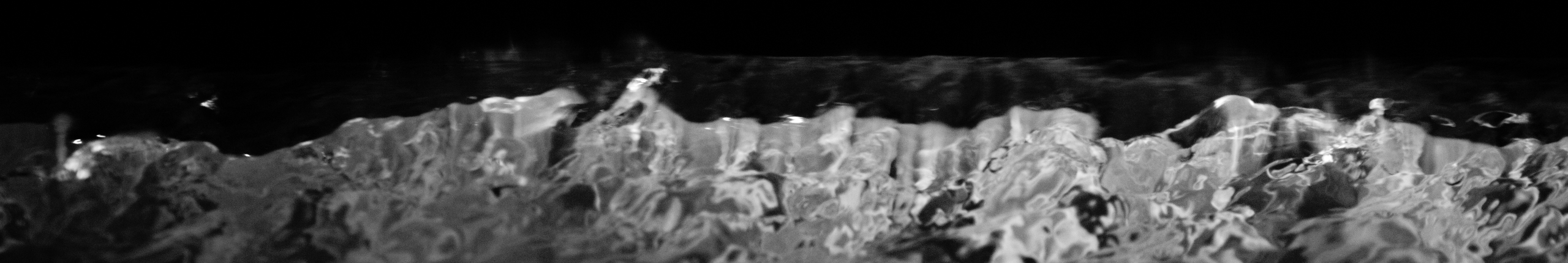}\\
				(b)\\
				\includegraphics[trim=0 0.0in 0 0.00in,clip=true,width=3in]{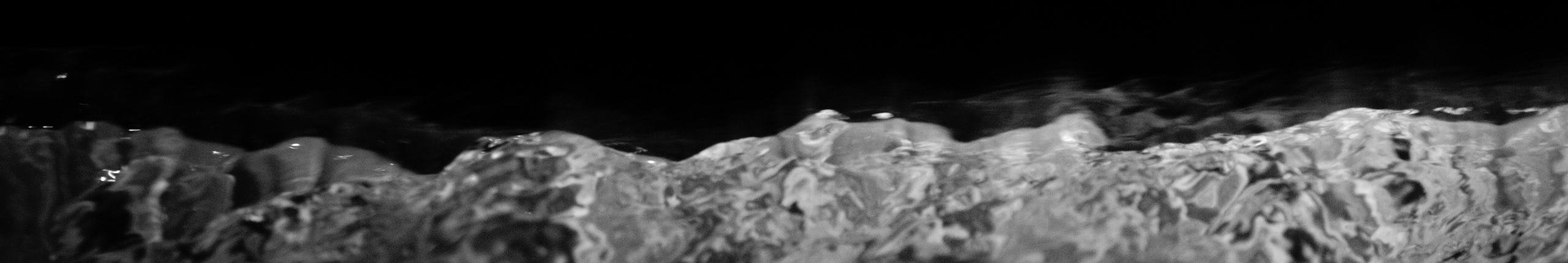}\\
				(c)\\
				\includegraphics[trim=0 0.0in 0 0.00in,clip=true,width=3in]{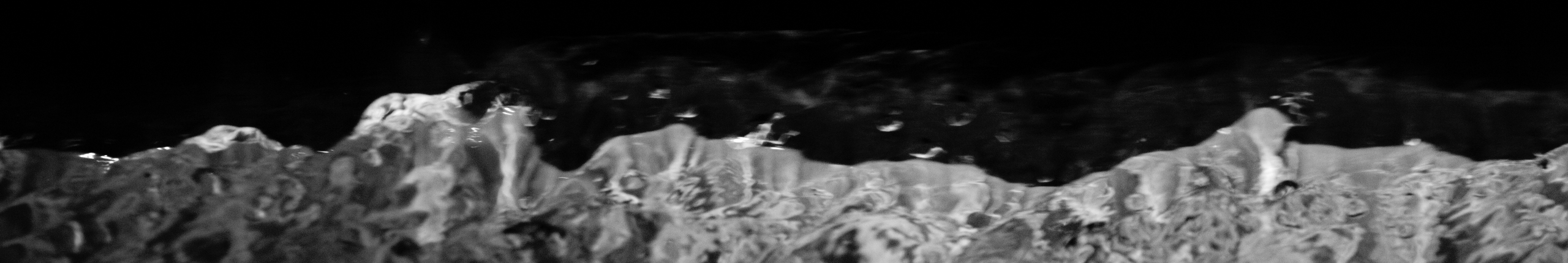}\\
				(d)\\
				\includegraphics[trim=0 0.0in 0 0.00in,clip=true,width=3in]{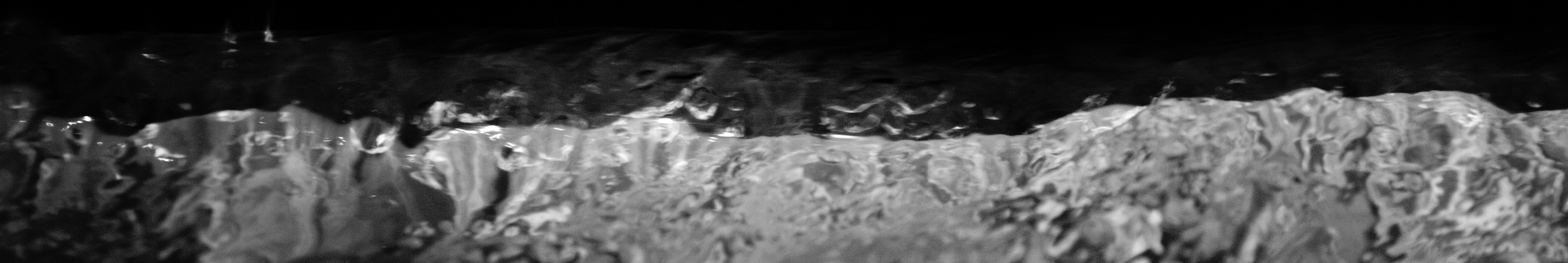}\\
				(e)\\
			\end{tabular}
		\end{center}
		\vspace{-0.2in} \caption{A sequence of five images, similar to the ones shown in Figure \ref{fig:overall}, from a high speed movie of the free surface during a belt launch to 5~m/s with the light sheet parallel to the belt and located at approximately $y = 1.25$ cm away from the belt. These images are taken at equivalent belt travel lengths of (a) 0~m (b) 5~m (c) 10~m (d) 15~m and (e) 20~m relative to the bow of the ship. The horizontal field of view for these images is approximately 15~cm.}\label{fig:LighSheetParallel}
	\end{figure} 
	
	As discussed in the experimental details section, in our previously reported measurements \cite[]{Washuta2016} the plane of the vertical light sheet was oriented normal to the belt surface and two cameras, from upstream and downstream, looked down at an angle of approximately 20 degrees at the intersection of the light sheet and the free surface. The images in Figure~\ref{fig:overall} are from the downstream camera in one of these wall-normal water surface profile measurements, and these images have been flipped horizontally for convenience in order to match the coordinate system of later plots, so that the belt is near the left side of each image and is moving out of the page. The position of the belt is marked on the left side of image (a) and the intensity pattern to the left of this location is a reflection of the light pattern on the right. This line of symmetry gives a good indication of the position of the belt in each image. The sharp boundary between the upper dark and lower bright region of each image is the intersection of the light sheet and the water surface. The upper regions of the later images contain light scattered from roughness features on the water surface behind the light sheet. These roughness features include bubbles that appear to be floating on the water surface and moving primarily in the direction of the belt motion. The bright area below the boundary is created by the glowing fluorescent in the underwater portion of the light sheet.  The complex light intensity pattern here is created by a combination of the refraction of the laser light sheet as it passes down through the water surface and the refraction of  the light from the glowing underwater dye as the light passes up through the water surface between the light sheet and the camera, on its way to the lens. It can be seen from these images that surface height fluctuations (ripples) are created close to the belt surface, at the left side of each image, and propagate away from the belt (to the right). As time passes, the surface height fluctuations grow dramatically and eventually surface breaking and air entrainment events begin to occur, resulting in bubble and droplet production.
	
	From these images, it is observed that the free surface remains nearly quiescent during a period of belt travel at the beginning of each run; during this time period, the LIF images appear similar to what is seen in Figure~\ref{fig:overall} (a).  After a short time,  the surface suddenly bursts with activity near the belt surface, creating free surface ripples. After this point, see Figure~\ref{fig:overall} (b), the free surface fluctuations are continually generated close to the belt and this generation region grows in time. As the belt travel length continues to increase, free surface ripples begin to appear to the right side of the image, away from the belt, see Figure~\ref{fig:overall} (c-d). Qualitatively, from looking at Figure~\ref{fig:overall} (c-d) it is evident that the free surface ripples are most intense closest to the belt and decay in intensity as they move away from the belt.
	
	In more recent experiments with the light sheet perpendicular to the belt, the two cameras were both placed downstream in a side-by-side configuration with an overlap in their fields of view. By combining the profiles from the two cameras, a higher resolution was achieved (approximately 15 pixels/mm) while viewing a similar horizontal distance away from the belt (approximately 30 cm). Results shown in Figures \ref{fig:surface_profiles} (a) and Figure \ref{fig:rms_height} are obtained using this new configuration.
	
	A corresponding set of images with the plane of the light sheet parallel to the belt are shown in Figure \ref{fig:LighSheetParallel}. In these images the laser light sheet was located at a distance of approximately $y = 1.25$ cm away from the belt surface and the camera is looking toward the belt and down at the free surface at a small angle from horizontal. The  belt is in the black background traveling from left to right in the series of images shown in the figure. As in the images of the surface profiles perpendicular to the belt, shown in Figure~\ref{fig:overall}, the sharp boundary in the images in Figure \ref{fig:LighSheetParallel} is the intersection of the light sheet with the water surface. The images in Figure \ref{fig:LighSheetParallel} (a-e) correspond the the same lengths of belt travel as the images in  Figure~\ref{fig:overall}.
	
	\begin{figure}[!htb]
		\begin{center}
			\begin{tabular}{cc}
				\includegraphics[height=.43in]{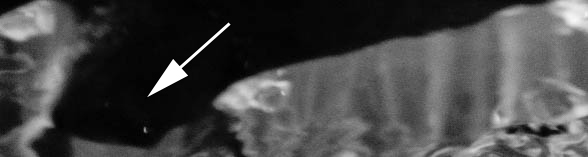} &
				\includegraphics[height=.43in]{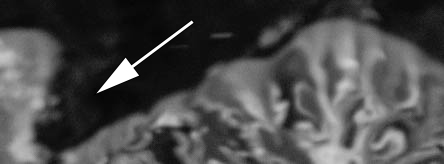}\\
				(a) & (f)\\
				
				\includegraphics[height=.43in]{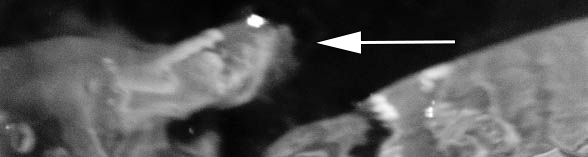} &
				\includegraphics[height=.43in]{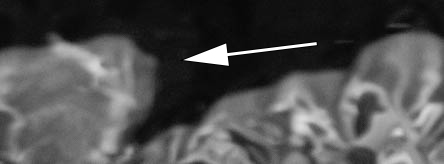}\\
				(b) & (g)\\
				
				\includegraphics[height=.43in]{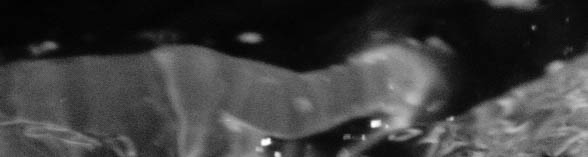} &
				\includegraphics[height=.43in]{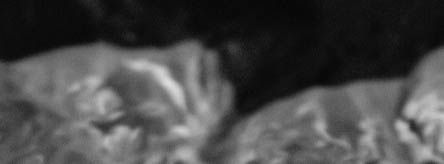}\\
				(c) & (h)\\
				
				\includegraphics[height=.43in]{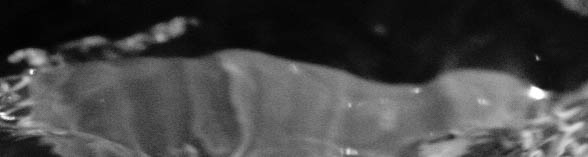} &
				\includegraphics[height=.43in]{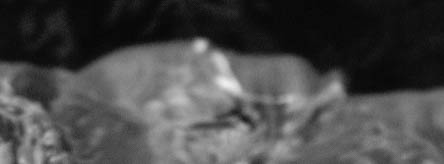}\\
				(d) & (i)\\
				
				\includegraphics[height=.43in]{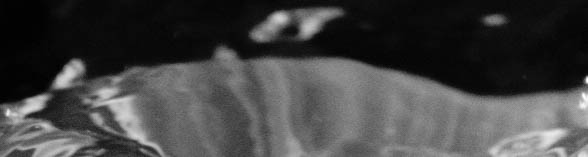} &
				\includegraphics[height=.43in]{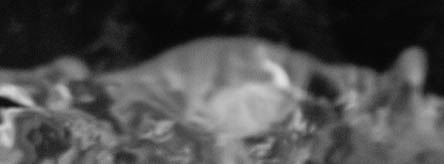}\\
				(e) & (j)\\
			\end{tabular}
		\end{center}
		\vspace*{-0.2in} \caption{Two series of images capturing entrainment events parallel to the motion of the belt $y = 1.25$ cm away from the belt at a belt speed of $5$ m/s with the belt moving from the left to the right. Images (a) through (e) correspond to $12.98$~m, $13.03$~m, $13.07$~m, $13.10$~m, and $13.13$~m of belt travel. In this series of images we see a jet of water, indicated by an arrow in image (b), overtake a surface depression, indicated by an arroaw in image (a), the jet eventually overtakes the air cavity below (c) and entrains a pocket of air in the process (d-e). In images (f-j), corresponding to belt travel of $15.71$~m, $15.73$~m, $15.75$~m, $15.78$~m, and $15.80$~m  respectively, we see a similar event where a water jet overtakes a cavity in the free surface, presumably entraining air into the water below. The surface depression is indicated by an arrow in image (f) while the jet is indicated by the arrow in image (g). Both of these series of images evolve on the order of tens of milliseconds} \label{fig:parallel_entrainment_events}
	\end{figure}
	
	From watching the movies of the free surface profiles parallel to the belt, it is clear that the free surface experiences a sudden burst of activity somewhere between images (a) and (b) in Figure~\ref{fig:LighSheetParallel}. This burst of activity on the free surface is associated with the growing size of the turbulent boundary layer in the water. Once the free surface become rough, wave-like breaking events can be observed moving from left to right (in the same direction as the motion of the belt) in the midst of other surface features moving parallel and perpendicular to the light sheet. These wave-like breaking events, presented in Figure~\ref{fig:LighSheetParallel} (b - d), are persistent and can be observed frequently once the free surface becomes rough. 
	
	Sometimes, the above-mentioned breaking events appear to entrap pockets of air into the water below.  Two of these breaking events are shown in the images in Figure~\ref{fig:parallel_entrainment_events}, which were taken from a wall parallel LIF movie for a  belt speed of 5~m/s and with the light sheet 1.25~cm from the belt surface. The non-uniformity of the light intensity at the free surface is partially due to the curvature of the free surface, which reflects laser light and focuses it underneath the surface, and partly because water has obstructed the laser light from reaching the areas below. In Figure \ref{fig:parallel_entrainment_events} (a) to (e) we see a a sequence of images taken at $12.98$~m, $13.03$~m, $13.07$~m, $13.10$~m, and $13.13$~m of belt travel. To the left in Figure \ref{fig:parallel_entrainment_events} (a) we see a jet forming with an air cavity directly below. The jet then proceeds to plunge forward and above the air cavity, from left to right in the direction of the belt motion, in images (b-c). The jet then splashes on the free surface and closes off the pocket of air in (d) . Finally, air is presumably entrained in the flow by image (e) and the jet is no longer present on the free surface. The time span between images (a) through (e) is 30 milliseconds. Images (f-j) in Figure \ref{fig:parallel_entrainment_events} tell a similar story. A jet of water is moving from left to right over a cavity in (f). The jet proceeds to overtake the air cavity in (g-h) and the jet splashes on the free surface in (i-j), again, presumably entraining air. The sequence of images in (f-j) takes place over a time span of 18 milliseconds. There are many similar wave-like breaking events in the movies of the free surface profiles parallel to the belt surface.
	
	\begin{figure*}[!htb]
		\begin{center}
			\begin{tabular}{ccc}
				(a) & (b) & (c) \\
				\includegraphics[trim=0 0 0.7in 0in, clip=true,width=2in]{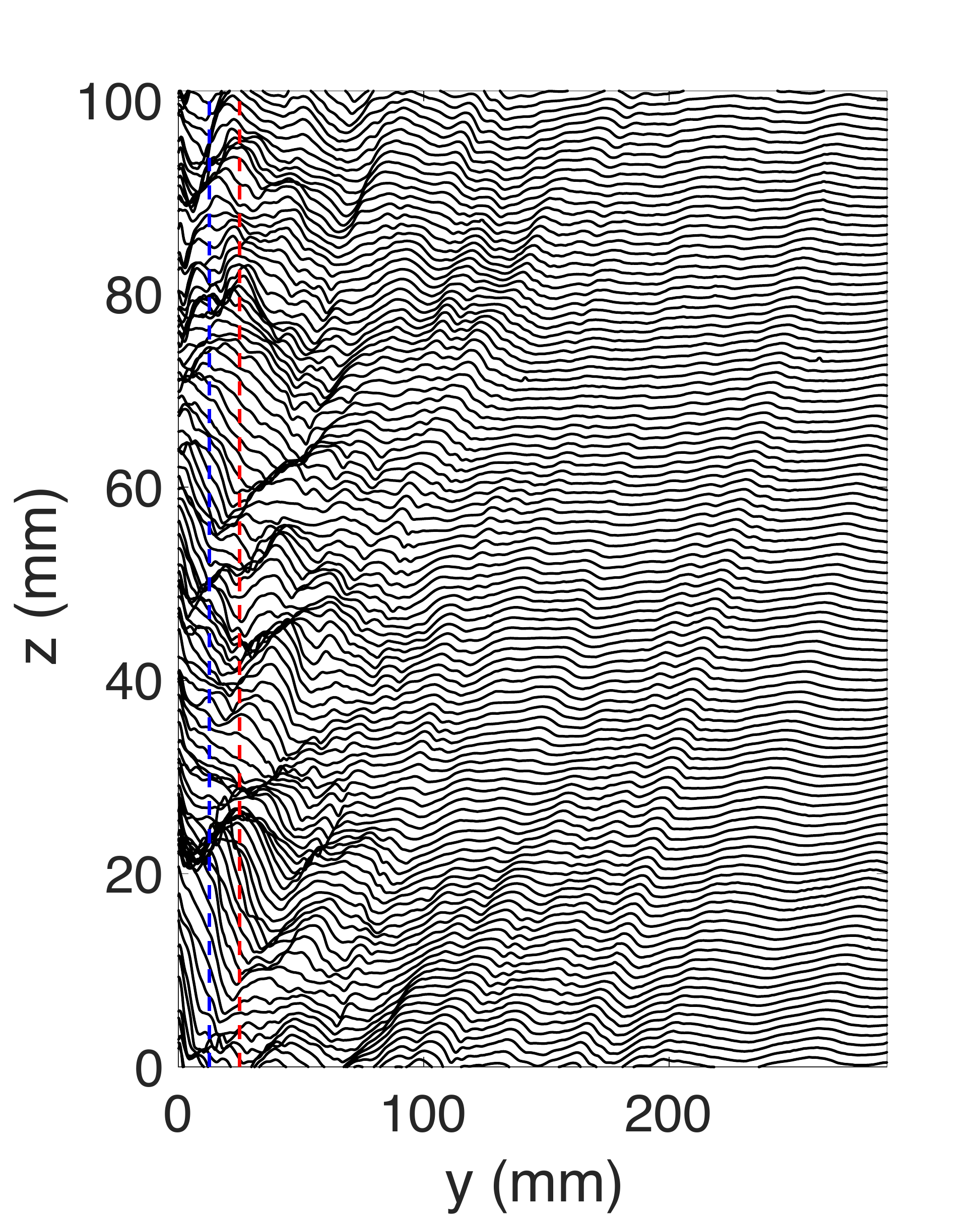}&
				\includegraphics[trim=0 0 0.7in 0in, clip=true,width=2in]{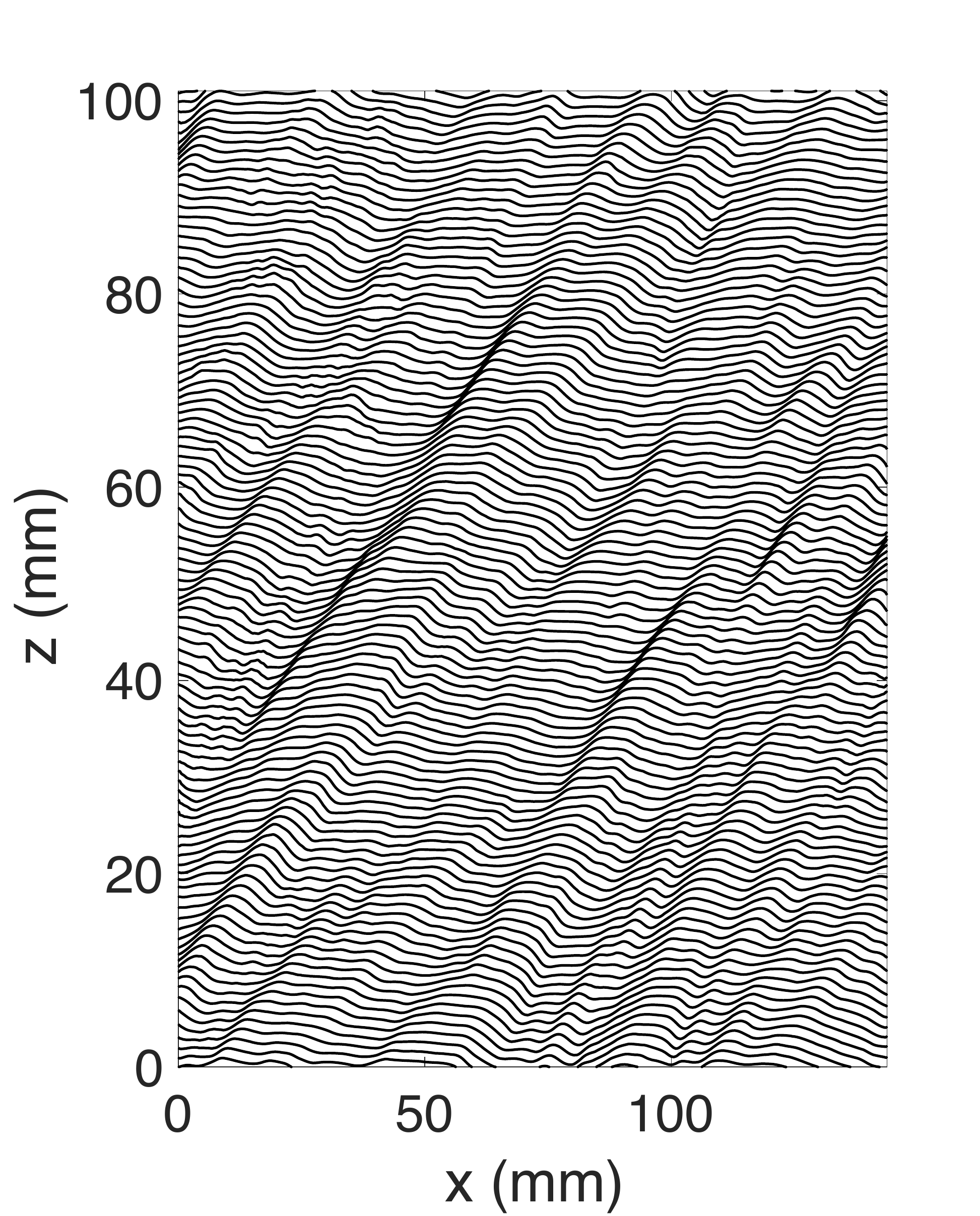}&
				\includegraphics[trim=0 0 0.7in 0in, clip=true,width=2in]{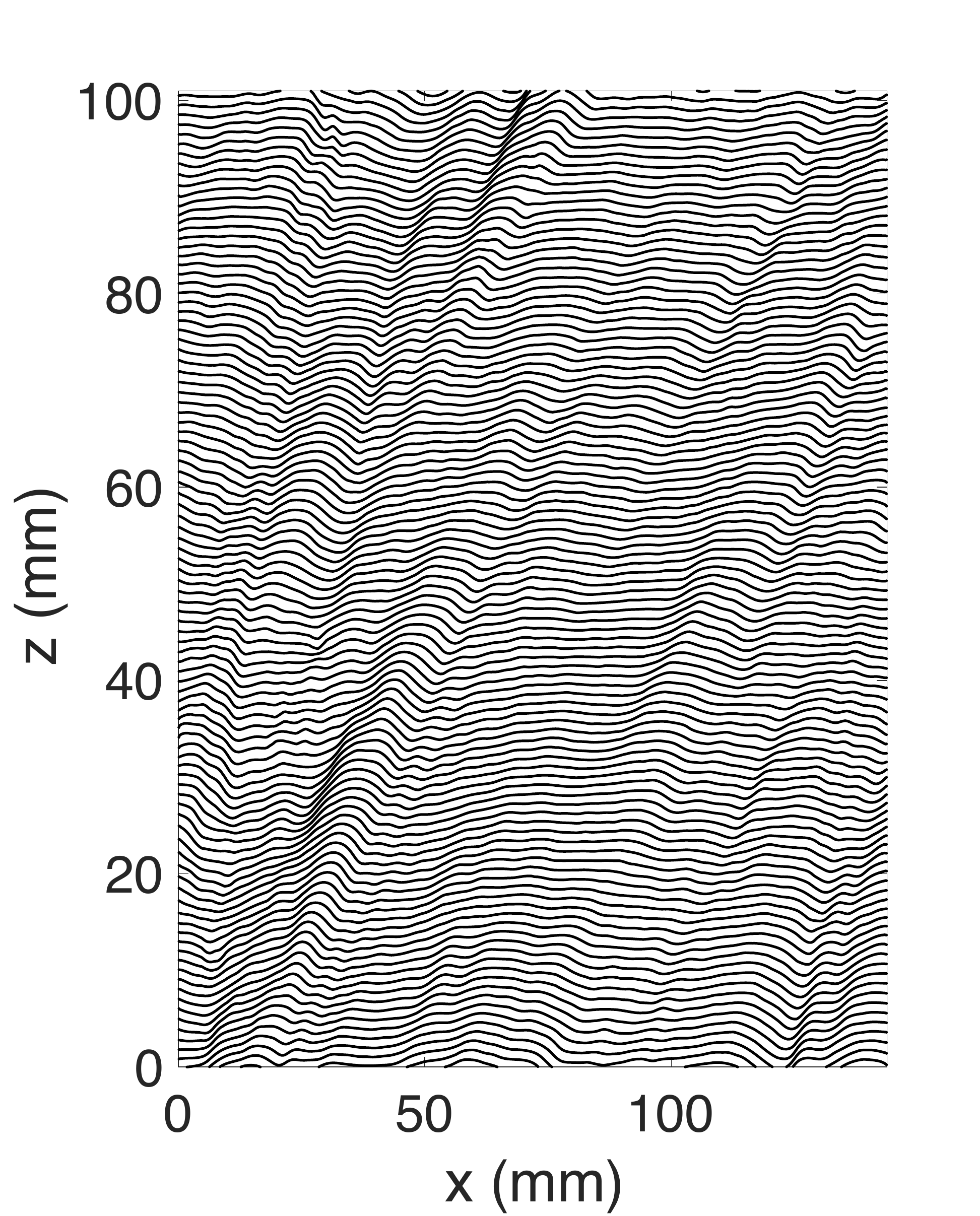}
			\end{tabular}
		\end{center}
		\caption{Sequence of profiles of the water surface during belt launch to 3~m/s. (a) Light sheet normal to the belt. (b) Light sheet parallel to the belt at $y=1.25$ cm. (c) Light sheet parallel to the belt at $y=2.5$ cm. The dashed lines in (a) show the location of the light sheet in (b) and (c). The time between profiles is 4 ms in (a) and 1 ms in (b) and (c). Each profile is shifted by 1 mm from the previous profile to reduce overlap and show propagation of surface features through time. The belt is located at y=0 in (a). The estimated propagation speed of the surface features in y-direction and for a region far from the belt for (a) is 0.34 m/s, and in x-direction for (b) and (c) is 1.0 m/s and 0.75 m/s, respectively. The sequence of images are taken when the belt has travelled approximately 10 m.}
		\label{fig:surface_profiles}
	\end{figure*}
	
	In addition to qualitative observations of free surface motions, quantitative surface profiles can be extracted from each frame of the LIF movies through the use of gradient-based image processing techniques. Figure \ref{fig:surface_profiles} shows an example of the surface profiles extracted from LIF images using image processing in MATLAB for a belt speed of $3$ m/s. In these plots the horizontal axis is the horizontal distance in each set of movies. The surface profiles in Figure \ref{fig:surface_profiles} (a) come from movies of the laser light sheet perpendicular to the belt, similar to the images show in Figure \ref{fig:overall}, hence the horizontal axis is in the y direction with the belt located at $y = 0$ mm. The surface profiles in Figure \ref{fig:surface_profiles} (b) and (c) come from movies of the surface profile parallel to the belt at two different distances from the belt (b, $y = 1.25$ cm and c, $y = 2.5$ cm), similar to the surface profile images shown in Figure \ref{fig:LighSheetParallel}; hence the horizontal axis is in the $x$ direction. The profiles are spaced in time by $4$~ms in (a) and by $1$~ms in  (b) and (c),  Each new surface profile is shifted 1 mm up from the previous profile to reduce overlap and show the propagation of surface features through time. The earliest profile in time is shown at the bottom.
	
	Using this plotting technique surface features like ripple crests can be tracked over a number of successive profiles and the slopes of imaginary lines connecting these features  indicates their horizontal speed. Analyzing \ref{fig:surface_profiles} (a) we can estimate the speed of surface features moving away from the belt at a speed of $0.34$ m/s, which is much lower than the belt speed of $3$ m/s. It should be noted that there is a constant train of surface features propagating outwards in plot (a). Plot (b) shows the parallel surface profiles at $y = 1.25$ cm away from the belt. The location of the light sheet is shown as a blue dashed line in plot (a). Surface features propagating along the direction of the belt can be seen and their speed is estimated to be around $1$ m/s in the $x$ direction. Similarly, plot (c) shows parallel surface profiles at $y = 2.5$ cm away from the belt (its location shown in a red dashed line on plot (a)), with surface features speed estimated to be about $0.75$ m/s. Theses surface feature speed estimates from plots (b) and (c) are taken fairly close to the surface of the belt, yet their speeds are significantly less than the belt speed of 3~m/s. 
	It's interesting to note that surface features traveling parallel to the belt ($1$ m/s), measured at a distance of $y = 1.25$ cm away from the belt, travel about three times faster than features moving away from the belt ($0.34$ m/s).  It should be kept in mind these velocity estimates are the $y$ or $x$ components of the phase speed.\par 
	
\begin{figure}[!htb]
		\centering
		\begin{overpic}[width=0.8\linewidth]{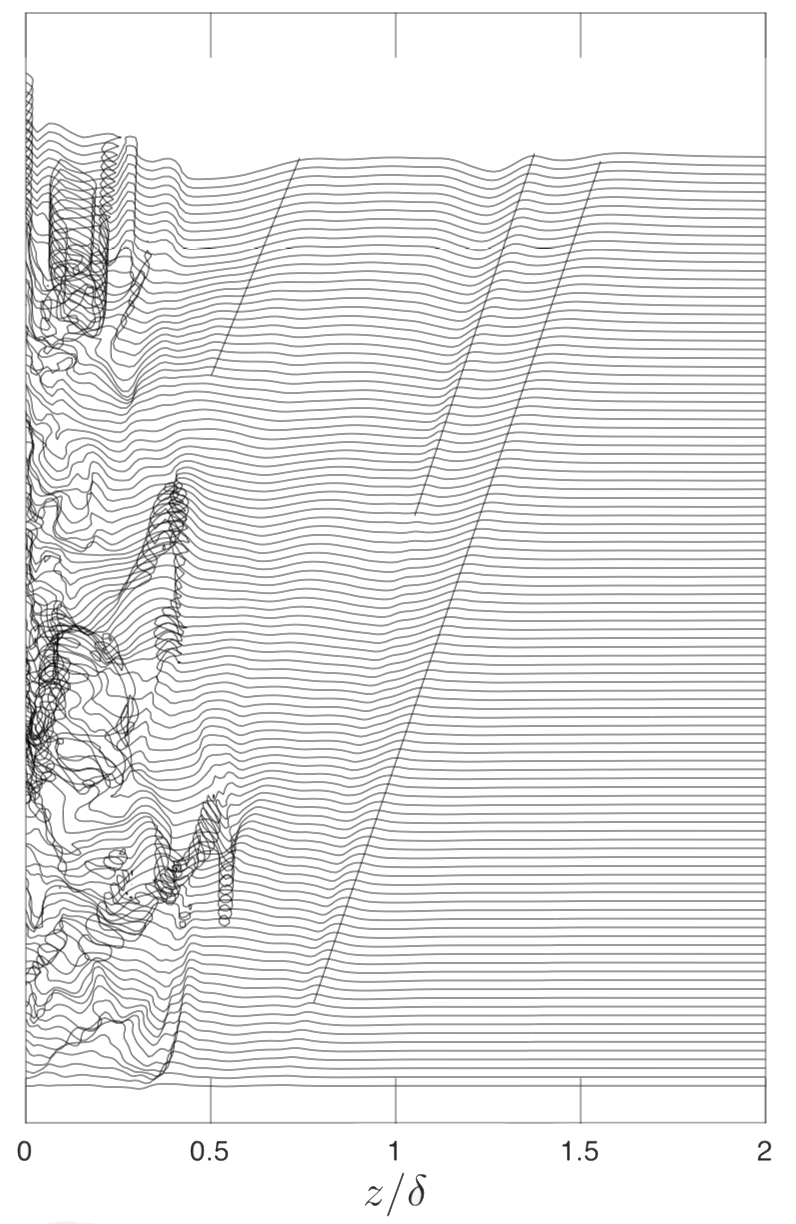}
			\put(30,1){\colorbox{white}{$y/\delta$}}
		\end{overpic}
		\caption{Evolution of the free surface near a surface piercing belt as calculated by a DNS. Profiles are taken at mid-streamwise location at $Fr_\theta=12$ and are normalized in the y direction by the length of the computational domain. }
		\label{fig:instant_DNS_profiles}
	\end{figure}

\begin{figure}[!htb]
		\begin{center}
			\begin{tabular}{c}
				(a)\\
				\includegraphics[width=3in]{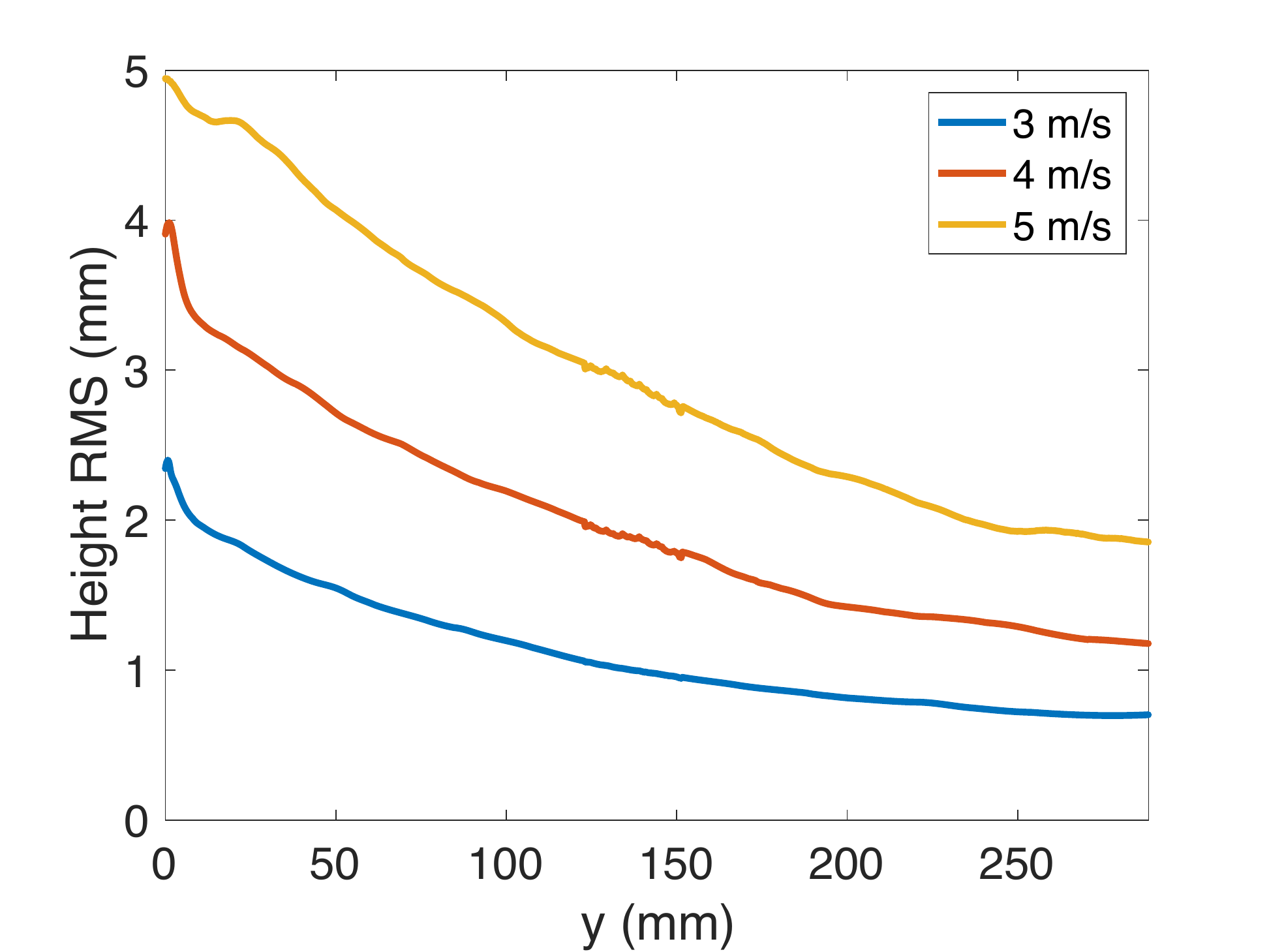} \\
				(b)\\
				\includegraphics[width=3in]{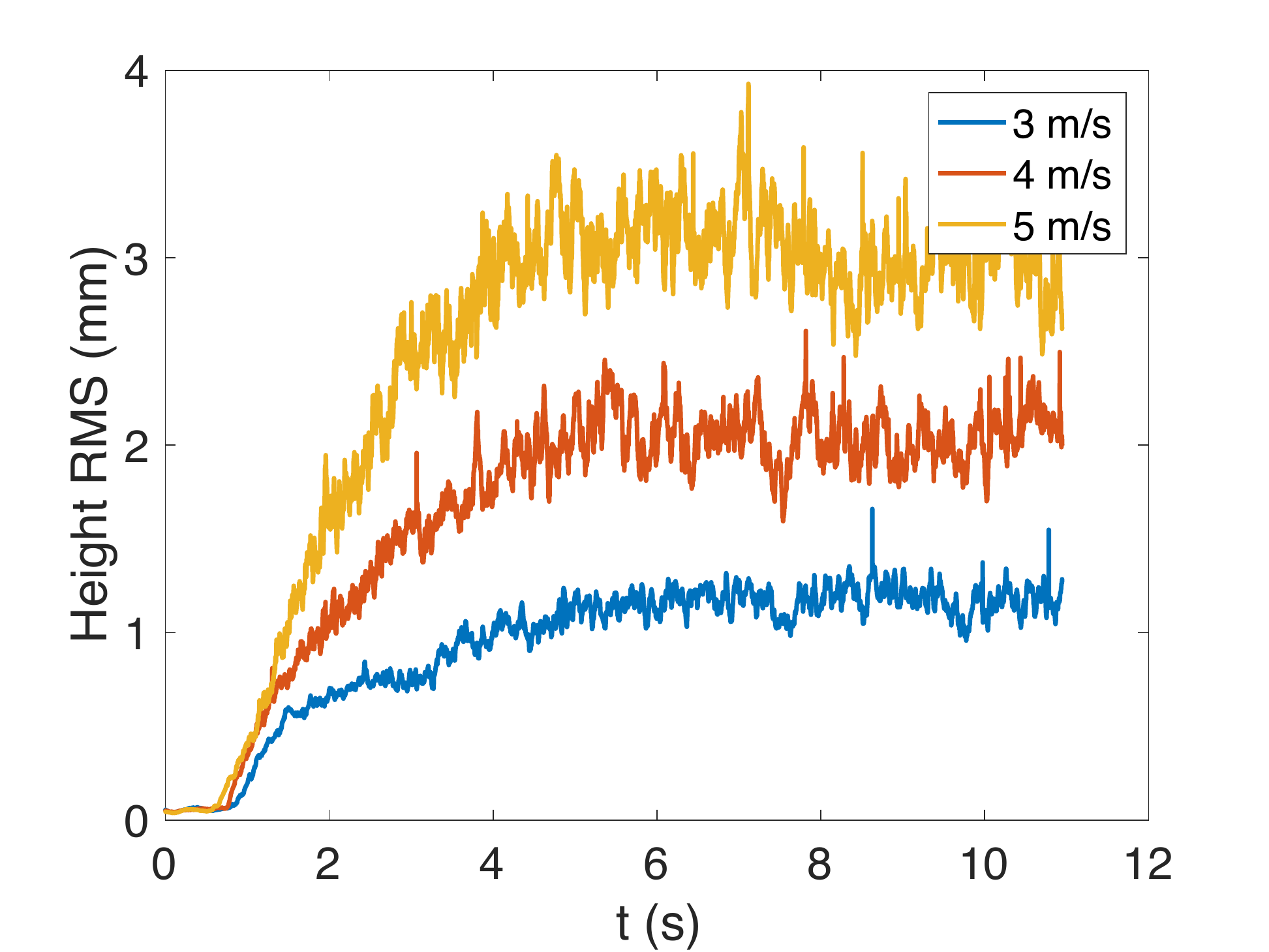}\\
			\end{tabular}
		\end{center}
		\vspace*{-0.1in} \caption{RMS of surface height fluctuations for light sheet perpendicular to the belt. (a) Height RMS in time versus distance from the belt for speeds of $U$ = 3, 4, and 5 m/s, and (b) RMS surface height averaged over $0\leq y \leq 30$~cm versus time for the same three speeds. Each curve is an ensemble average over 20 identical runs for each belt velocity.}
		\label{fig:rms_height}
	\end{figure}

	Comparison of computational results to experiments can be made by considering a succession of surface profiles at the mid-streamwise location of the numerical domain for $Fr=12$ (Figure \ref{fig:instant_DNS_profiles}), analogous to the experimental data in figure \ref{fig:surface_profiles}. The profiles are plotted in the same manner in both figures.  In Figure~\ref{fig:instant_DNS_profiles}, the lowermost profile corresponds to $Re_\theta=900$ and the uppermost profile corresponds to $Re_\theta=1400$. The surface disturbances have greater amplitude closer to the moving wall as compared to the outer regions, in agreement with the experiments. In the immediate vicinity of the moving wall ($0<y/\delta<0.25$) the disturbances appear to be uncorrelated and persist for only a few profiles, suggesting that in this region waves are heavily influenced by the underlying turbulent boundary layer flow. In the regions away from the wall however, the waves persist for much longer periods and maintain their shape, similar to the results in the experiments as discussed in the previous paragraph. The straight black lines track the crests of a few of the outer region waves. This shows that the propagation speed of these waves is very nearly constant and that the waves exhibit the behavior of freely moving waves. On average the propagation speed normalized by wall speed is 0.08 which is on the same range as the experiments (the corresponding experimental value for a belt speed of 3 m/s is 0.11). Overall from a qualitative point of view, the computations are in agreement with the experimental results and capture the surface dynamics of the two-phase turbulent boundary layer.\par
	The profile data from the experiments discussed above was used to obtain distributions of the root-mean-square (RMS) water surface height fluctuation as a function of time and space dimensions.  
	The RMS height at any $x$ or $y$  location is obtained as the square root of the average of the squares of the differences between the height and the average height, where the average is taken over the run time and over all experimental runs with the same belt speed.   Figure~\ref{fig:rms_height} (a) is a plot of the RMS free surface height versus $y$  at belt speeds  $U = $5, 4, and 3~m/s averaged over 20 runs for each speed. The belt is located at $y = 0$ mm. The RMS height reaches a maximum near the belt region where the free surface fluctuations were visibly the most violent as seen in Figure~\ref{fig:surface_profiles} (a). Further away from the belt the free surface RMS fluctuations decay for all three belt speeds. Figure \ref{fig:rms_height} (b) shows the RMS height, averaged over all $y$, versus time for all three belt speeds. The belt starts to move at $t = 0$~s, however the RMS height for all three speeds does not change until a little bit before $t = 1$~s. The RMS height then increases at similar constant rate  for all three belt speeds until about $t = 1.5$~s when the three curves start to diverge. After approximately $t = 5$~s, all three height RMS curves reach a constant value.
	
	
\begin{figure}[!htb]
		\begin{center}
			\includegraphics[width=3in]{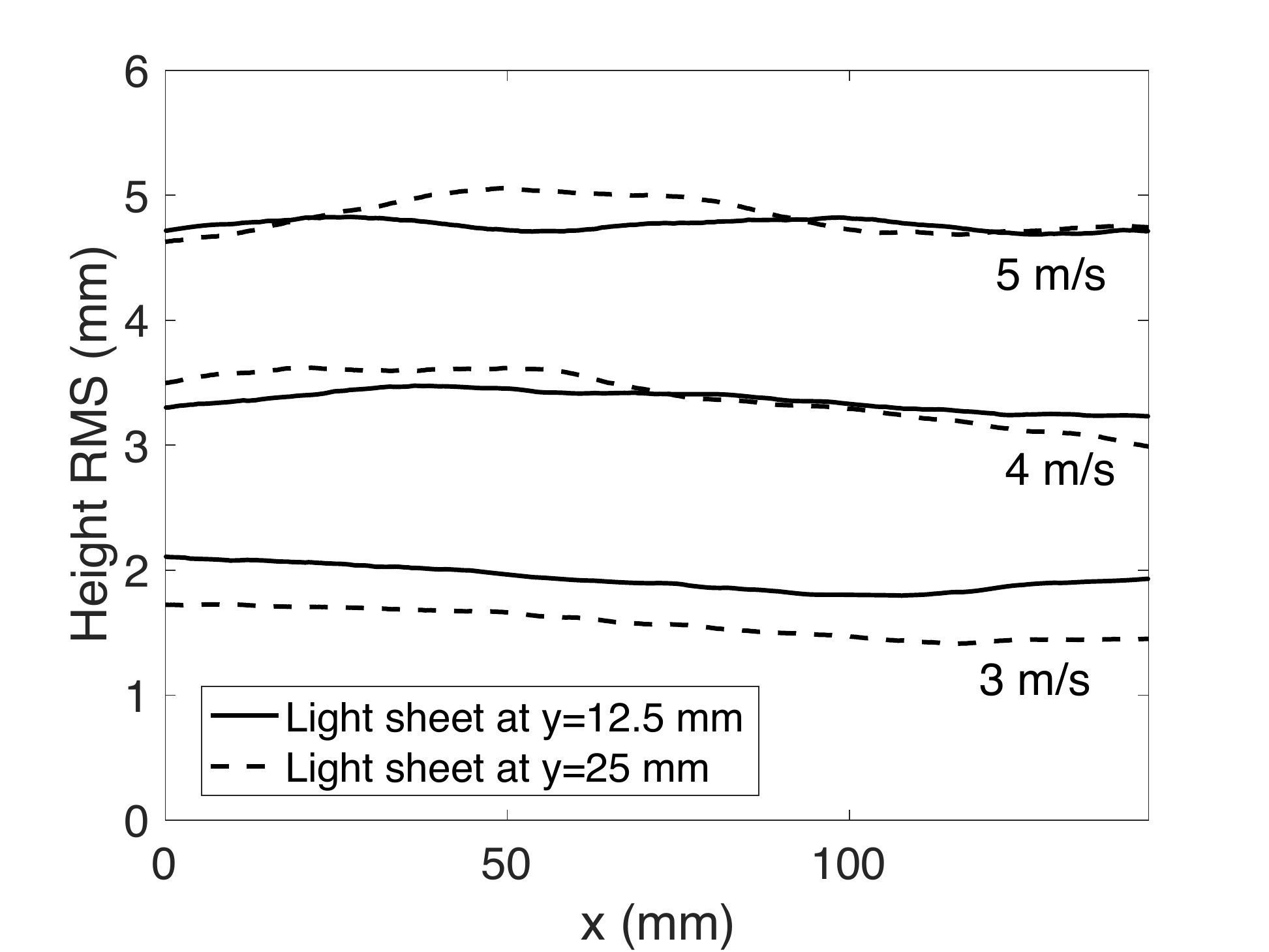} \\
		\end{center}
		\vspace*{-0.1in} \caption{RMS of surface height fluctuations in time versus x for light sheet parallel to the belt at $y = 12.5$ and $25.0$~mm away from the belt surface.}
		\label{fig:rms_height_parallel}
	\end{figure}

	A similar analysis was carried out on the surface profiles parallel to the belt. Figure~\ref{fig:rms_height_parallel} shows the RMS height versus $x$ for belt speeds of $U =$ 3, 4, and 5~m/s at two different distances from the belt for each speed. The RMS height varies with $x$ in a random manner with a relatively small amplitude; it is thought that this variation would decrease with increasing numbers of runs.  
	In agreement with Figure~\ref{fig:rms_height} (a), the data in Figure~\ref{fig:rms_height_parallel} indicates a strong increase in the RMS height with belt speed and little change (except for the $U=$ 3~m/s case) between the two measurement locations, $y = $ 12.5 and 25~mm.

	\subsection{Bubble Statistics}
	
	In this section, preliminary estimates of bubble statistics are reported for a belt speed of 5~m/s. A single camera set up, as shown in Figure~\ref{fig:bubble_setup_planar}, was used to record images of entrained bubbles as the belt starts from rest and travels a distance of $24.88$~m. The camera has a field of view of 47 by 47 mm at the plane of the belt surface and a resolution of 67~$\mu$m per pixel.   The belt is illuminated with diffuse white light and the movies are recorded at 1,000~fps with a total of 5,000 images taken for each run. The resulting series of images were processed using a MATLAB code which identifies all the bubbles in each image. Bubbles are measured down to a diameter of ~0.5~mm.  With the lens f-number used in the measurements, bubbles are in focus over a horizontal distance of about 35 mm from the belt surface along the line of sight  of the camera lens and this region contains all the bubbles present in the imaged region of the flow.  Since most of the larger bubbles are not spherical, an equivalent radius is calculated based on the two-dimensional projected shape imaged by the camera. Each bubble is then subsequently tracked based on the series of frames in which it appears and its two-dimensional position and average equivalent radius are recorded. 
	
\begin{figure}[!htb]
		\begin{center}
			\includegraphics[trim=0.0in 0.0in 0.0in 0.00in,clip=true,width=3in]{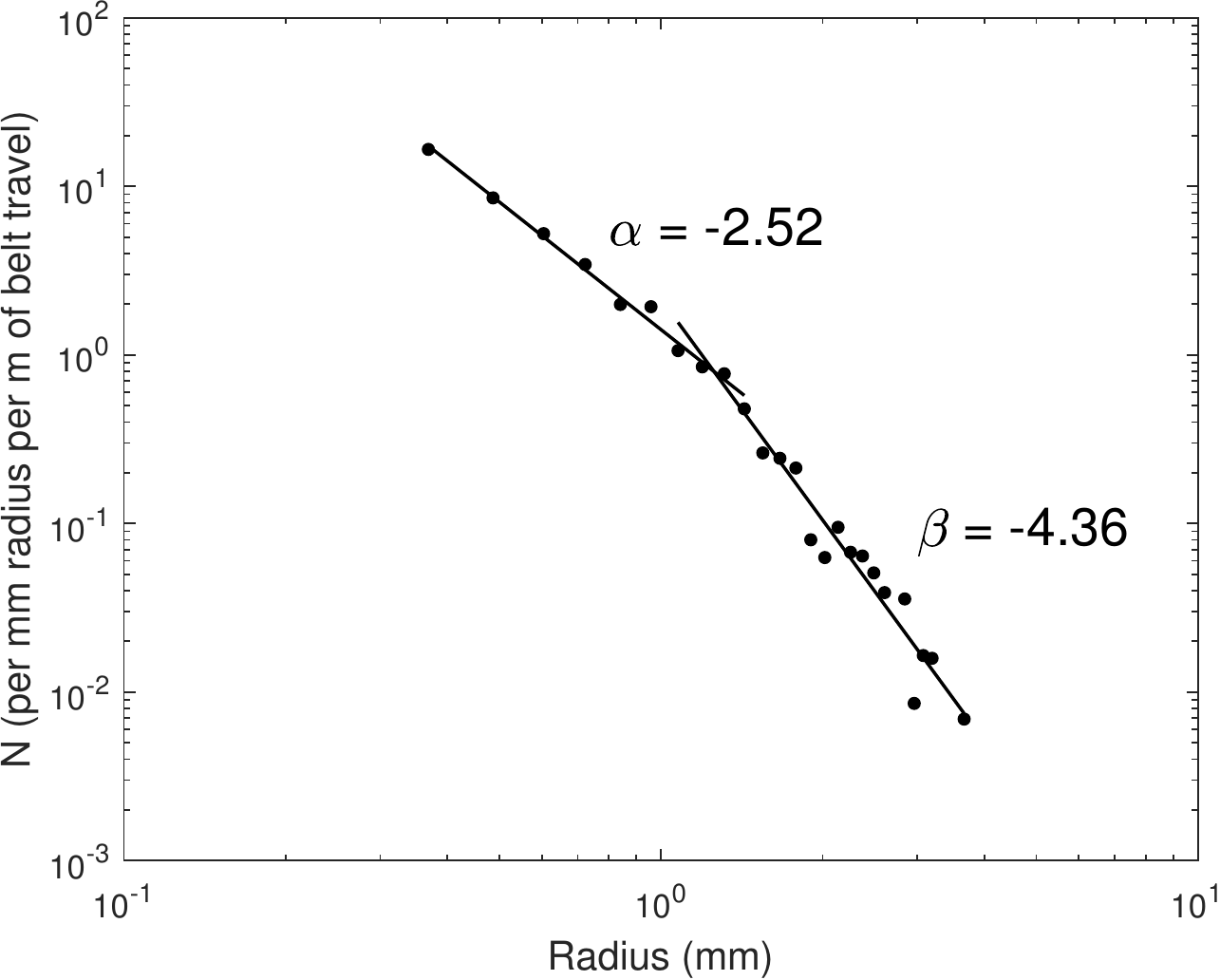}\\
		\end{center}
		\vspace{-0.2in} \caption{Log-log plot of the number of bubbles per unit radius bin width $dr$ versus bubble radius for a belt speed of 5~m/s.  The data set includes all the bubbles that passed through the upstream side of the measurement region in an experimental run with a length of belt travel of 24.88~m.   A constant value of radius bin width $dr = $ 0.118~mm was used and the centers of the bins ranged from $ r = $ 0.369 to 3.669~mm.   Two distinct linear regions are observed with a distinct break in slope identified at a bubble radius of $r_i = 1.265$ mm. The slope of the upper region is given by $\alpha =  -2.52$~mm$^{-1}$ and the lower region by $\beta = -4.36$~mm$^{-1}$.}
		\label{fig:prob_dist_bubbles_5mps}
	\end{figure}

	Figure~\ref{fig:prob_dist_bubbles_5mps} is a log-log plot of the experimentally measured number of bubbles per radius bin width versus bubble  radius for a belt speed of 5~m/s.  The sample population is all of the bubbles that passed through the upstream side of the measurement region during the belt travel of 24.88~m. A uniform bin spacing of $dr = $ 0.118~mm is used and the centers of the bins range from $ r = $ 0.369 to 3.669~mm. 
	Separate linear regions are observed for small-diameter and large-diameter bubbles. The two linear regions are fitted separately using linear regression to a function of the form $Ar^{\alpha}$ for the region of smaller bubbles and $Br^{\beta}$ for the region of larger bubbles. These functions plot as  straight lines in Figure~\ref{fig:prob_dist_bubbles_5mps} and the  optimum position  for the break in slope between the two regions was determined by an iterative bisection-like routine outlined as follows: First, an initial guess, $r_0$, is estimated as the radius where the break in slope is to occur, the data is split into two distinct sets and a power law, of the form described above, is fitted to each set. The intersection, $r_i$, of the two fitted lines is then found.  If the difference between $r_0$ and $r_i$ does not fall within a specific tolerance, a new guess for $r_0$ between the previous values of $r_0$ and $r_i$ is assigned and the processes is repeated until the tolerance is reached. Using this method with an initial guess of $r_0 = $ 1.3~mm, the break in slope is estimated to be approximately $r_i = $ 1.265~mm.
	The break in slope in the bubble size distribution has long been observed and identified as the Hinze scale, see for example the work of  \cite{Deane2002} on bubble size distributions in breaking waves. Generally speaking, the Hinze scale implies that different physical mechanisms influence the two different sides of the bubble size spectrum. 
	Dean and Stokes suggest that  bubbles that were larger than the Hinze scale were fragmented by turbulent flow with a -10/3 power-law scaling, while bubbles smaller than the Hinze scale are stabilized by surface tension and show a -3/2 power-law scaling with the radius \citep{Deane2002}.  The Hinze scale is defined as
	
	\begin{equation}
	r_H=2^{-8/5}\epsilon^{-2/5}(\sigma We_c/\rho)^{3/5}
	\end{equation}
	where $\epsilon$ is the turbulent dissipation rate and $We_c$ is the critical Weber number and typically takes on a value of 4.7 (see for example \cite{Deane2002}).\par

\begin{figure}[!htb]
		\centering
		\includegraphics[width=\linewidth]{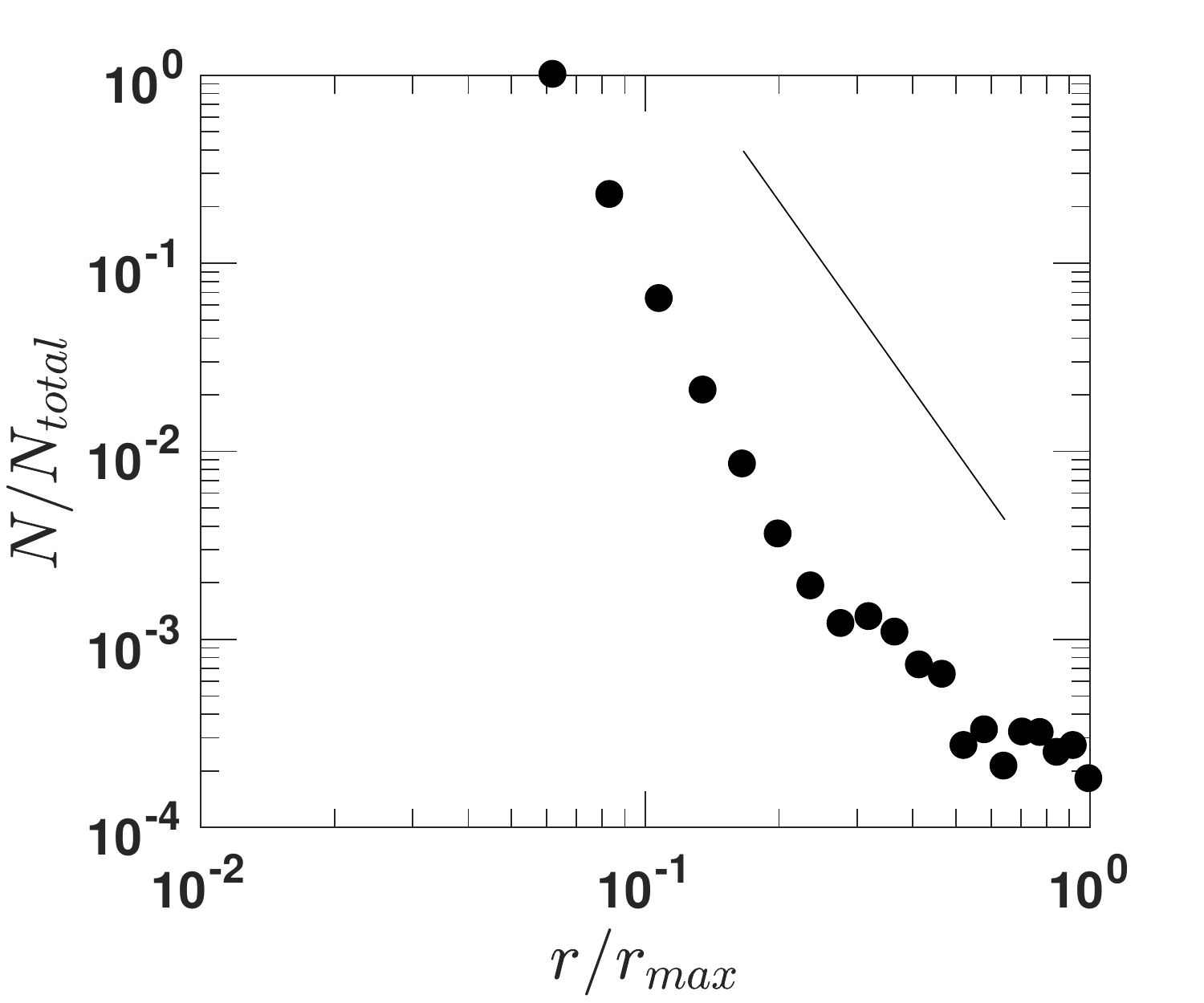}
		\caption{Relative bubble population versus bubble radius from a DNS simulation of the problem. Horizontal axis has been normalized by the maximum bubble radius and the vertical axis by the total number of bubbles observed.  The bubble radius is the radius of a spherical bubble with the same volume as the irregularly shaped bubble in the calculations.}
		\label{fig:NvsRad}
	\end{figure}

	In our numerical simulations we found that $r_H /\delta \approx3.5\times10^{-3} $. The largest bubbles in our simulations have a radius of about $\delta/20 \sim 0.16$ with 25-30 computational points across the bubble whereas the smallest observed bubbles have a radius of about $\delta/160\sim 0.02$ with 3-4 points across. The latter is an order of magnitude greater than the Hinze scale. Further refinement in the numerical resolution to capture smaller bubbles is out of the scope of the present DNS, which focused on the primary entrainment events as a result of the turbulent boundary layer interacting with the free surface. Figure \ref{fig:NvsRad} shows the number of observed bubbles against bubble radius from the DNS simulation. 
	The radius of a bubble is calculated by considering
	the equivalent spherical bubble with the same volume.
	The vertical axis has been normalized with the total number of observations and the radius has been normalized by the largest radius in the data set. A line with a slope of $-10/3$ is plotted to the top right of the data for reference. The scaling agrees fairly well with the $-10/3$ law and shows further qualitative agreement to the experiments (Figure \ref{fig:prob_dist_bubbles_5mps}).
	
\begin{figure}[!htb]
		\begin{center}
			\includegraphics[trim=0.0in 0.0in 0.0in 0.00in,clip=true,width=3in]{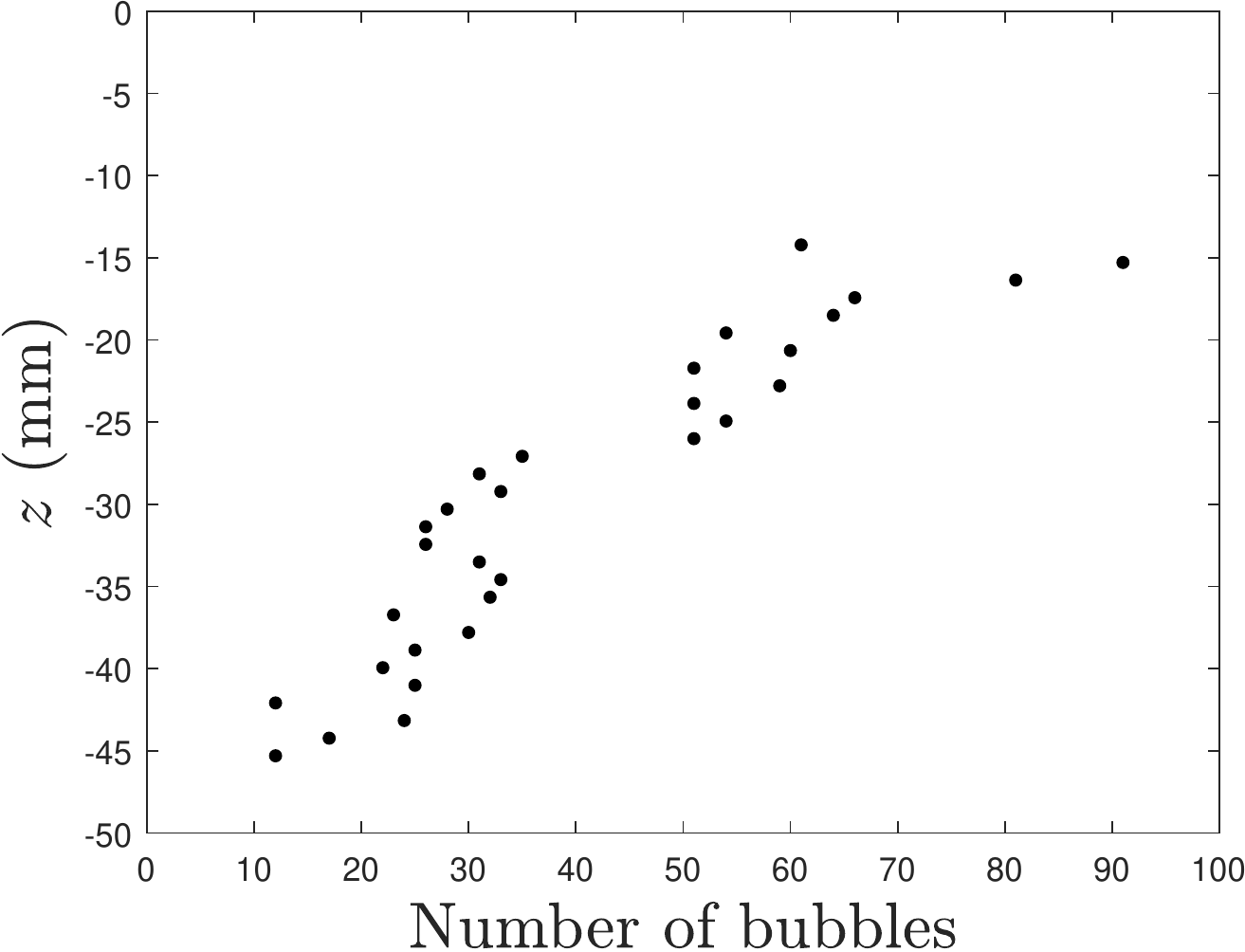}\\
		\end{center}
		\vspace{-0.2in} \caption{Total number of bubbles observed in the z-direction (depth) from launch of the belt to $24.88$ meters of belt traveled at a speed of 5 m/s. Each increment in the z-direction, $dz = 1.072$ mm from a depth of $z = -14.2$ mm to $z = -45.3$ mm relative to the height of the free surface before the belt launch.}
		\label{fig:num_of_bubbles_vs_z_5mps}
	\end{figure}

	Figure \ref{fig:num_of_bubbles_vs_z_5mps} shows the mean number of unique bubbles vs. depth measured in the experiments for a belt travel of $24.88$~m at a belt speed of 5~m/s. Before the launch of the belt, the calm free surface is positioned at $z = $ 0 mm. Once the belt is launched, the free surface fluctuates dramatically in the $z$ direction, making it difficult to measure bubbles close to $z = $ 0 mm. Generally, the free surface does not fluctuate more than 15~mm below it's original depth, which is the reason why the measurements presented start at $z \approx 14$~mm. Bubbles are tracked in the series of images in which they appear, the average depth of the bubble is obtained by averaging the $z$ position over all the tracked particle trajectories. The $z$ direction is divided from $z = -14.2$~mm to $z = -45.3$~mm by increments of $dz = 1.072$~mm. Each bubble's mean $z$ position is then placed into each appropriate bin. From Figure \ref{fig:num_of_bubbles_vs_z_5mps}, it can be seen that the number of bubbles slowly decreases from a few hundred bubbles in the area around $z = 15$~mm to tens of bubbles near $z = 50$~mm indicating that the majority of these large bubbles in the boundary layer tend to stay near the surface.
	
\begin{figure}[!htb]
		\centering
		\includegraphics[width=\linewidth]{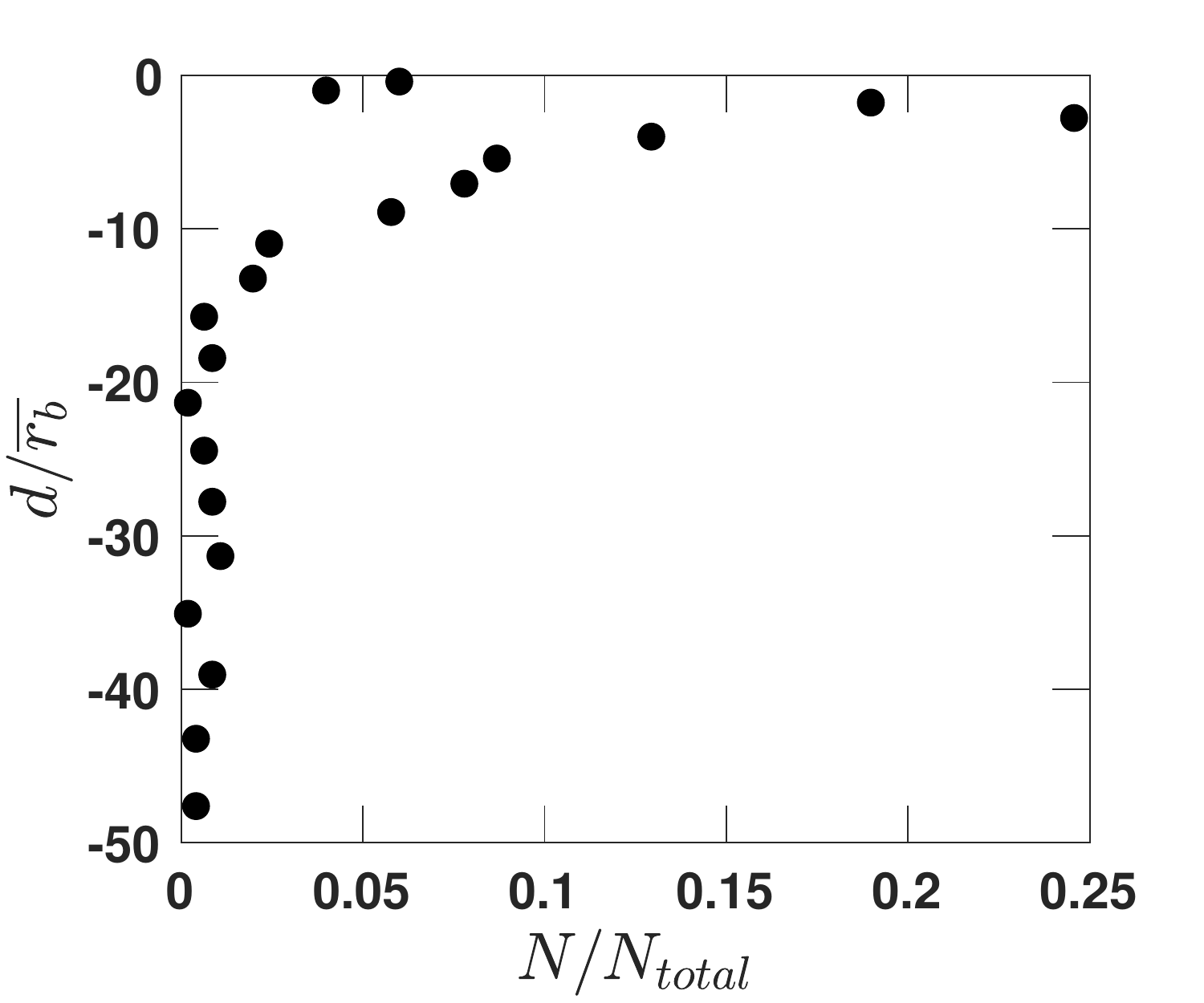}
		\caption{Relative bubble population with respect to depth (DNS). Depth has been normalized by average bubble radius.}
		\label{fig:numBubbleVsDepth}
	\end{figure}

	Figure \ref{fig:numBubbleVsDepth} shows depth of observed bubbles in DNS against their relative population and is qualitatively analogous to figure experimental data in \ref{fig:num_of_bubbles_vs_z_5mps}. The depth of the bubbles has been normalized by the average bubble radius and has been broken up into twenty equally sized bins. The overall trend is similar to that of the experiment where the majority of the bubbles are found closer to the free surface. It must be noted that the two plots are not directly comparable given the limitations explained earlier
	
\begin{figure}[!htb]
		\begin{center}
			\includegraphics[trim=0.0in 0.0in 0.0in 0.00in,clip=true,width=3in]{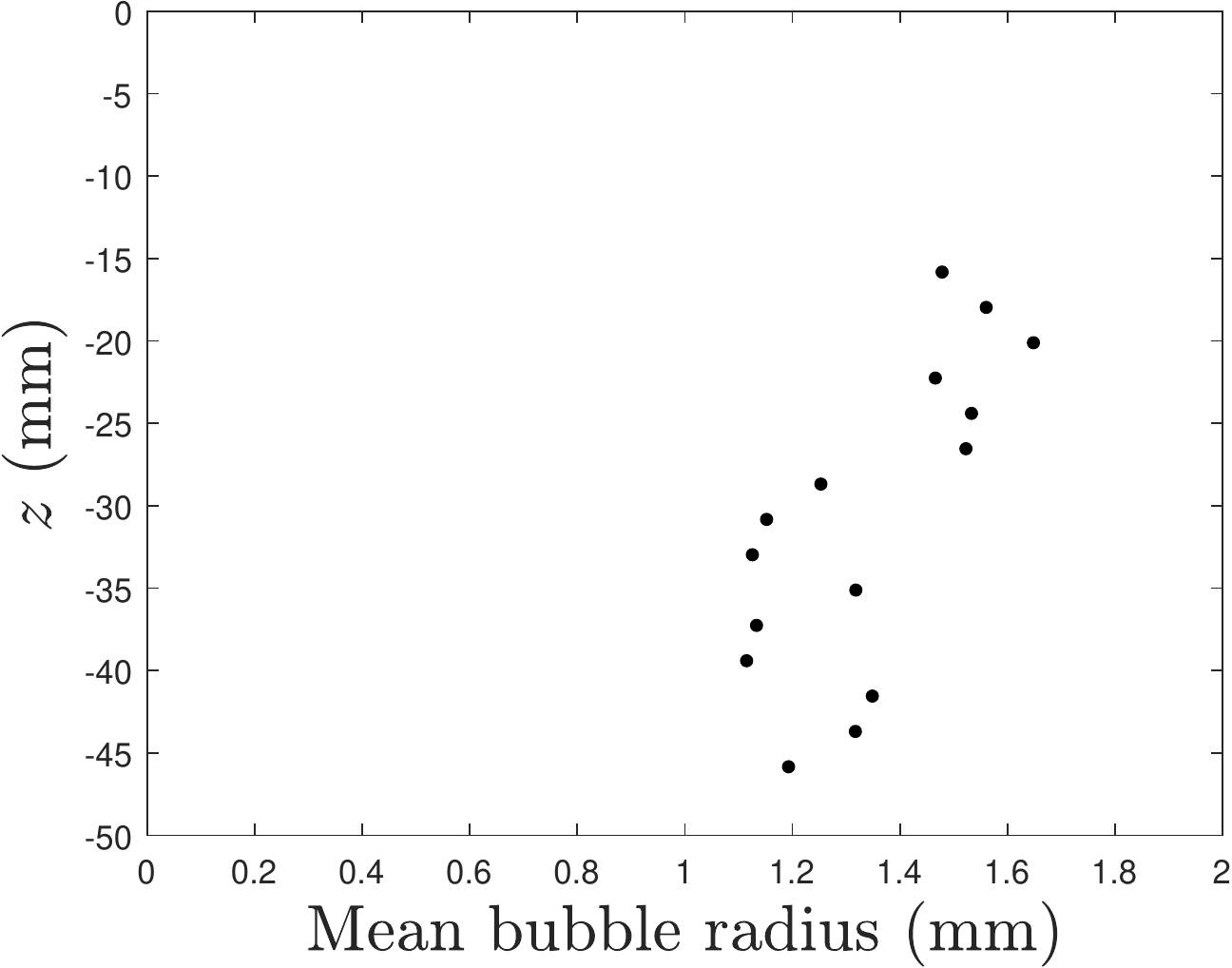}\\
		\end{center}
		\vspace{-0.2in} \caption{Average bubble radius measured in the experiments from launch of the belt to 49.22 meters of belt traveled at a speed of 5 m/s. Each increment in the z-direction $dz = 2.14$ mm from a depth of $z = -13.68$ mm to $z = -45.84$ mm relative to the height of the free surface before the belt launch.} \label{fig:mean_bubble_radius_vs_z_5mps}
	\end{figure}

	Given that most of the large bubbles reside near the free surface, it may be of interest to look at the average equivalent bubble radius versus depth. Figure~\ref{fig:mean_bubble_radius_vs_z_5mps} shows the experimentally measured average bubble radius versus depth from $z = -13.68$~mm to $z = -45.84$~mm in increments of $dz = 2.14$~mm. The average depth of each unique bubble is calculated, in a similar way as described in the previous paragraph, and the bubble is placed into the appropriate $z$ bin. Once all bubbles are assigned the the proper bin, the average radius of the bubbles in each bin is calculated. It should be noted that data points close to the free surface are averaged over significantly more bubbles than ones farther away, perhaps accounting for the noisier data at greater depths. The average bubble radius increases by about 0.4~mm (about 30\%) from the deepest and shallowest measurement positions.  
	
\begin{figure}[!htb]
		\begin{center}
			\includegraphics[trim=0.0in 0.0in 0.0in 0.00in,clip=true,width=3in]{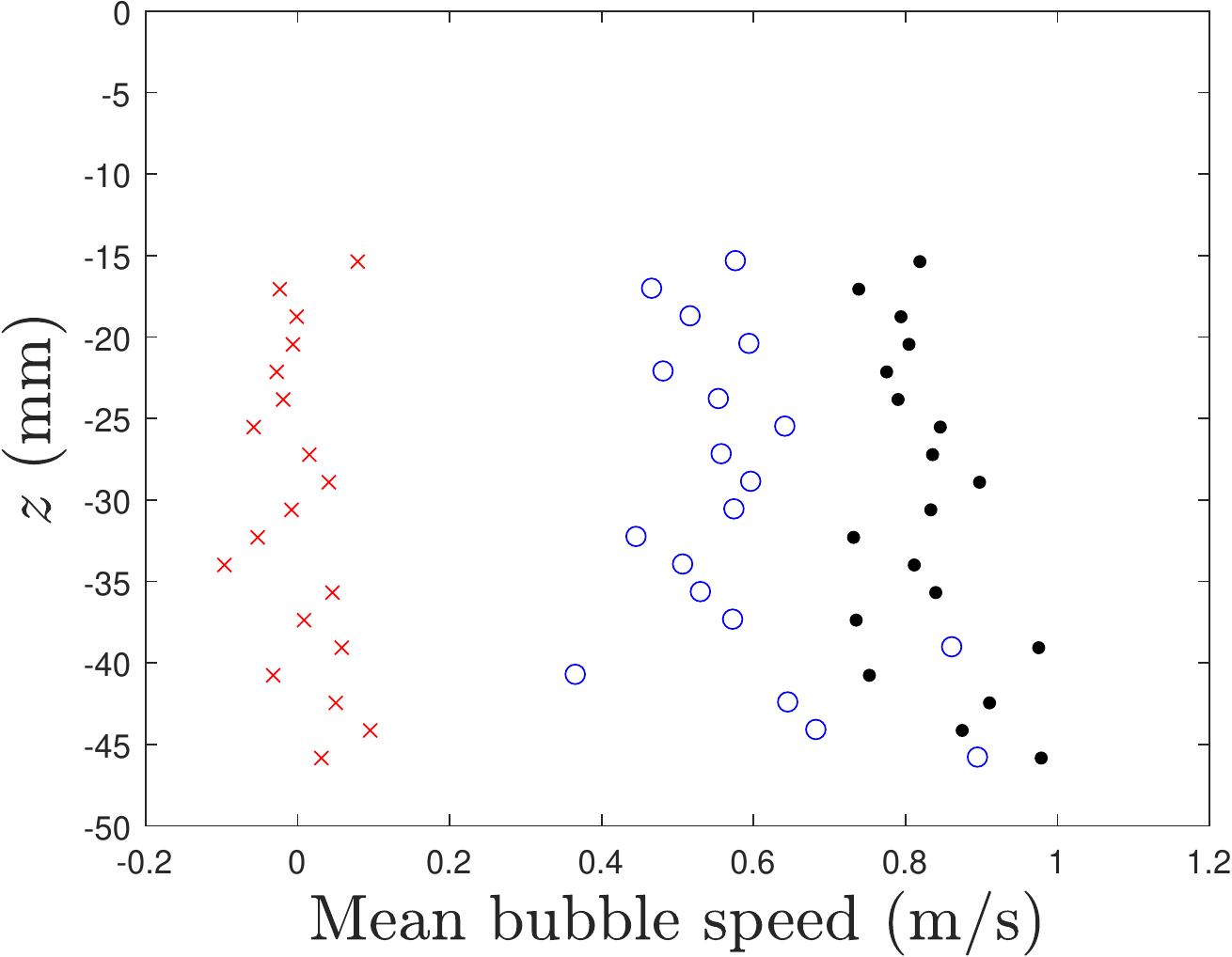}\\
		\end{center}
		\vspace{-0.2in} \caption{Average bubble speed and velocity components as a function of depth $z$. $\bullet$ indicates the average speed (in the $x$-$z$ plane). $\color{blue} \circ$ is the $u$ velocity component and  $\color{red} \times$ is the $w$ velocity component. Each data point is from bubbles whose average depth was in a bin width of $dz = 1.693$~mm with bin positions ranging from a depth of $z = -13.68$ mm to $z = -45.84$~mm relative to the height of the free surface before the belt launch.} \label{fig:mean_bubble_speed_vs_z_5mps}
	\end{figure}

	Finally, from the tracked bubble trajectories,  we can fit a second order polynomial to estimate the speed of the bubbles as they enter the camera's field of view. Second order polynomials were fitted to the $x$ and $z$ positions versus time data for each unique bubble. Then the $u$ and $w$ components of velocity were computed for each bubble as it entered the upstream side of the camera's field of view. The speed was calculated as $|\vec{u}| = \sqrt{u^2+w^2}$. The results of these calculations, including the bubble speeds and the values of the $u$ and $w$ velocity components, are shown in Figure \ref{fig:mean_bubble_speed_vs_z_5mps}.  It is interesting to note that the mean bubble speed does not seem to change dramatically over the range of depths in which the measurements were taken.  The $u$ component of velocity is 3 to 4 times larger then the $w$ component.
	
	\subsection{Analysis of turbulent structures}
\begin{figure}[!htb]
		\begin{center}
			\begin{tabular}{cc}
				\includegraphics[height=.65in]{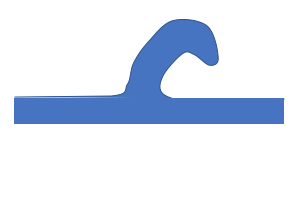} \hspace{.25in} &
				\includegraphics[height=.65in]{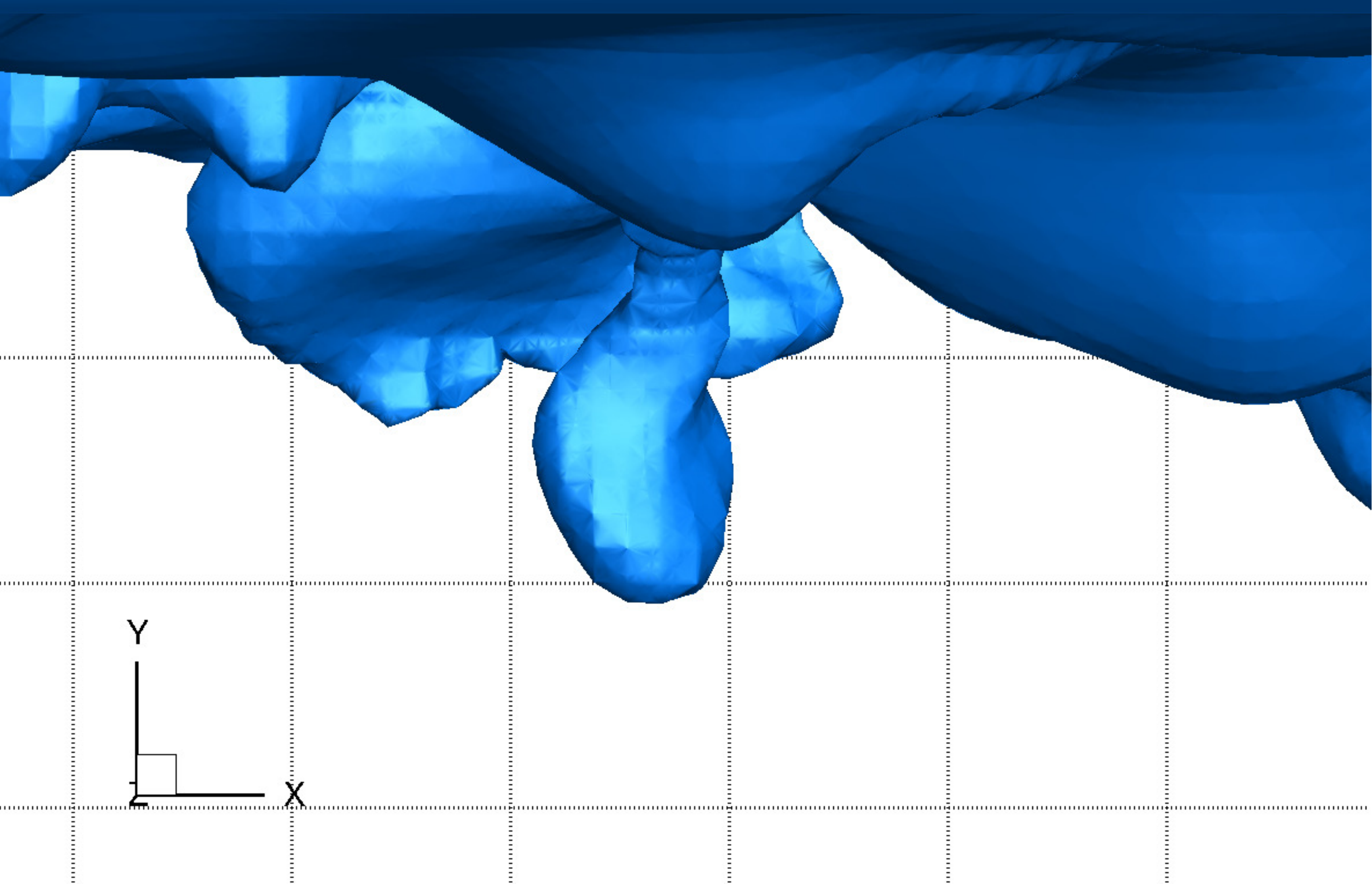}\\
				(a) & (d)\\

				\includegraphics[height=.65in]{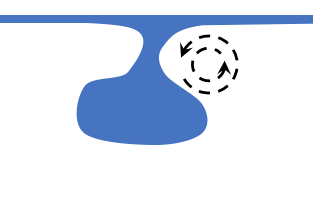} \hspace{.25in} &
				\includegraphics[height=.65in]{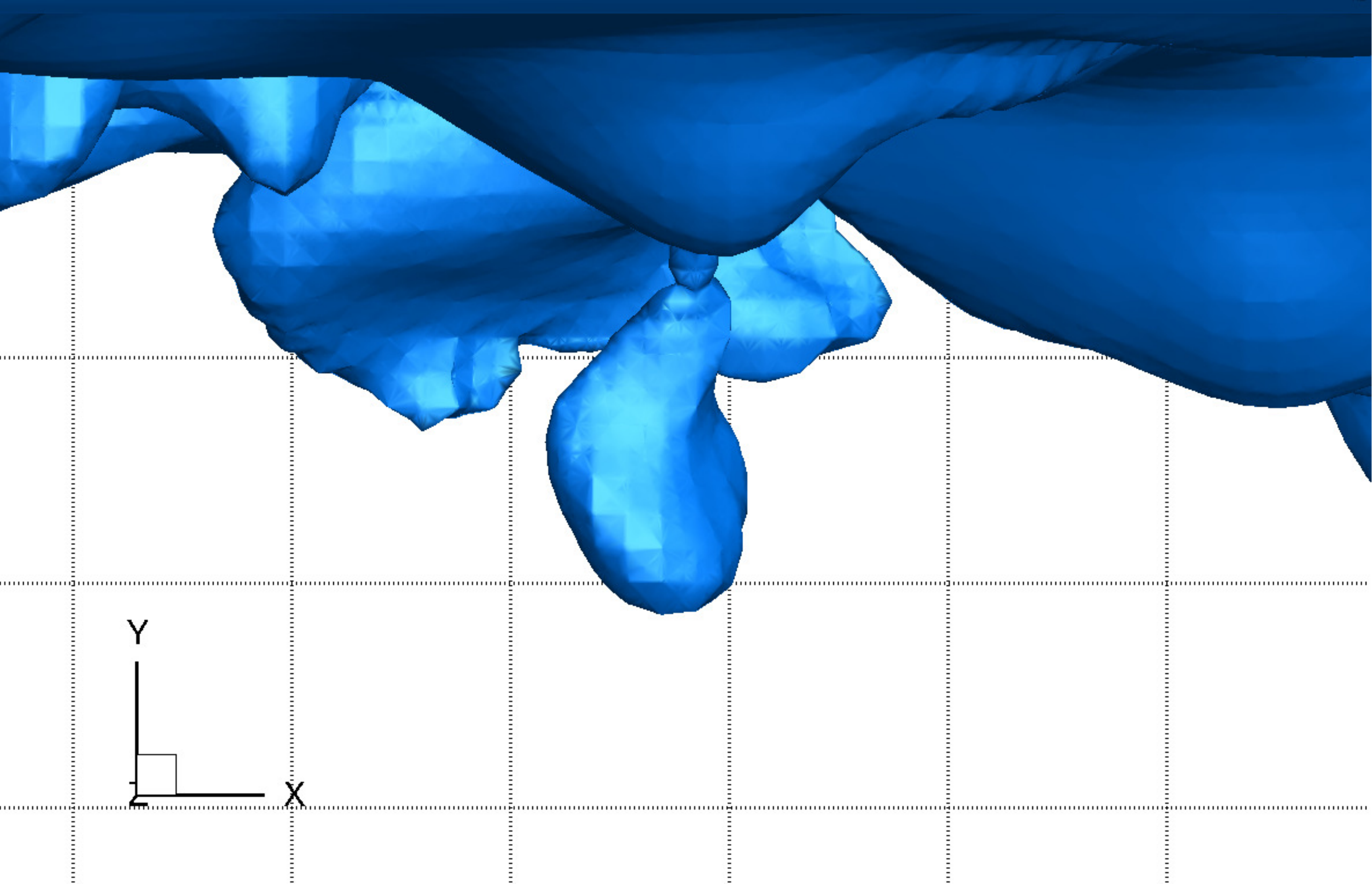}\\
				(b) & (e)\\
				
				\includegraphics[height=.65in]{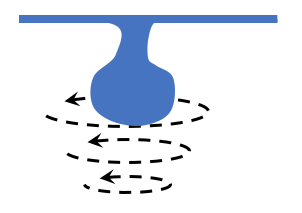} \hspace{.25in} &
				\includegraphics[height=.65in]{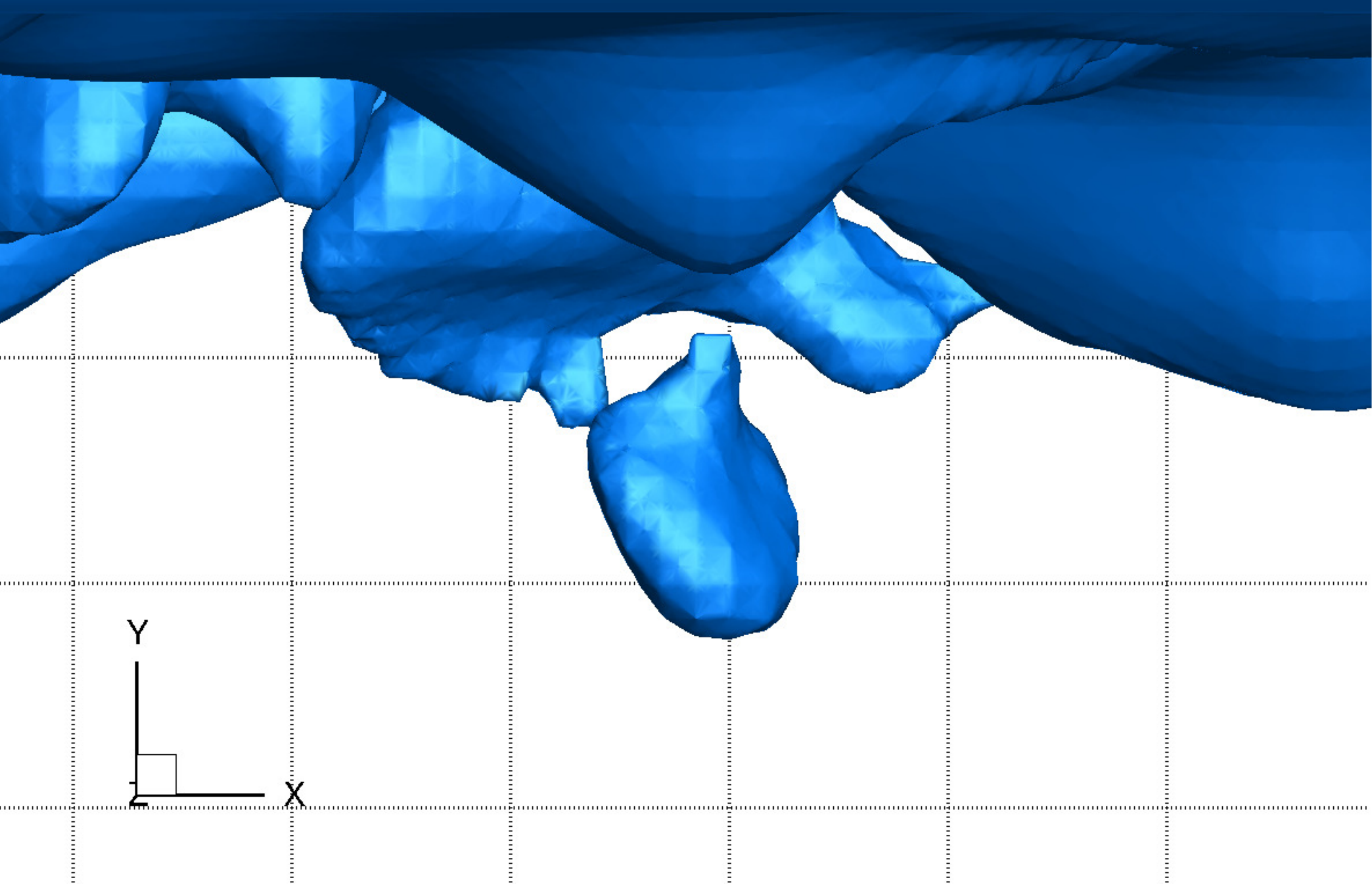}\\
				(c) & (f)\\
			\end{tabular}
		\end{center}
		\vspace*{-0.2in} \caption{Proposed air entrainment mechanism schematics, a breaking wave (a), vortices oriented parallel to the free surface (b), and vortices oriented perpendicular to the free surface (c). (d-f) shows an example sequence of a single bubble being entrained from the DNS results.} \label{fig:ent_mech}
	\end{figure}

	In this section we will discuss the mechanisms of air entrainment in the context of the numerical simulations. In general, there are three different entrainment mechanism. Water droplets, for example, can break off ligaments and entrain air upon impact with the free-surface (see Figure \ref{fig:ent_mech}a). This type of air entrainment has been studied extensively primarily in simplified configurations \citep{Esmailizadeh1986, Oguz1990, Hasan1990, TOMITA2007, Ray2015, Hendrix2016}. In such case, a droplet falling towards the free surface traps air between it and the water surface. A crater forms on the surface and upon impact of the droplet, the air inside the crater is entrapped. In the numerical study, about 12\% of the air entrainment incidents are from surface impact. Alternatively, entrainment is also caused by turbulent motions underneath the surface. There are two types of vortices that result in entrainment and are distinguished by their orientation with respect to the surface. The first are the vortices that are mainly oriented parallel to the free surface (Figure \ref{fig:ent_mech}b). The second are those that are perpendicular to the free surface (Figure \ref{fig:ent_mech}c). In our numerical study, we found that the latter type of vortices are rare ($< 1\%$) and most of the turbulent entrainment comes from the former type ($\sim 88\%$). A small portion of the entraining vortices lie between the two where the orientation with respect to the free surface is not clear.\par

\begin{figure}[!htb]
		\centering
		\begin{overpic}[width=\linewidth]{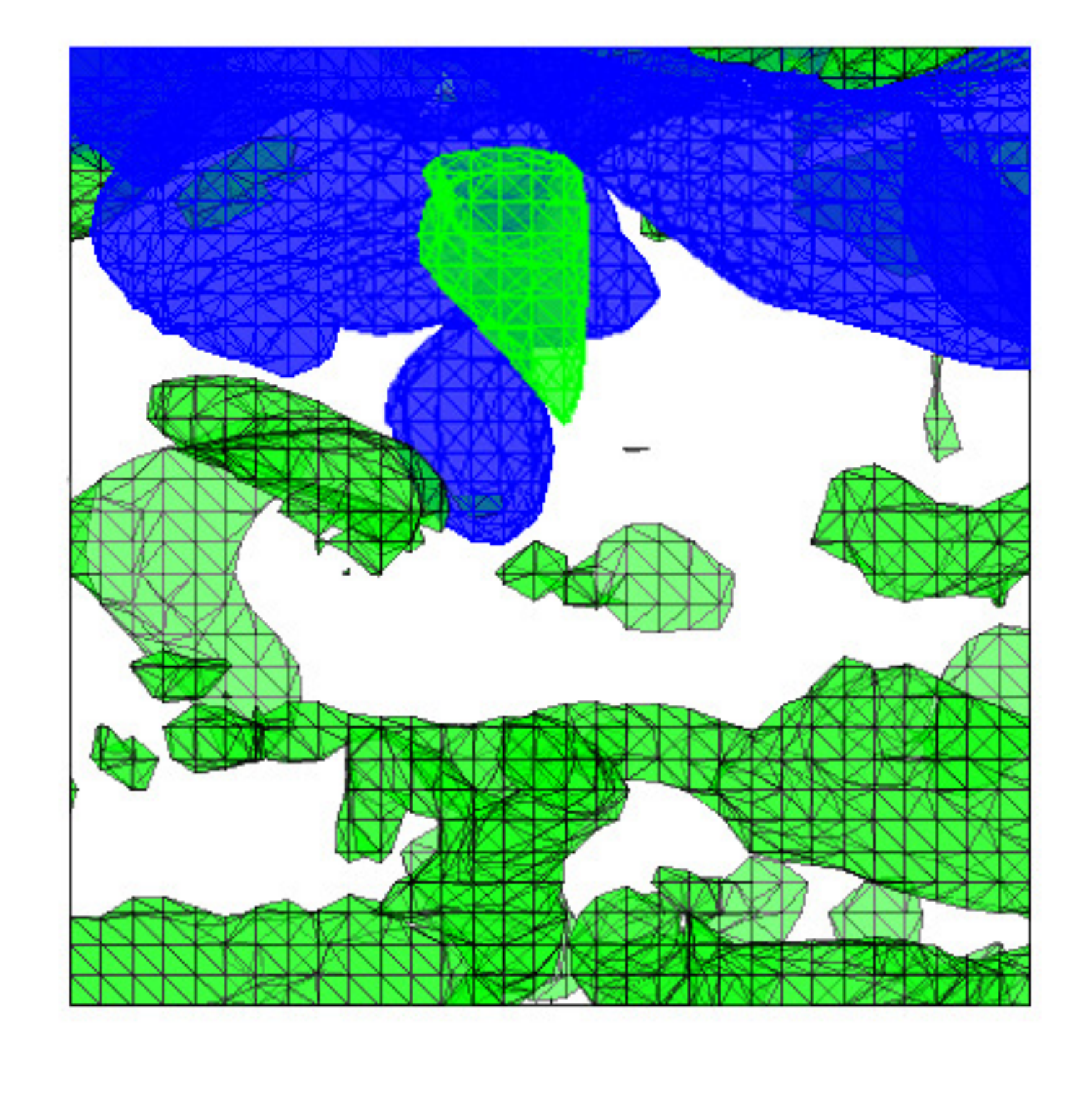}
			\put(75,48){\vector(-1,1){20}}
			\put(50,40){\colorbox{white}{Primary Vortex}}
		\end{overpic}
		
		\caption{An example of vortical structures at an entraiment site from the DNS simulation}
		\label{fig:all_vorts}
	\end{figure}

	The typical process of this entrainment regime begins with the interface being pulled into the flow beneath, creating a "neck" where the ligament attaches to the free surface (Figure~\ref{fig:ent_mech} d). The neck continues to narrow until it breaks and an air bubble is released into the flow. In the numerical studies, we focus our attention on bubble generation and will not discuss the fate of the bubbles later on. We can examine turbulent air entrainment by considering the local turbulent eddies using the vortex identification scheme introduced by \citet{Hunt1988} known as the Q-criterion. Any spatial point where the Eulerian norm of the vorticity tensor dominates that of the rate of strain is designated as being part of the vortical structure. Figure \ref{fig:all_vorts} shows the vortical structures in the vicinity of an entrainment event. The vortex core closest to the neck of the air ligament is identified in the figure. The low pressure center of the vortex has pulled in the interface and is narrowing the neck.
	
	The Q-criterion can be useful in identifying the vortex cores but is not a reliable method for establishing the size of the vortices and hence their local length scale. We employ the following 8 step procedure to identify and quantify the local scale of vortices that entrain air.\\
	
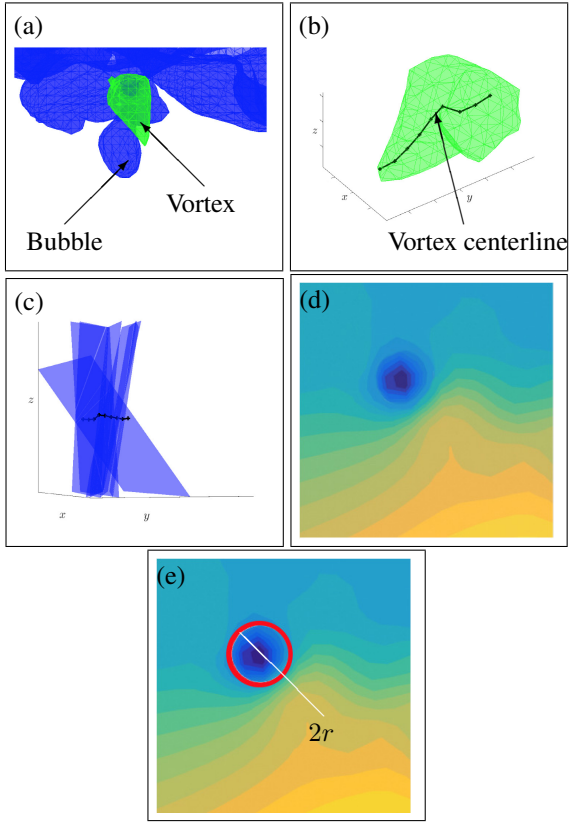
\begin{figure}[!htb]
		\centering
		\fbox{\begin{overpic}[width=0.425\linewidth]{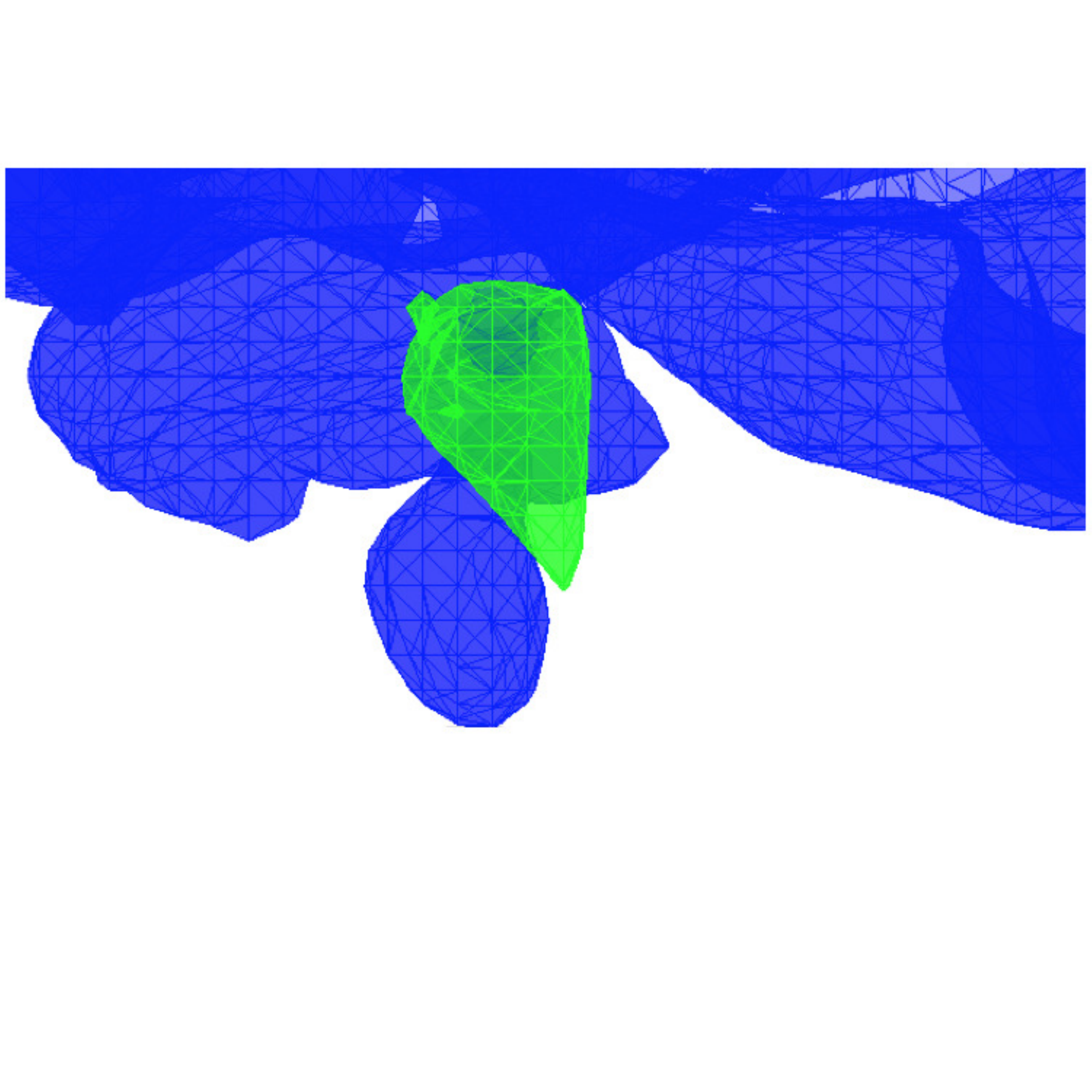} 
				\put(00,90){\text{(a)}}
				\put(75,30){\vector(-1,1){25}}
				\put(60,20){\text{Vortex}}
				\put(20,15){\vector(1,1){25}}
				\put(5,5){\text{Bubble}}
		\end{overpic} }
		\fbox{\begin{overpic}[width=0.425\linewidth]{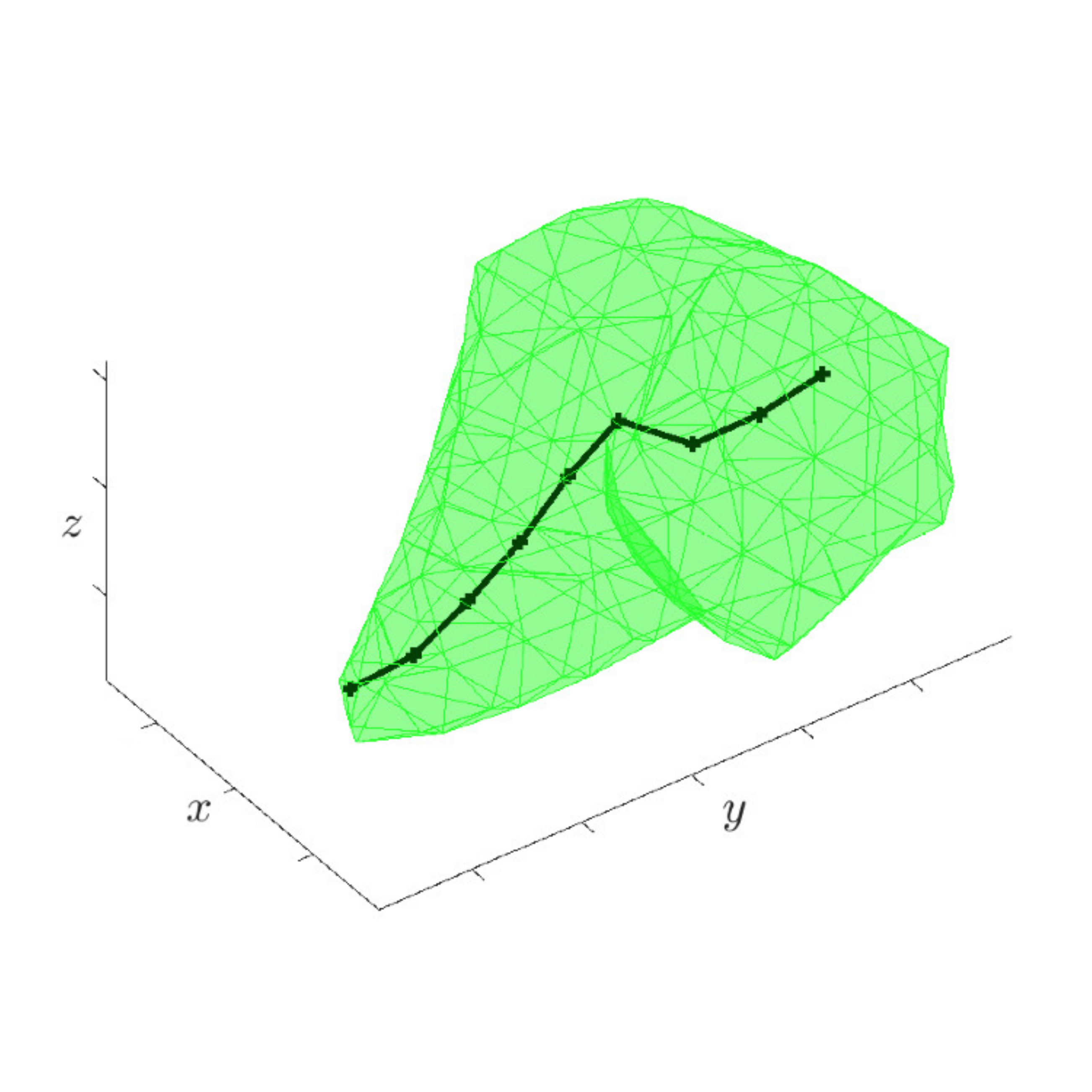}
				\put(0,90){\text{(b)}}
				\put(65,15){\vector(-1,4){11}}
				\put(35,5){\text{Vortex centerline}}
		\end{overpic} }\\
		\fbox{\begin{overpic}[width=0.425\linewidth]{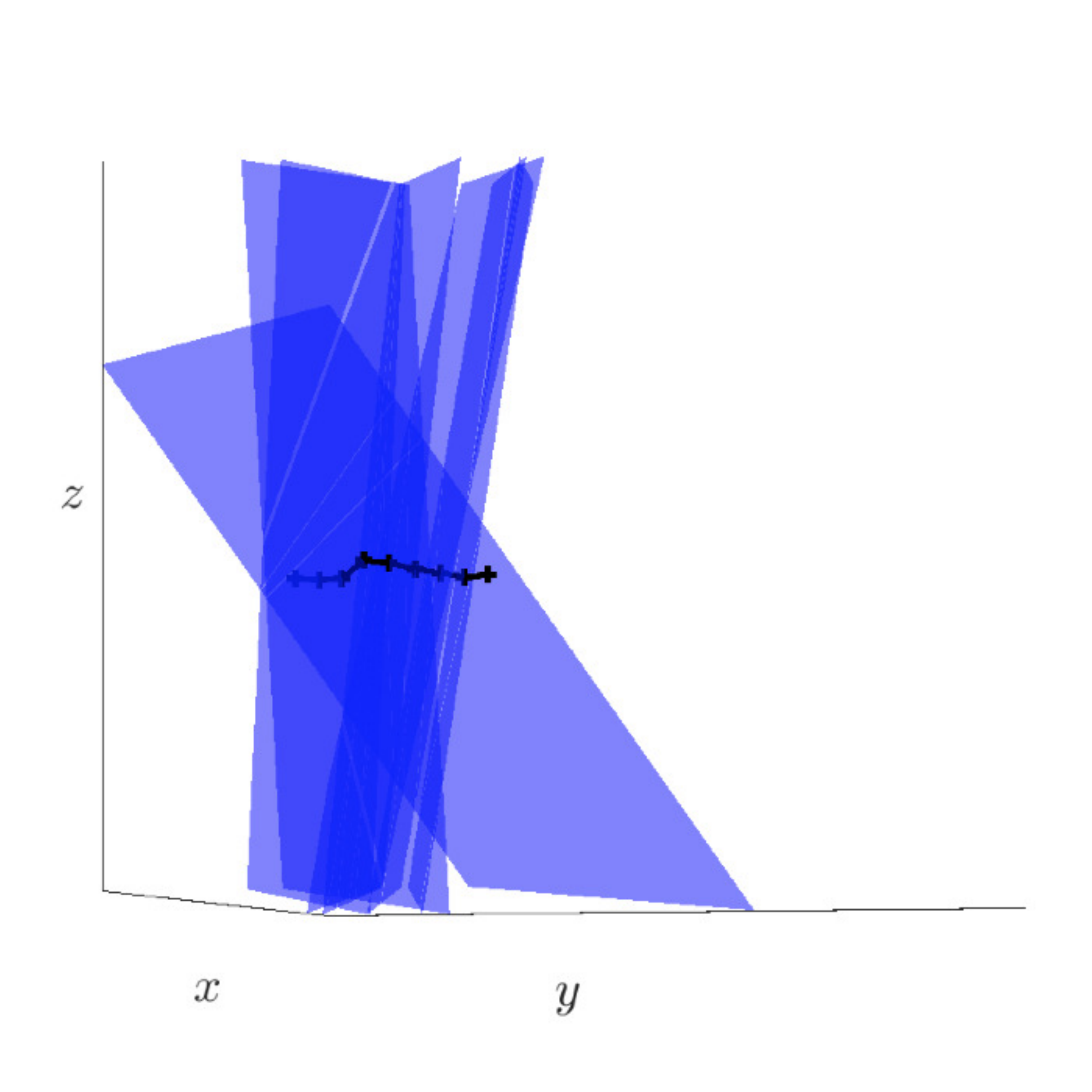}
				\put(00,90){\text{(c)}}
		\end{overpic} }
		\fbox{\begin{overpic}[width=0.425\linewidth]{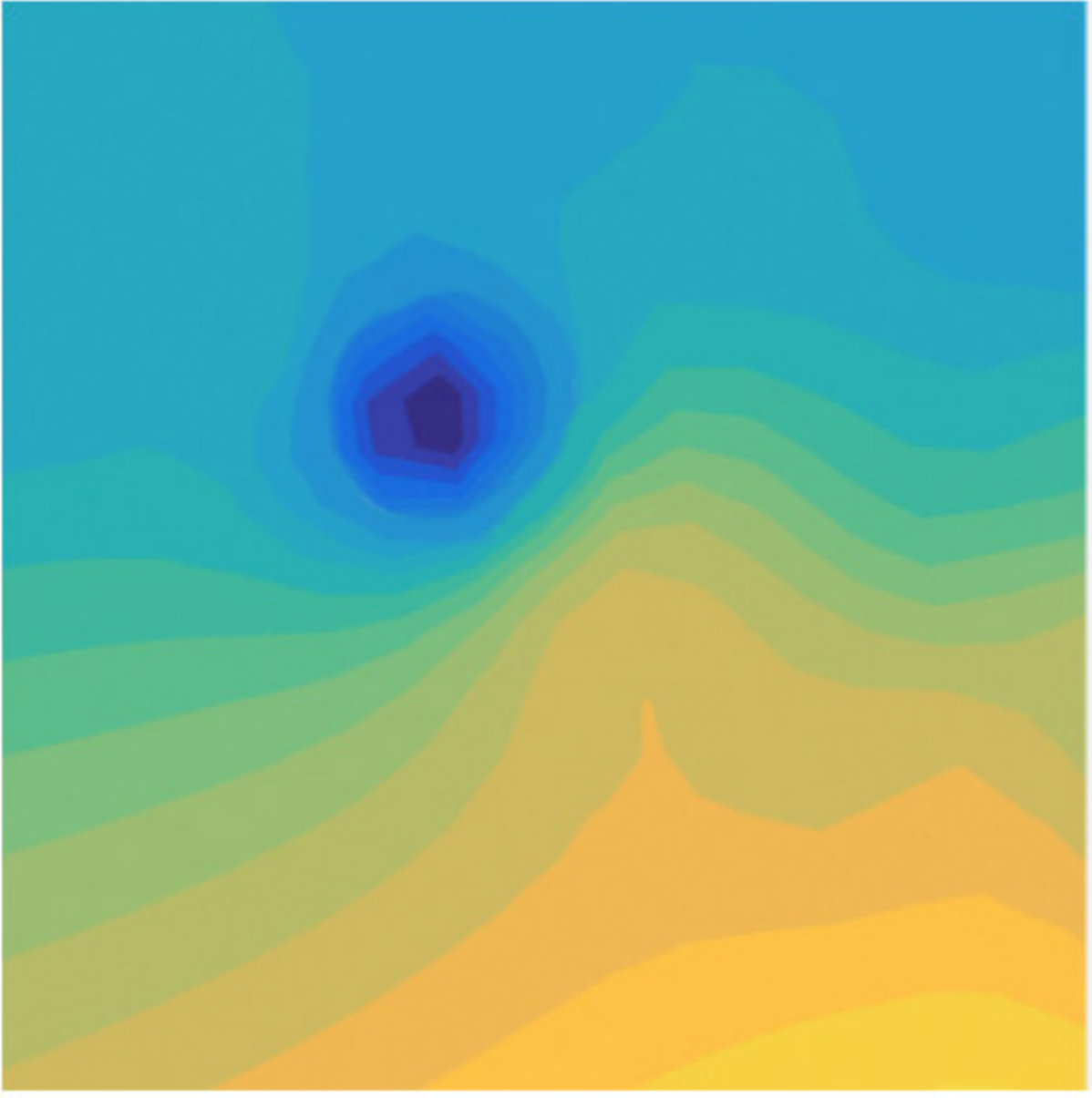}
				\put(00,90){\text{(d)}}
		\end{overpic} }
		\fbox{\begin{overpic}[width=0.425\linewidth]{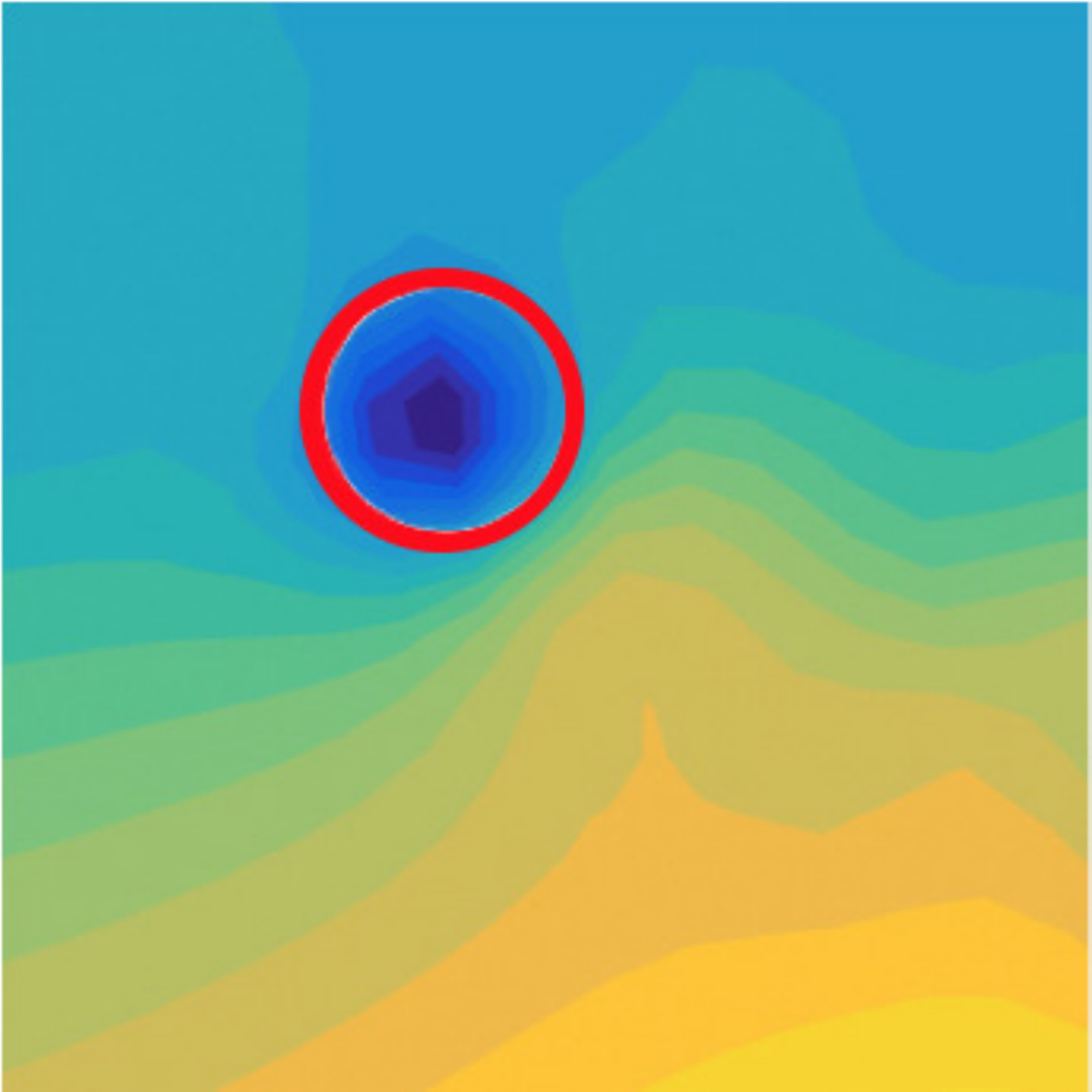}
				\put(00,90){\text{(e)}}
				\put(66,38){\color{white}{\line(-1,1){33} }}
				\put(60,28){\text{$2r$}}
		\end{overpic} }
		\caption{Vortex quantification method: 
			(a) Vortex identification using Q-criterion. (b) Defining vortex spine. (c) Planes perpendicular to the spine. (d) Vorticity magnitude on a sample plane. (e) Local vortex radius. In (a), the blue surface is the air-water interface.  In (b) to (e), blue is a contour level of the vorticity magnitude. }
		\label{fig:vortexQuantMethod}
	\end{figure}

	\noindent
	\textbf{Step 1:} Identify an entrainment event.\\
	\textbf{Step 2:} Visually identify the vortex responsible for entrainment by setting the appropriate value for $Q_c$ (Figure \ref{fig:vortexQuantMethod}a).\\
	\textbf{Step 3:} Define the centerline of the vortex by the line of minimum pressure along the length of the vortex (Figure \ref{fig:vortexQuantMethod}b).\\
	\textbf{Step 4:} Create planes perpendicular to and along the spine (Figure \ref{fig:vortexQuantMethod}c) and calculate the vorticity magnitude on the planes.\\
	\textbf{Step 5:} For each plane, identify the isoline for $\omega_c$.\\
	\textbf{Step 6:} For each plane calculate the circulation within the area designated by the $\omega_c$ isoline. Define $\Gamma$ as the average of circulation across all planes.\\
	\textbf{Step 7:} Find the maximum distance from the vortex center to points of the $\omega_c$ isoline for each plane ($r$). The average of $r$ across all planes as the average radius of the vortex $R$ which is also the local length scale of the vortex.\\
	\textbf{Step 8:} Define local Reynolds, Froude and Weber numbers as $Re_l=\frac{\Gamma}{\nu}$, $Fr_l=\frac{\Gamma}{\sqrt{gR^3}}$ and $We_l=\frac{\rho\Gamma^3}{\sigma R}$, respectively.

\begin{figure}[!htb]
		\centering
		\begin{overpic}[width=0.8\linewidth]{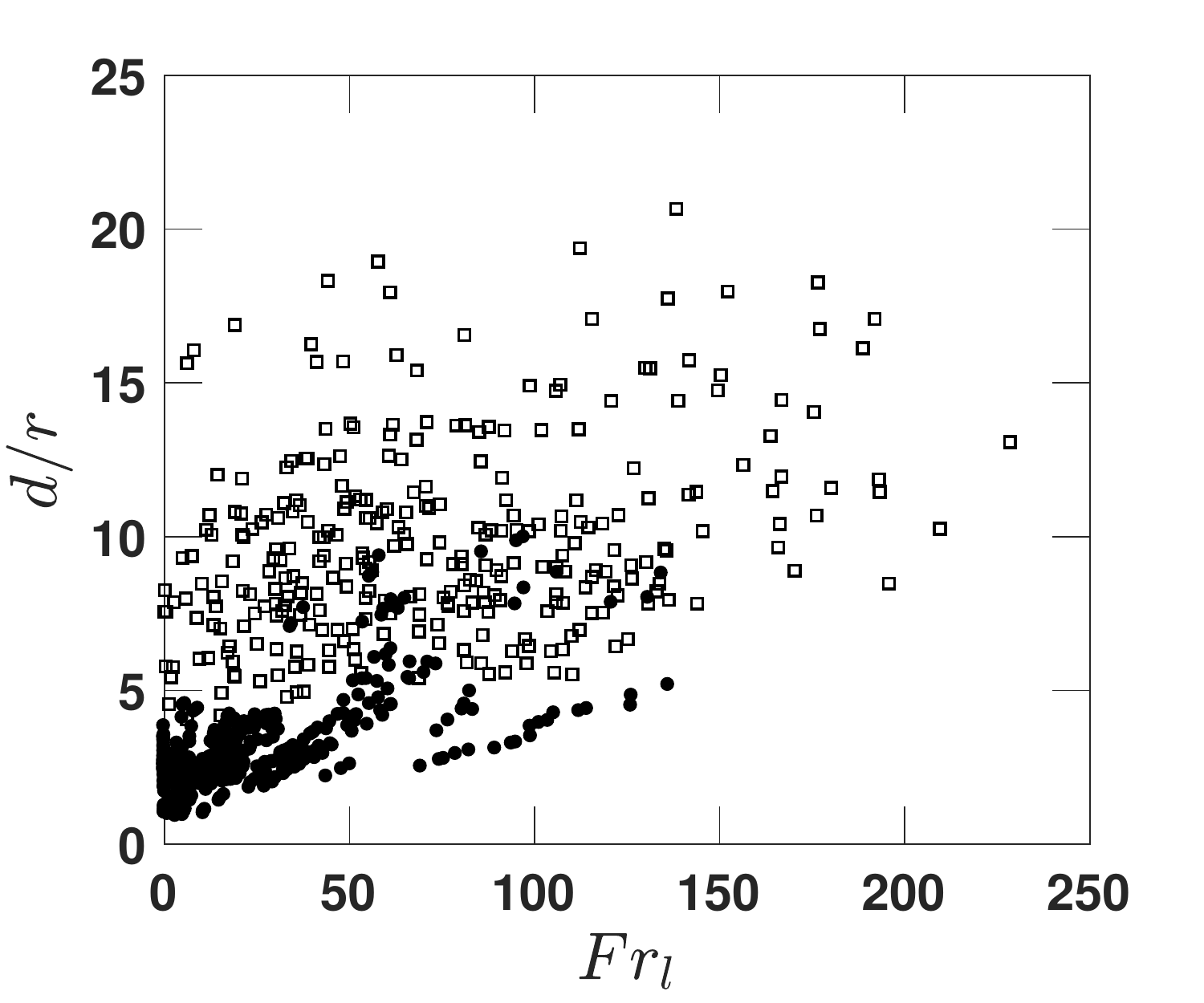} 
			\put(15,70){\text{(a)}}
		\end{overpic}
		
		\begin{overpic}[width=0.8\linewidth]{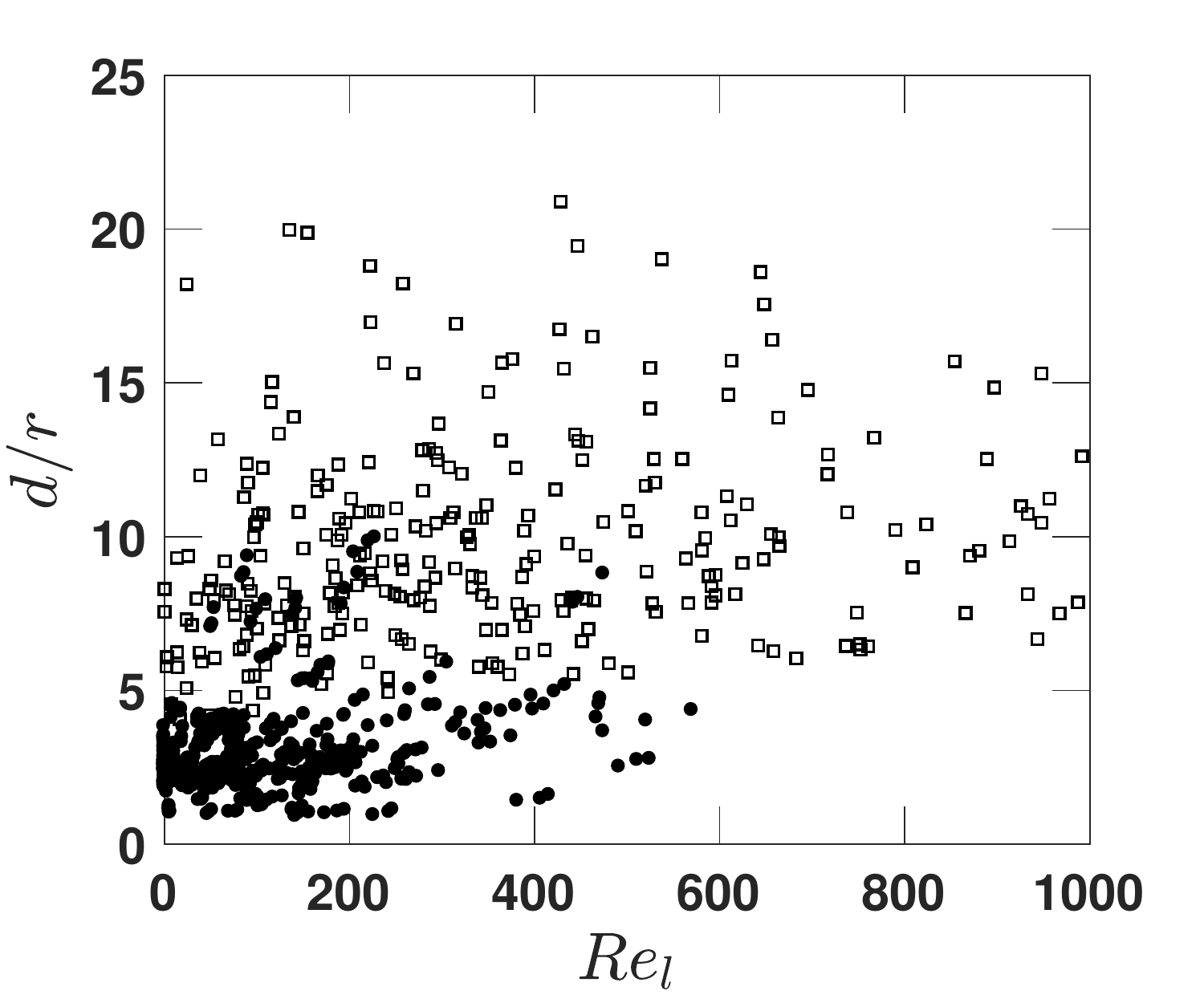}
			\put(15,70){\text{(b)}}
		\end{overpic}
		
		\begin{overpic}[width=0.8\linewidth]{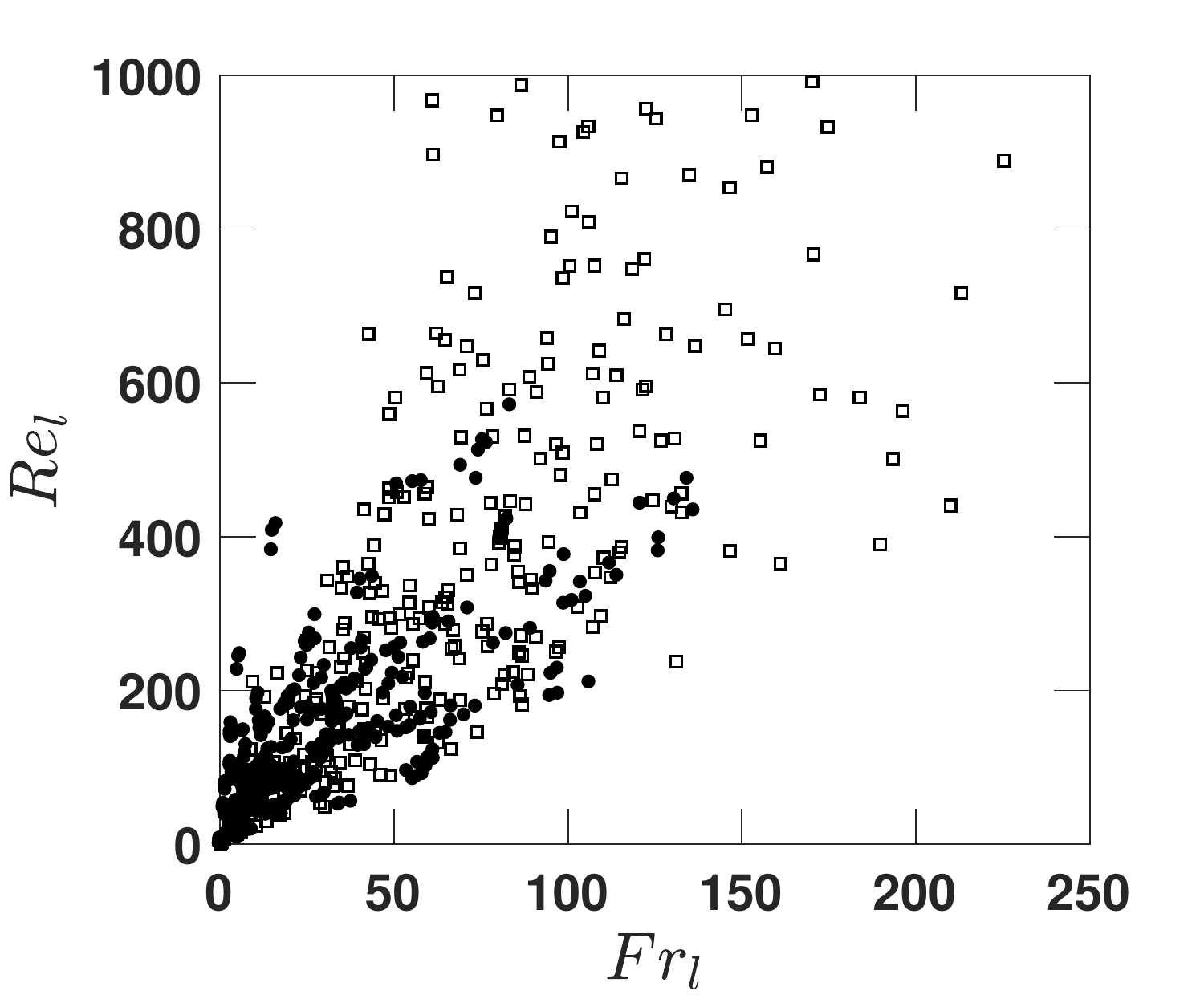}
			\put(20,70){\text{(c)}}
		\end{overpic} 
		
		\caption{Entrainment parametric space: (a) $Fr_l$ vs. $d/r$ (b) $Fr_l$ vs. $Re_l$ (c) $d/r$ vs. $Re_l$. \textbullet, entraining; \square, non-entraining.}
		\label{fig:parameterSpace}
	\end{figure}

\begin{figure}[!htb]
		\centering
		\begin{overpic}[width=0.8\linewidth]{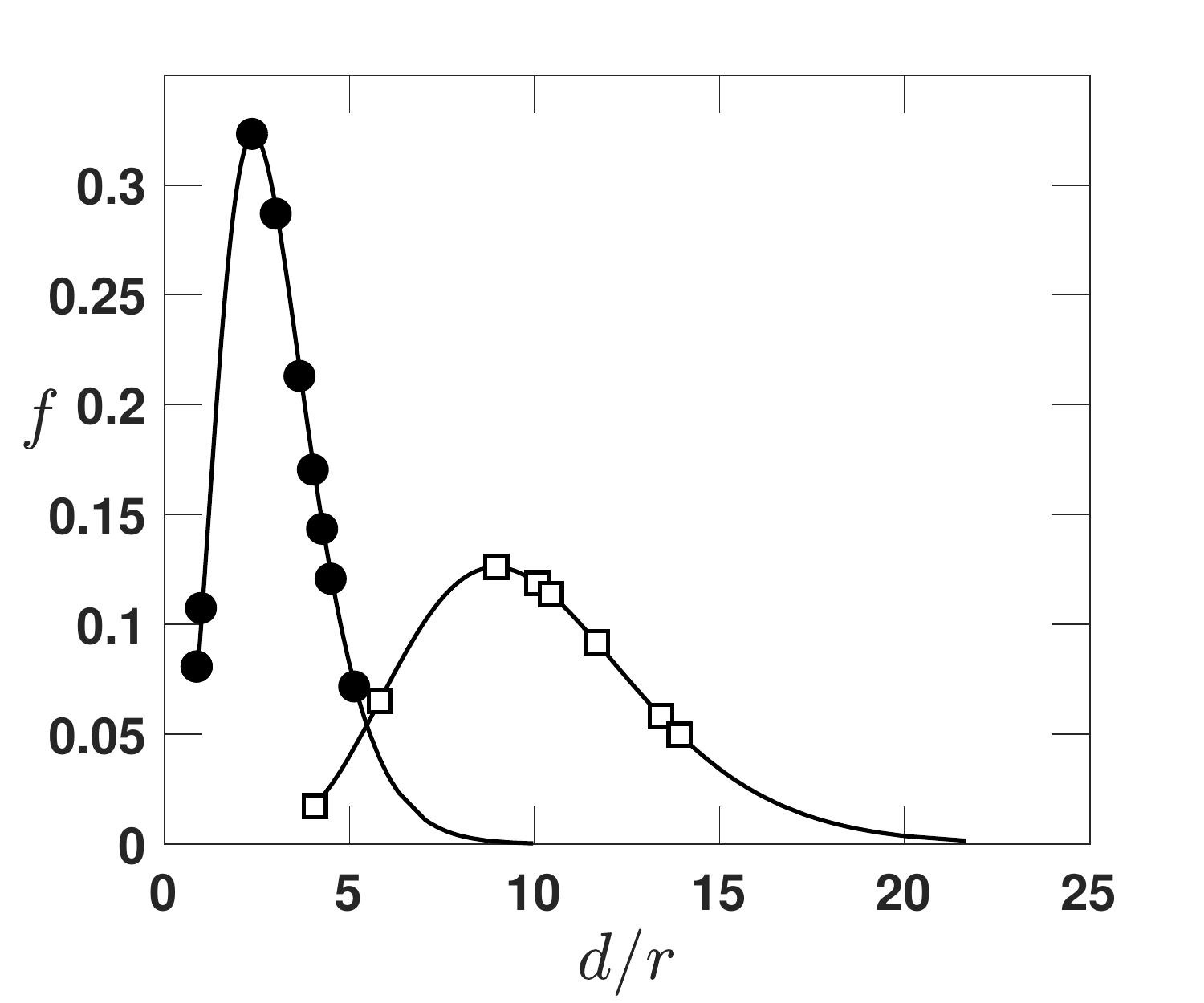} 
			\put(78,70){\text{(a)}}
		\end{overpic}
		
		\begin{overpic}[width=0.8\linewidth]{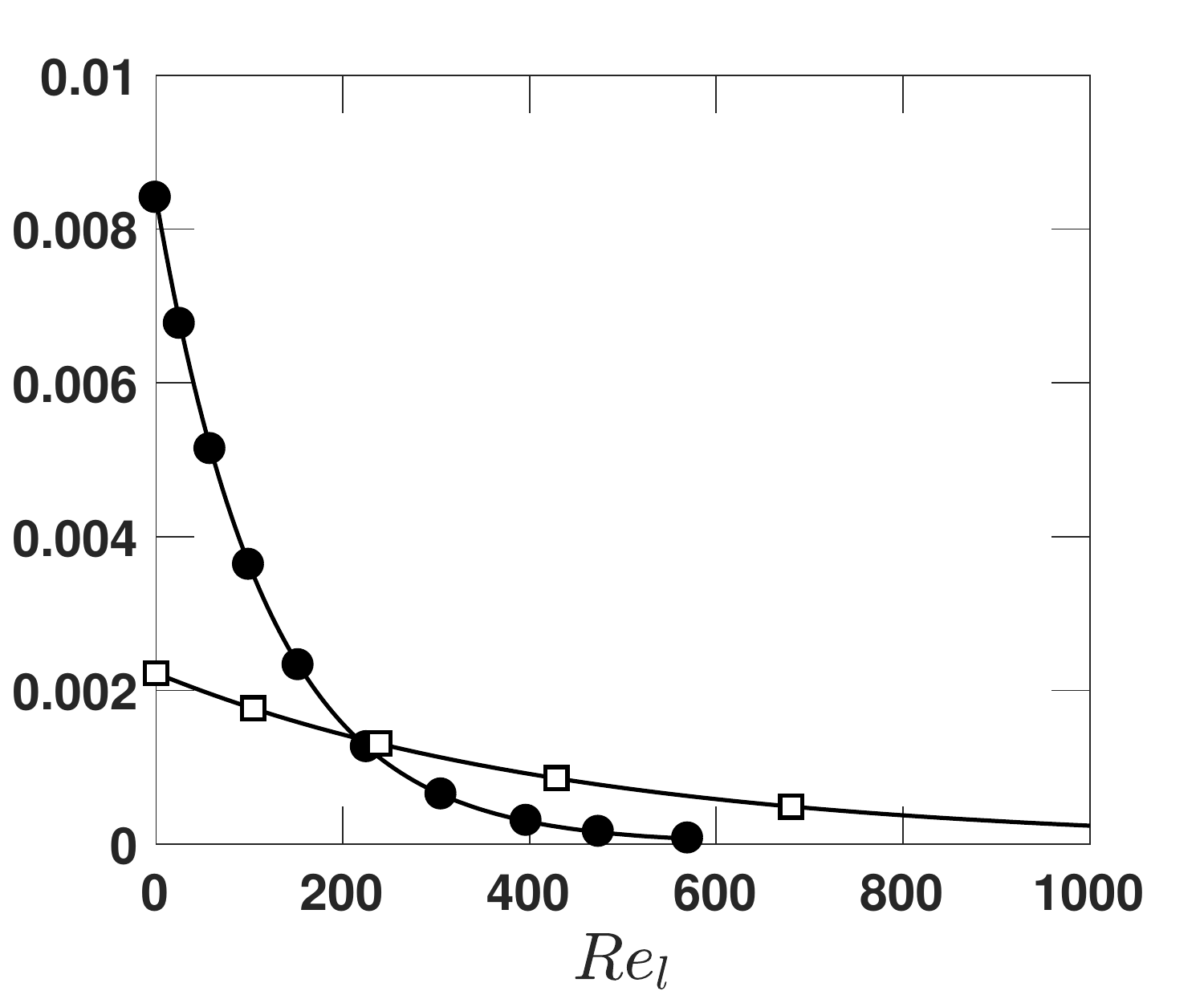}
			\put(78,70){\text{(b)}}
		\end{overpic}
		
		\begin{overpic}[width=0.8\linewidth]{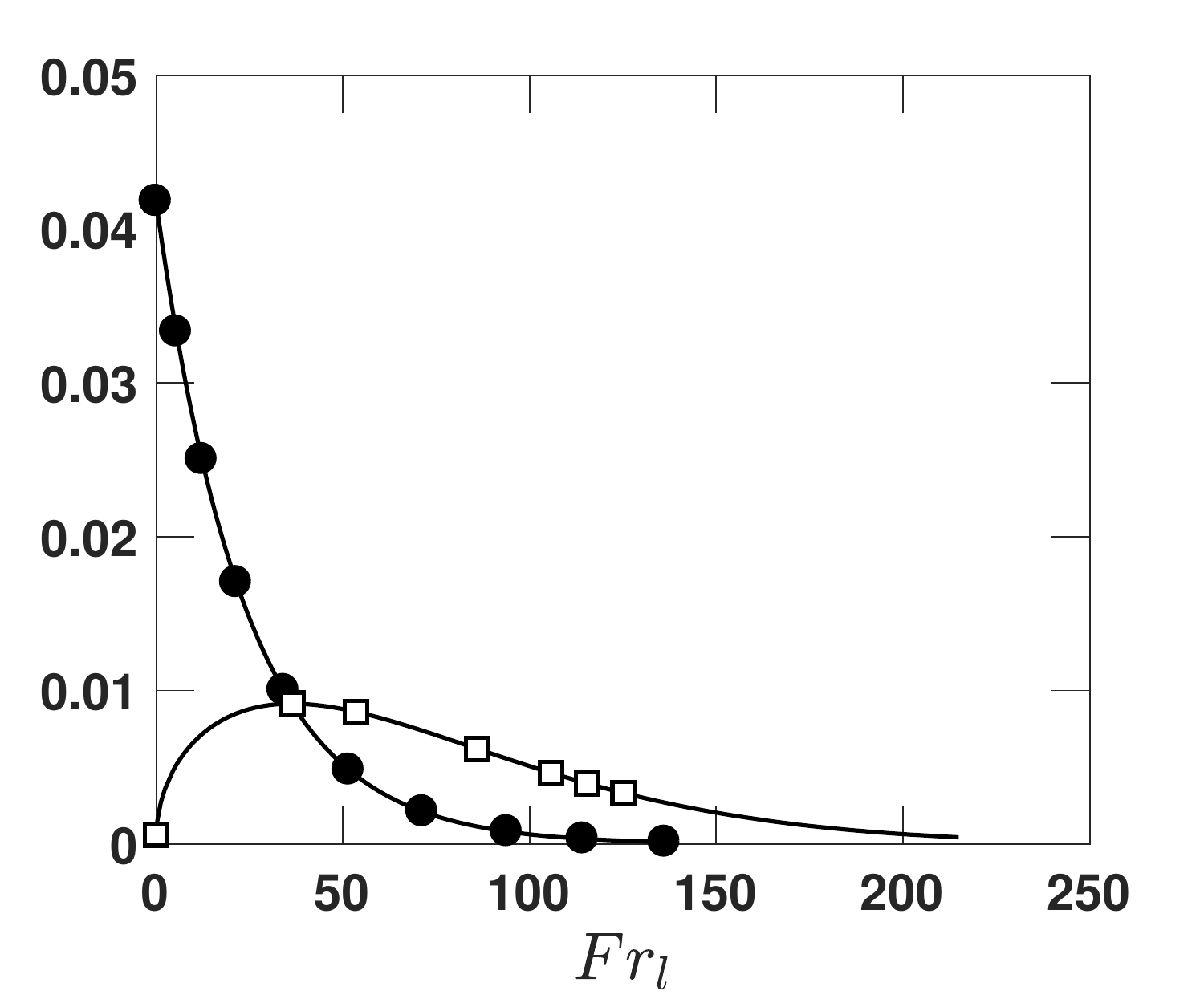}
			\put(78,70){\text{(c)}}
		\end{overpic} 
		
		\caption{Probability density function: (a) $d/r$  (b) $Re_l$  (c) $Fr_l$. \textbullet, entraining; \square, non-entraining.}
		\label{fig:pdf}
	\end{figure}

	These three dimensionless groups along with the distance of the vortex from the free surface are the independent variables of turbulent entrainment. Owing to the small value of surface tension, the typical local Weber numbers are very large (O($10^4$)) compared to the other three variables and will not be discussed further. We also non-dimensionalized the distance of the vortex from the surface $d$ using the radius of the vortex. As a result we have a three-dimensional parametric space $(d/r,Re_l,Fr_l)$ for all the entraining and non-entraining vortices. By placing these vortices on the parametric space, we seek to find the dominant factors separating the two types of vortices. We gathered data for 447 entraining and 450 non-entraining vortices for all computations at the two aforementioned Froude numbers. Figure \ref{fig:parameterSpace}a shows the scatter plot of normalized distance against local Froude number for both entraining and non-entraining vortices. Most of the entraining vortices are clustered in the corner with $Fr_l<50$ and their population thins out as Froude number is increased. Non-entraining vortices however are spread much more uniformly with 
	respect to the Froude number. For $d/r>6$ only a few entraining vortices exist whereas non-entraining vortices are numerous. The opposite is true for $d/r<6$ indicating that a vortex is not able to entrain air if it sits more than six times its own radius below the surface. This is irrespective of Froude number or the Reynolds number as can be seen in Figure \ref{fig:parameterSpace}b. The scatter plot of Reynolds number against Froude number is depicted in \ref{fig:parameterSpace}c. There is a significant overlap of entraining and non-entraining vortices and therefore the $(Re_l,Fr_l)$ pair cannot predict entrainment. It is worth noting that most of the entraining vortices collocate in the corner of the plot and tend to assume smaller values of $Re_l$. 
	Figure \ref{fig:pdf}a depicts the probability density function of the normalized distance $d/r$. The entraining vortices are closer to the surface with a mean of 3.01 and standard deviation of 1.55. The non-entraining vortices have a mean equal to 9.85 with a standard deviation of 3.25. The two density functions cross at $(5.45,0.05)$ and do not show significant overlap. The value $d/r=5.45$ seems reasonable as the critical distance for entrainment.
	Figure \ref{fig:pdf}b shows the probability density function of the local Reynolds number $Re_l$. The two profiles cross at $Re_l\approx 200$ and there is significant overlap of the two distributions for Reynolds numbers smaller than this value. However, for Reynolds numbers larger than 200, the overlap is very small in comparison and the probability of finding entraining vortices drops dramatically, much faster than the non-entraining vortices.  Figure \ref{fig:pdf}c shows the probability density function of $Fr_l$. The curves coincide at $Fr_l\approx 37$ and there is significant overlap below and above this value. Both entraining and non-entraining vortices cluster towards the smaller Froude numbers however this is more pronounced in the case of the entraining vortices.

\begin{figure}[!htb]
		\centering
		\includegraphics[width=\linewidth]{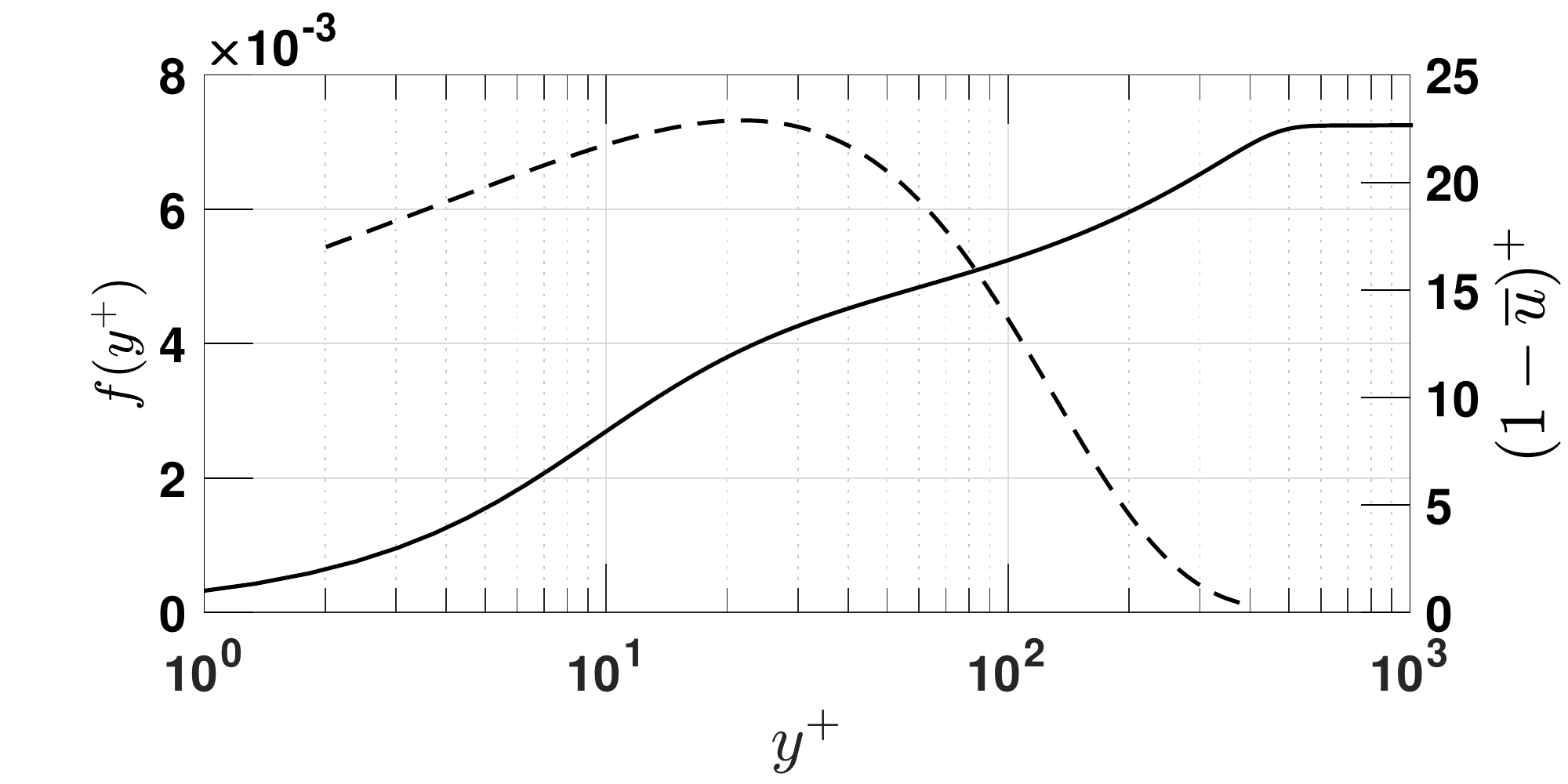} 
		\caption{Wall-normal distance PDF for entraining vortices. Deep water mean velocity is given for comparison.}
		\label{fig:zplusPDF}
	\end{figure}

	Figure \ref{fig:zplusPDF} depicts the probability distribution function for the wall-normal location of the entraining vortices. The deep water velocity profile has also been included for comparison. The mean is at $z^+\approx100$ and the standard deviation is about 83 meaning that the majority of the entraining vortices reside in buffer layer and log-law regions of 
	the boundary layer.

	\section{CONCLUSIONS}
	
	In this research, the problem of the interaction of a turbulent boundary layer with a free surface is studied experimentally using a laboratory-scale device and a DNS simulation of a similar problem. The experimental device utilizes a stainless steel belt, driven by two powered vertically oriented rollers, as a surface piercing vertical wall of infinite length. This belt accelerates in under 0.7 seconds to constant speed $U$ in an effort to mimic the sudden passage of a flat-sided ship. Utilizing the full length and velocity scales of large naval ships, this device creates a temporally-evolving boundary layer analogous to the spatially-evolving boundary layer along the length of a ship, using the transformation $x=Ut$, where $x$ is distance from the leading edge and $t$ is time. Water surface profiles measured along lines perpendicular and parallel to the belt surface were recorded with a cinematic LIF system to study the generation of surface height fluctuations by the sub-surface turbulence. Entrained bubbles were measured using a high-speed camera setup which was able to  measure and track bubbles down to a radius  of $r = 0.5$~mm. To complement the experiments, DNS of the two-phase boundary layer problem were carried out where a section of the belt was considered. The boundary conditions of the computational domain are similar to that of the experiments. The DNS results allow access to the entire flow field that is otherwise inaccessible through experiments.
	
	From qualitative observation of the free surface profiles in the LIF movies with the light sheet parallel to the belt, it was found that surface features that resemble breaking water waves  travel downstream, parallel to the belt surface. Two of these surface features are observed in detail qualitatively and it is hypothesized that they are one of the mechanisms through which air is entrained into the free surface boundary layer. The speed of these free surface features is measured parallel to the belt and it is found that they move about three times faster than similar features moving away from the belt.  Also, it is found that the downstream speed of these features decays quickly when they are measured further away from the belt. The experimental free surface profiles are compared to the computational results and are found to agree qualitatively.
	
	Experimental measurements of bubbles are also reported. The bubble size spectrum is found to have a break in slope at around $r = 2.46$ mm with two characteristic linear regions. The break in slope suggests a Hinze-like scale that dominates the bubble radius spectrum. The experimental results are compared to the bubble radius spectrum from the DNS and they seem to agree qualitatively. The number of bubbles per depth increment is  found to decrease with depth, in agreement with the  DNS results. The mean bubble radius and mean bubble speed versus depth are found to t decrease slowly  with depth.
	
	Three entrainment mechanisms, breaking waves, vertically oriented vortices, and horizontally oriented vortices, are studied in detail from the DNS results. It is found that horizontally oriented vortices account for $\approx 88 \%$ of the entrainment events in the DNS, while entrainment events by breaking  waves and vertically oriented vortices account for $\approx 11 \%$  and $< 1 \%$ of the total entrainment events, respectively.
	
	The support of the Office of Naval Research under grant number N000141712081 (Program Managers: Ki-Han Kim and Thomas Fu) is gratefully acknowledged.

	\bibliography{Library} 
	\bibliographystyle{29ONR}

	\subsection{Discussion}
	
	\noindent\textbf{Discusser I:}
	
	\textbf{Discusser:} This is an excellent paper that presents an experimental and numerical study of air entrainment and surface fluctuations in a developing boundary layer. Besides the interest on a fundamental fluid mechanics problem, the work is significant in that provides further insight into entrainment through the boundary layer/free surface contact line and may help develop appropriate models for air entrainment.
	
	\textbf{Reply}:  We thank for the discusser for his careful reading of our paper and his insightful questions and comments.

	\textbf{Discusser:} Detailed questions and comments:
	\begin{enumerate}
		\item \textbf{Comment/Question}: 
		It would be desirable to add to the introduction some discussion on how the work in this paper and other contributions by the authors could help development of air entrainment models for bubbly wake applications.
		
		\textbf{Reply}:  Thank you for pointing this out.  We have added some references to the bibliography and made a few modifications along these lines to the beginning of the introduction.

		%
		%
		
		%
		
		\item  \textbf{Comment/Question}: The discussion on air entrainment mechanisms is interesting. Can the authors really resolve Mesler entrainment, to the extent of 12~\% of the bubbles? Bubbles resulting from impact are usually very small.
		
		\textbf{Reply}: This portion of the bubbles is mainly due to the larger cavities collapsing and entrapping large blobs of
		air which subsequently break up. We rarely see entrainment due to narrow columns of splashing water.
		
	\end{enumerate}

	\noindent\textbf{Discusser II:}
	
	We thank for the discusser for his careful reading of our paper and his insightful questions and comments.

	\noindent \textbf{Detailed comments and questions:}
	
	\begin{enumerate}
		\item \textbf{Comment/Question}: The authors mention these experiments should be carried out in salt water.   Were these experiments done in fresh or salt water, unclear.

		\textbf{Reply}:  All the experiments presented in this paper were carried out in filtered fresh water. The paper  was modified to clarify this point.  Once a complete set of bubble measurements is performed in fresh water, we  hope to perform an identical set of measurements in salt water.

		\item  \textbf{Comment/Question}: The authors state: It is worth noting that most of the entraining vortices collocate in the corner of the plot and tend to assume smaller values of $Re_l$.   However, from Fig 27b, it looks like at higher $Re$ ($ > 600$) almost no vortices have $d/r<6$.    It may be worth commenting/discussing this.  Is this due to the nature of higher $Re$ or could your grid resolution be playing into it?
		
		\textbf{Reply}:  The absence of vortices in that area of the plot ($d/r < $ 6 ; $Re > $ 600) is an issue still under investigation. 
		We are at the moment acquiring additional data to rule out the the possibility that this is due to a small dataset. The resolution of the grid is unlikely to be the issue since we keep the near-interface region at the finest refinement level at all times.
		
	\end{enumerate}
\end{document}